\documentclass[10pt,aps,prb,longbibliography,twocolumn]{revtex4-2}

\usepackage{array}
\usepackage{amsmath}
\usepackage{comment}
\usepackage{amssymb} 
\usepackage{amsthm} 
\usepackage{textcomp}  
\usepackage{upgreek}
\usepackage{units}
\usepackage[pdftex]{graphicx}
\usepackage{microtype}
\usepackage{enumitem}
\usepackage{xcolor}
\usepackage{gnuplottex}    
\usepackage{multirow} 
\usepackage{sansmath}
\usepackage{txfonts}
\usepackage{float}
\usepackage{adjustbox}
\usepackage{subcaption}
\usepackage{tabularx}
\usepackage{booktabs}

\newcommand{\ket}[1]{\left| #1 \right>} 
\newcommand{\bra}[1]{\left< #1 \right|} 
\renewcommand{\vec}[1]{\ensuremath{\boldsymbol{#1}}}

\usepackage{hyperref}
\hypersetup{
    colorlinks=true,
    linkcolor=blue,
    filecolor=magenta,      
    urlcolor=cyan
    }

\usepackage{ragged2e}
\usepackage[font=small,labelfont=bf]{caption}
\DeclareCaptionJustification{justified}{\justifying}
\usepackage{orcidlink}

\begin{document}

\title{Angular momentum splitter effect of \texorpdfstring{$d$}{d}-wave axial phonons in orbital altermagnets}

\author{Dimos Chatzichrysafis\, \orcidlink{0009-0000-6755-4057}}
\affiliation{Institute of Physics, Johannes Gutenberg University Mainz, 55128 Mainz, Germany}

\author{Alexander Mook\,\orcidlink{0000-0002-8599-9209}}
\affiliation{Institute of Solid State Theory, University of Münster, 48149 Münster, Germany}

\begin{abstract}
    We theoretically demonstrate that axial phonons, lattice vibration quanta carrying finite angular momentum, can host a $d$-wave angular momentum texture in orbital altermagnets in the absence of spin-orbit coupling. We consider a minimal electronic tight-binding model with $d$-wave loop-current order that breaks time-reversal symmetry. Within the Born-Oppenheimer approximation, we incorporate electron-phonon coupling via the molecular Berry curvature and show that the underlying $d$-wave orbital magnetic moment texture of the electronic state is transferred to the phonons without requiring the relativistic spin-orbit coupling. Our results expand the range of platforms available for engineering axial phonons and point to functionality unique to $d$-wave textures, including angular-momentum Seebeck and splitter effects, corresponding to longitudinal and transverse angular-momentum currents driven by a temperature gradient.
\end{abstract}

\maketitle 

\section{Introduction}
Axial (or chiral) phonons are quantized lattice vibrations that carry finite angular momentum \cite{Juraschek2025}. They have attracted significant attention not only for fundamental reasons but also for their application potential \cite{zhang2025advancesphononsbandtopology, Zhang2025, doi:10.1021/acs.nanolett.4c00606, chen2026topologicalphononics}. Furthermore, they have emerged as a unifying theme across diverse areas of condensed matter physics, including ultra-fast dynamics \cite{Mrudul_2025, Tauchert2022, Davies2024}, magnetism \cite{Luo2023, PhysRevMaterials.3.064405, PhysRevB.110.094401, PhysRevB.111.134414}, orbitronics \cite{yao2026dynamicalorbitalangularmomentum, nabei2026orbital}, strongly correlated systems \cite{bfll-sdrb, Grissonnanche2020, tyf9-thv8} and topological matter \cite{PhysRevLett.127.125901}. Consequently, substantial effort has been devoted to understanding their microscopic origin. While external stimuli such as magnetic fields or laser pulses can induce rotational motion in phonons \cite{THE_Lifa_Zhang, Lifa_phonon_De_Hass_effect, PhysRevB.103.214302, Romao2024}, it is particularly compelling to identify intrinsic, material specific mechanisms that generate such angular momentum.

To understand how axial phonons can arise intrinsically, we recall that a finite angular momentum $\vec{J}_{\vec{k}}$ of a quasiparticle requires the breaking of either inversion or time-reversal symmetry \cite{PhysRevB.108.134307}. In non-centrosymmetric but time-reversal symmetric systems, angular momentum is odd in momentum $\vec{J}_{\vec{k}} = -\vec{J}_{-\vec{k}}$. Such phonons have been predicted to occur in kagome \cite{PhysRevB.100.094303} and hexagonal lattices \cite{PhysRevLett.115.115502} and were subsequently observed in the latter through transient infrared spectroscopy \cite{doi:10.1126/science.aar2711}. They also exhibit experimental signatures in thermal transport measurements, such as those performed in quartz \cite{PhysRevLett.132.056302, nabei2026orbital}. By contrast, realizing axial phonons with even parity, characterized by $\vec{J}_{\vec{k}} = \vec{J}_{-\vec{k}}$, requires broken time-reversal symmetry while preserving inversion symmetry. This scenario is more subtle because phonons cannot intrinsically break time-reversal symmetry; instead, they acquire effective time-reversal symmetry breaking through coupling to a suitable magnetic background.

One route to breaking time-reversal symmetry is through coupling the lattice to electronic degrees of freedom. In the context of axial phonons, a prominent framework is based on the molecular Berry curvature (MBC) \cite{RevModPhys.64.51, PhysRevLett.113.263004, PhysRevLett.126.225703, PhysRevX.15.011036}. Within the Born-Oppenheimer approximation, the electronic wave function depends parametrically on the lattice coordinates. Adiabatically evolving the set of the lattice coordinates leads to geometric effects, whose gauge-invariant description is provided by the MBC. Acting as an effective real-space gauge field, the MBC modifies the phonon dynamics and can generate even-parity axial phonons when the electronic system breaks time-reversal symmetry \cite{PhysRevB.108.134307}. This mechanism has been explored in both model studies \cite{7f8x-fkx9,PhysRevX.15.011036, Chiral_Phonons_Angular_Momentum, MBP_Phonon_THE_Nagaosa, li2025phonondichroismsrevealingunusual, PhysRevLett.119.075301, PhysRevLett.134.206701, das2026antiferrochiralphononsmathcalpmathcaltsymmetricantiferromagnets} and first-principles calculations \cite{tpjd-dh1m, dhakal2025theoryintrinsicphononthermal, PhysRevLett.130.086701, PhysRevX.14.011041, zhang2025generalabinitioframework}. Notably, with the exception of Ref.~\cite{bendin2026dwavephononangularmomentum, wang2026alteraxialphononscollinearmagnets} existing studies have focused on electronic systems compatible with ferromagnetic point groups, where the induced axial phonons exhibit an $s$-wave angular-momentum texture in momentum space and can therefore possess a finite angular momentum at the Brillouin-zone center \cite{tpjd-dh1m, zhang2025generalabinitioframework,Chiral_Phonons_Angular_Momentum}. 

Beyond $s$-wave symmetry, higher-order even-parity angular-momentum textures have recently emerged as a hallmark of altermagnetism \cite{PhysRevX.12.040501, PhysRevX.12.031042}. Altermagnets are magnetically compensated materials whose electronic bands exhibit non-relativistic spin-splitting with a $d$-, $g$-, or $i$-wave form factor \cite{PhysRevX.12.031042}. They have attracted considerable interest for spintronic applications because they combine key advantages of ferromagnets and antiferromagnets \cite{PhysRevX.12.031042}. In particular, their spin-polarized band structure enables non-relativistic transport phenomena such as spin-dependent Seebeck and spin-splitter effects \cite{PhysRevLett.126.127701} \cite{PhysRevB.110.094427}, corresponding to longitudinal and transverse spin (thermal)currents. 

Notably, many established and candidate altermagnets are insulating \cite{PhysRevX.12.031042, mmdm-hrj4}, placing their magnetic excitations---magnons---at the center of thermally driven spin transport. These magnons inherit the unconventional spin-splitting textures of the underlying electronic structure \cite{fgc1-5blp, PhysRevX.12.031042, PhysRevLett.131.256703, PhysRevLett.131.186702, PhysRevLett.133.156702}. Recent theoretical work \cite{bendin2026dwavephononangularmomentum, wang2026alteraxialphononscollinearmagnets} further demonstrated that relativistic spin-orbit coupling can transfer these textures to phonons, generating corresponding angular-momentum textures. This finding naturally raises the question: \emph{can phonons, as the primary heat carriers in insulating materials, acquire beyond s-wave even-parity angular momentum textures through mechanisms that are not limited by the typically weak scale of relativistic effects?}

In this work, we demonstrate that $d$-wave axial phonons, characterized by a $d$-wave angular momentum texture in reciprocal space, can arise in \emph{orbital altermagnets} without relying on spin-orbit coupling. Orbital altermagnets realize beyond-$s$-wave even-parity symmetries through orbital rather than spin degrees of freedom. In particular, they break time-reversal symmetry in real space via loop-current order, where individual lattice plaquettes carry orbital magnetic moments whose spatial arrangement encodes the underlying altermagnetic symmetry \cite{PhysRevLett.134.146001, chakraborty2025orbitalaltermagnetismkagomelattice, pan2026orbitalaltermagnetism, Yu2025, ydjj-6r9l}. Motivated by this emerging material class, we consider a minimal tight-binding model on the checkerboard lattice that hosts loop currents and exhibits an altermagnetic $d$-wave orbital magnetic moment texture in the electronic spectrum. Because time-reversal symmetry is broken at the real-space level it is directly transferred to the phonons through the molecular Berry curvature without the need for relativistic interactions. Consequently, both optical and acoustic phonon branches acquire a $d$-wave angular momentum texture in reciprocal space. To detect these axial phonons, we propose angular-momentum Seebeck and splitter effects, corresponding to longitudinal and transverse angular-momentum currents driven by a temperature gradient. Our results establish orbital altermagnets as a robust platform for engineering axial phonons beyond $s$-wave symmetry and reveal a direct link between altermagnetic order and phonon angular momentum. 

The rest of the paper is organized as follows. In Sec.~\ref{main sec: electrons}, we present a minimal tight-binding model of an orbital altermagnet with a $d$-wave loop-current order on the checkerboard lattice. We compute the momentum texture of the electronic orbital magnetic moments, demonstrating their altermagnetic character. In Sec.~\ref{main sec: MBC}, we review the concept of the molecular Berry curvature and specify it to our orbital altermagnet system. 
In Sec.~\ref{main sec: MBC dynamics}, we discuss how the MBC modifies the phonon dynamics of the checkerboard lattice and calculate the phonon angular momentum that also exhibits $d$-wave symmetry. Finally, in Sec.~\ref{main sec: splitter}, we propose the phonon angular-momentum Seebeck and splitter effects as characteristic thermal transport signatures of $d$-wave axial phonons. We conclude in Sec.~\ref{sec:Conclusion} and provide details in several appendices.

\begin{figure}
    \includegraphics[width=1\linewidth]{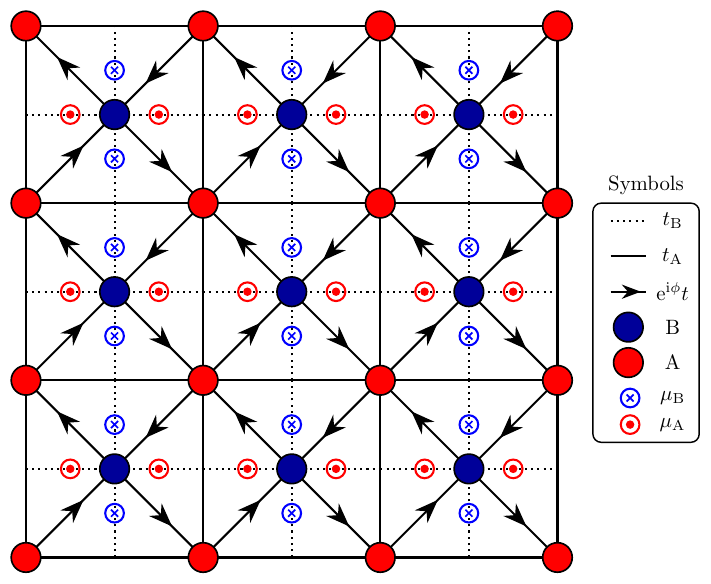}
    \caption{Orbital altermagnet on the checkerboard lattice illustrating the tight-binding Hamiltonian in Eq.~\eqref{eq: TB-H real space}. Nearest-neighbor hopping amplitudes are complex, $\mathrm{e}^{\mathrm{i} \phi}t$, where the arrows indicate the direction corresponding to a positive phase $\phi$. Next-nearest-neighbor hopping amplitudes on the A and B sublattices are denoted by  $t_\text{A}$ and $t_\text{B}$, respectively, with $t_\text{A} \neq t_\text{B}$. The circulating currents can be thought to generate out-of-plane orbital magnetic moments $\mu_{\text{A}}$ and $\mu_{\text{B}}$, which are guides to the eye and make the $d$-wave symmetry apparent.}
    \label{fig:checkerboard lattice}
\end{figure}

\section{Orbital \texorpdfstring{$d$}{d}-wave altermagnet: a minimal tight binding model}
\label{main sec: electrons}

We begin by introducing the electronic system and considering a tight-binding Hamiltonian on the checkerboard lattice that captures the orbital altermagnet with $d$-wave loop-current order shown in Fig.~\ref{fig:checkerboard lattice}. In real space, the Hamiltonian takes the form
\begin{equation}
    H_{\text{el}} = -t\sum_{\langle ll' \rangle} \sum_{\substack{\kappa, \kappa'}} \mathrm{e}^{\mathrm{i} \phi_{l \kappa}^{l' \kappa'}} c_{l\kappa}^{\dagger} c_{l' \kappa'}  -\sum_{\langle \langle ll' \rangle \rangle} \sum_{\kappa} t_{\kappa} c^{\dagger}_{l \kappa} c_{l' \kappa} + \sum_{l, \kappa} V_{\kappa} c^{\dagger}_{l \kappa} c_{l\kappa}, 
    \label{eq: TB-H real space}
\end{equation}
where $c^\dagger_{l \kappa}$ and $c_{l \kappa}$ are spinless canonical fermionic creation and annihilation operators, respectively. Here, $l, l'$ are unit cell indices, and $\kappa, \kappa' = \text{A, B}$ refer to the two sublattices of the system.  We choose the basis vectors of the two sublattices within the unit cell as $\vec{\tau}_\text{A} = 0 $ and $\vec{\tau}_\text{B} = a \left(\hat{\vec{x}} + \hat{\vec{y}}\right) / 2$, where $a$ is the lattice constant, which is set to unity unless stated otherwise. Here, $\hat{\vec{x}}$ and $\hat{\vec{y}}$ are unit vectors along the $x$- and $y$-direction, respectively. Furthermore, $t$ is the nearest-neighbor hopping amplitude, $t_\kappa$ that of the second neighbors, with $t_\text{A} \neq t_\text{B}$, and $V_\text{A} = - V_\text{B} = V$ is a staggered potential. The phase $\phi_{l\kappa}^{l' \kappa'} = \eta_{l \kappa}^{l' \kappa'} \phi$, whose sign $\eta_{l \kappa}^{l' \kappa'} = \pm1$ depends on the direction of the nearest-neighbor hopping, encodes the loop current order.
From here on, we choose for simplicity  $t_{\text{A}}=-t_{\text{B}}$ that renders the system particle-hole symmetric. The numerical values of the parameters used throughout the text  are summarized in Table~\ref{tab:numerical-values}.

We Fourier transform the fermionic operators,
\begin{equation}
\begin{split}
& 
    c_{l \kappa} = \frac{1}{\sqrt{N}} \sum_{\vec{k}} \mathrm{e}^{\mathrm{i} \vec{k} \cdot  \vec{R}_l^0} c_{\vec{k} \kappa}, \\
& 
 c_{l \kappa}^{\dagger} = \frac{1}{\sqrt{N}} \sum_{\vec{k}} \mathrm{e}^{-\mathrm{i} \vec{k} \cdot  \vec{R}_l^0} c_{\vec{k} \kappa}^{\dagger}, 
\end{split}
\label{eq: FT electron operators}
\end{equation}
where $N$ is the total number of unit cells, and $\vec{R}_l^0$ the equilibrium position of the $l$th unit cell.
We obtain the following electronic Hamiltonian in momentum space, 
\begin{equation}
    H_{\text{el}} = \sum_{\vec{k}} \vec{\Psi}^{\dagger}_{\vec{k}} H_{\vec{k}} \vec{\vec{\Psi}}_{\vec{k}}, 
    \label{eq: TB in k-space}
\end{equation}
where 
$ \vec{\vec{\Psi}}_{\vec{k}} = 
\begin{pmatrix}
     c_{\vec{k} \text{A}}  & c_{\vec{k} \text{B}}
\end{pmatrix}^\text{T}
$ is the two-dimensional sublattice basis vector. The Bloch Hamilton kernel 
\begin{equation}
    H_{\vec{k}} 
     = 
    \begin{pmatrix}
        \lambda_{\vec{k}} & \Delta_{\vec{k}} \\
        \Delta_{\vec{k}}^* & -\lambda_{\vec{k}}
    \label{eq: TB k-space Kernel}
    \end{pmatrix},
\end{equation}
contains
\begin{equation}
\begin{split}
   \lambda_{\vec{k}} &
    = 2 t_\text{A} \left(\cos k_x + \cos k_y   \right) + V
    , \\
    \Delta_{\vec{k}} & = 
    t 
    \left(
    \mathrm{e}^{-\mathrm{i} \phi}
    \left( 
    1 + \mathrm{e}^{-\mathrm{i} \left(k_x + k_y\right)}
    \right)
    + 
    \mathrm{e}^{\mathrm{i} \phi}
    \left(
    \mathrm{e}^{-\mathrm{i} k_x} + 
    \mathrm{e}^{-\mathrm{i} k_y}
    \right)
    \right). 
\end{split}
  \label{eq:def-of-stuff}
\end{equation}

\begin{figure*}
		\centering
\includegraphics[width=\linewidth]{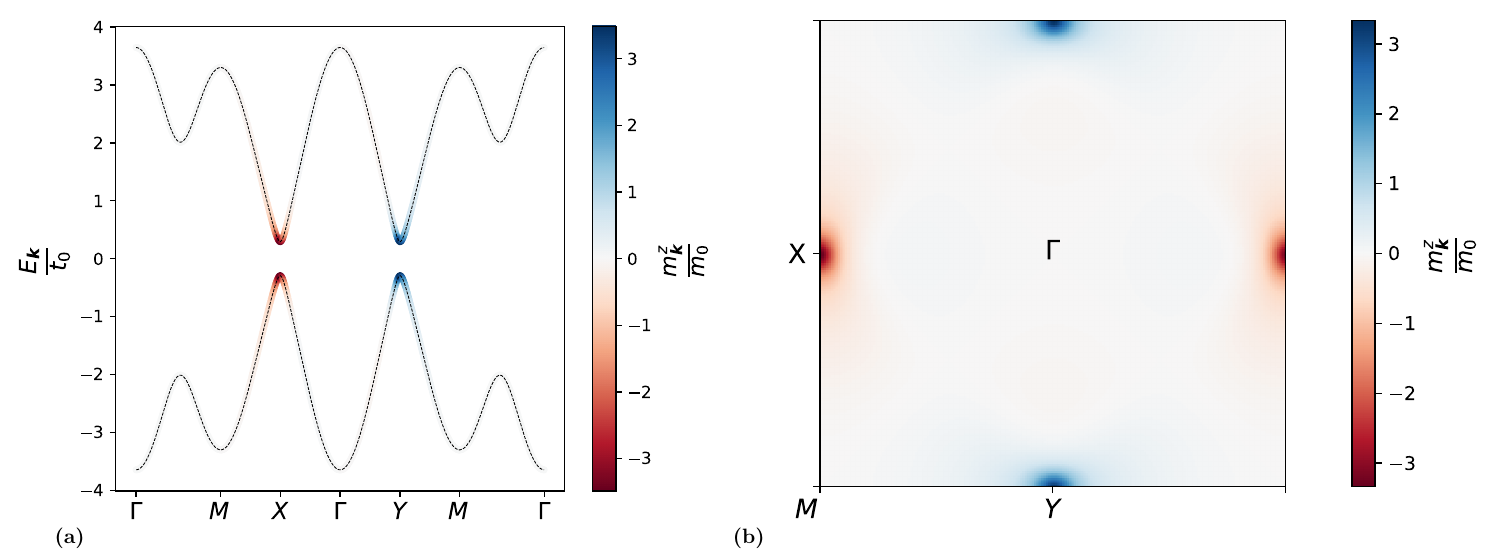}
		\label{result2}
\caption{(a) Electron spectrum plotted in reciprocal space along high-symmetry directions. Dashed lines indicate the electron energies in Eq.~\eqref{eq:electroniceigenvalues} and color indicates the orbital magnetic moment $m_{\vec{k}}^z$. (b) Momentum-resolved orbital magnetic moment $m_{\vec{k}}^z$, with $m_0 = e t_0 a_0^2 / (2c\hbar)$ being the unit magnetic moment.} 
\label{fig: electron spectrum and magnetization}
\end{figure*}
Diagonalizing $H_{\vec{k}}$ provides two electronic bands, with energies
\begin{equation}
    E_{\vec{k} \pm} = \pm \sqrt{\left|\Delta_{\vec{k}} \right|^2 + \lambda_{\vec{k}}^2},
    \label{eq:electroniceigenvalues}
\end{equation}
and corresponding eigenstates 
\begin{equation}
\begin{split}
    & u_{\vec{k} +} =\frac{1}{N_{\vec{k} +}}
    \begin{pmatrix}
        \lambda_{\vec{k}} + E_{\vec{k} +} \\
        \Delta_{\vec{k}}^* 
    \end{pmatrix},
    \\
    & 
     u_{\vec{k}-} =\frac{1}{N_{\vec{k}-}}
    \begin{pmatrix}
        \Delta_{\vec{k}} \\
        -\lambda_{\vec{k}} + E_{\vec{k} -}
    \end{pmatrix}, 
\end{split}
\label{eq: electron eigenstates}
\end{equation}
where $N_{\vec{k} +} = N_{\vec{k} -} = \sqrt{\left|\Delta_{\vec{k}} \right|^2 + \left(E_{\vec{k} +} + \lambda_{\vec{k}}\right)^2 }$ are normalization factors. 

For $V=0$, the spectrum is gapless, with band-touching points at the $X$ and $Y$ high-symmetry points. Since we focus on the insulating regime, we set $V \ne 0$ from here on, as shown in Fig.~\ref{fig: electron spectrum and magnetization}. 

The magnetic point group of $H_\text{el}$ is $4'/mm'm$ in Hermann-Mauguin notation. The following symmetries, which map red sites to red sites and blue sites to blue sites in Fig.~\ref{fig:checkerboard lattice}, 
are important:
\begin{enumerate}
    \item $\mathcal{T} \mathcal{C}_{4z,+}$: time-reversal symmetry $\mathcal{T}$ followed by a $90^{\circ}$ clockwise rotation around the out-of-plane $z$-axis perpendicular to the $xy$ plane,
    \item $\mathcal{T} \mathcal{M}_{x \left(y\right)}$: time-reversal symmetry $\mathcal{T}$ followed by reflection about the $x\left(y\right)$-axis,
    \item $\mathcal{M}_{xy}$: reflection against the $x=y$-axis.
\end{enumerate}
Note in particular that time-reversal $\mathcal{T}$ by itself is not a symmetry. Nor is $\mathcal{PT}$ symmetry (where $\mathcal{P}$ denotes inversion symmetry) as long as the two sublattices are different: $t_\text{A} \neq t_\text{B}$ or $V \ne 0$. 

These symmetries are compatible with a $d$-wave magnetic order. Following Refs.~\cite{li2026pwaveorbitalmagnetism, pan2026orbitalaltermagnetism} we make the $d$-wave texture explicit in momentum space by computing the intrinsic orbital magnetic moments $\vec{m}_{\vec{k} n}$ of the two bands within the modern theory of orbital magnetization \cite{PhysRevLett.95.137205, PhysRevLett.95.137204, THONHAUSER2011}. We assume an insulating zero-temperature limit at half filling, where the lower band with energy $E_{\vec{k} -}$ is fully occupied. In two dimensions, only the out-of-plane component of the intrinsic orbital moment,
\begin{equation}
    m_{\vec{k} n}^z = -\frac{e}{2 c \hbar} \sum_{n' \neq n} \text{Im} \left[v_{nn'}^{x} v_{n'n}^{y} \right] \frac{1}{E_{\vec{k} n}-E_{\vec{k} n'} },
\label{eq: orbital magnetization}
\end{equation} 
is nonzero. Here, $e>0$ is the elementary charge, and $c$ is the speed of light. Additionally, we have defined the velocity matrix elements as $v_{nn'}^{x} = \frac{1}{\hbar}\bra{n} \partial_{k_x} \overline{H}_{\vec{k}}\ket{n'}$ and $v_{n'n}^{y} = \frac{1}{\hbar}\bra{n'} \partial_{k_y} \overline{H}_{\vec{k}}\ket{n}$ with $n, n'= \pm$ and $\ket{n}, \ket{n'}$ being the eigenstates of the tight binding Hamiltonian kernel $\overline{H}_{\vec{k}}$ that is related to the one defined in Eq.~\eqref{eq: TB k-space Kernel} by a unitary transformation,
\begin{equation}
\begin{aligned}
    & \overline{H}_{\vec{k}} = U^{\dagger}_{\vec{k}} H_{\vec{k}} U_{\vec{k}}, \\
    & U_{\vec{k}} =  
    \begin{pmatrix}
        \mathrm{e}^{\mathrm{i} \vec{k} \cdot\left(\vec{\tau}_\text{A} + \vec{\tau}_\text{B} \right) /2 } & 0 \\
        0  &   \mathrm{e}^{-\mathrm{i} \vec{k} \cdot\left(\vec{\tau}_\text{A} + \vec{\tau}_\text{B} \right) /2 }
    \end{pmatrix}.
\end{aligned}
\label{eq: gauge for orbital magnetization}
\end{equation} 
This transformation is required because the intrinsic orbital magnetic moment in Eq.~\eqref{eq: orbital magnetization} is formulated in terms of the lattice-periodic part of the Bloch wave function. Within the tight-binding formalism, this quantity is obtained by diagonalizing the Bloch Hamiltonian constructed using the Fourier transform in Eq.~\eqref{eq: FT electron operators} only when the basis-atom positions $\vec{\tau}_{\kappa}$ are included in the Fourier phase.
We therefore use \(\overline{H}_{\vec{k}}\) and its eigenstates when evaluating the velocity matrix elements entering Eq.~(8) and retain the Fourier-transform gauge convention in Eq.~\eqref{eq: FT electron operators} for the remainder of the paper, since it simplifies our numerical evaluation of the MBC.

We plot $m_{\vec{k} +}^z = m_{\vec{k} -}^z = m_{\vec{k}}^z$ within the first Brillouin zone in Fig.~\ref{fig: electron spectrum and magnetization}(a) along high-symmetry lines. The $d$-wave texture becomes fully apparent in the full momentum dependence of $m_{\vec{k}}^z$ shown in Fig.~\ref{fig: electron spectrum and magnetization}(b).

\begin{table}[htbp]             
    \centering
    \caption{Numerical values of all physical parameters used in this work.}
    \label{tab:numerical-values}
    \begin{tabular}{c c}            
        \toprule
        Physical quantity & Numerical value \\
        \midrule
         Electron energy scale & $t_0=1 \text{ eV}$ \\[4pt]
        Ionic mass scale& $M_0 = 10 ^{-25} \text{ kg}$ \\[4pt]
        force constant scale & $n_0 = 0.1 \text{ N/m}$ \\[4pt]
        Lattice constant scale & $a_0 = 1\text{ \AA}$\\[4pt] 
        Phonon energy scale & 
        $\varepsilon_0 = \hbar \sqrt{\frac{n_0}{M_0}} = \hbar \omega_0 = 1  \text{ meV}$ \\ [4pt]
        Electron-phonon coupling scale & $t_0' = \frac{t_0}{a_0} = 1\text{ eV}/\text{\AA}$ \\ [4pt] 
        Nearest-neighbor hopping &$\displaystyle t/t_0 = -1$ \\[4pt]
        Second-neighbor hopping & $\displaystyle t_\text{A} /t_0=-t_{\text{B}}/t_0 = 0.5$
        \\[4pt]   
        Phase angle & $\phi = \pi / 4$ \\ [4pt] 
        On-site potential &$V/t_0=0.3$ \\ [4pt]
        Ionic masses &$M_\text{B}/M_0=2, \; M_{\text{A}}/M_0=1$ \\[4pt]  
        Nearest-neighbor force constants &$n_{11}/n_0 =-10^{2}, \; n_{12}/n_0=-50$ \\[4pt]
        Second-neighbor force constants &$\gamma_{11}/n_0 =-10$, $\gamma_{22}/n_0 =-30$\\[4pt]
        Electron-phonon coupling &$t'/ t'_0 =1$ \\
        \bottomrule
    \end{tabular}
\end{table}

\section{Molecular Berry curvature in d-wave orbital altermagnets} 
\label{main sec: MBC}

To describe phonon dynamics in an interacting electron-nuclear system, one typically employs the Born-Oppenheimer (BO) approximation.
More specifically, one assumes that due to the large difference in the energy scales of the two subsystems, the electrons follow changes brought to the lattice configuration $\left \{\vec{R}_l\right \}$ instantaneously, where $ \vec{R}_l = \vec{R}_{l \kappa}^0  + \vec{u}_{l \kappa}$ with $\vec{R}_{l \kappa}^0 = \vec{R}_{l}^0 + \vec{\tau}_{\kappa}$, and $\vec{u}_{l \kappa}$ being the displacement of the lattice site $\kappa$ in the $l$-th unit cell. 
Consequently, one can decouple the two systems and consider the dynamics of the many-body electron ground state $\Phi_{\text{el}}^0\left(\vec{r}, \{\vec{R}_l\} \right)$ independently of the lattice by treating the set of all ion positions, $\{\vec{R}_l\}$, as a parameter. 
Therefore, one considers the parameter space spanned by $\{\vec{R}_l\}$ and evolve the electronic wavefunction adiabatically by slowly varying the value of $\left \{ \vec{R}_l \right \}$. If one performs the above procedure in a closed loop, $\Phi^0_{\text{el}} \left(\vec{r},  \left\{\vec{R}_l \right \}\right)$ can only change by a phase, known as the molecular Berry phase (MBP) \cite{RevModPhys.64.51}
\begin{equation}
    \phi_{\text{MBP}} = \oint \vec{A}\left( \left\{\vec{R}_l \right \} \right) \cdot \vec{dR}. 
    \label{eq: Molecular Berry Curvature}
\end{equation}
Here, $\vec{A}\left( \left\{\vec{R}_l \right \} \right)$ is the molecular Berry connection. The component of $\vec{A}\left( \left\{\vec{R}_l \right \} \right)$ corresponding to the unit cell with equilibrium position $\vec{R}_{l}^0$, the $\kappa$th sublattice and the $\alpha$th component of $\vec{u}_{l \kappa}$ is given by \cite{PhysRevX.15.011036, Chiral_Phonons_Angular_Momentum, MBP_Phonon_THE_Nagaosa,  PhysRevLett.134.206701, das2026antiferrochiralphononsmathcalpmathcaltsymmetricantiferromagnets}
\begin{equation}
   A_{l \kappa \alpha} = \mathrm{i}\bra{\Phi_{\text{el}}^0} 
   \frac{\partial}{\partial u_{l \kappa \alpha}} \ket{\Phi_{\text{el}}^0}
    ,
    \label{eq: Molecular Berry Connection}
\end{equation}
where $\Phi_{\text{el}}^0$ is the electronic many-body ground state wave function. To declutter notation, we will call it $\Phi_{0}$ from here on. From the molecular Berry connection one can then construct a gauge-independent quantity, the molecular Berry curvature (MBC), defined as \cite{ Chiral_Phonons_Angular_Momentum, MBP_Phonon_THE_Nagaosa,  PhysRevLett.134.206701, das2026antiferrochiralphononsmathcalpmathcaltsymmetricantiferromagnets}
\begin{equation}
\begin{split}
     G_{\kappa' \beta}^{\kappa \alpha} \left( \vec{R}_l^0 - \vec{R}_{l'}^0\right) 
     & = \left[\frac{\partial A_{l' \kappa' \beta}}{\partial u_{l \kappa \alpha}} - \frac{\partial A_{l \kappa \alpha}}{\partial u_{l' \kappa' \beta}} \right]\bigg|_{\vec{u}\rightarrow 0}\\
     & = \mathrm{i} \left[ \left\langle \frac{\partial \Phi_0}{\partial u_{l \kappa \alpha}} \middle|\frac{\partial \Phi_0}{\partial u_{l' \kappa' \beta}} \right \rangle - 
     \left\langle \frac{\partial \Phi_0}{\partial u_{l' \kappa' \beta}} \middle|\frac{\partial \Phi_0}{\partial u_{l \kappa \alpha}} \right \rangle \right]\bigg|_{\vec{u}\rightarrow 0} \\
     &
     = 2 \text{Im}\left[ \left\langle \frac{\partial \Phi_0}{\partial u_{l \kappa \alpha}} \middle|\frac{\partial \Phi_0}{\partial u_{l' \kappa' \beta}} \right \rangle\right]\bigg|_{\vec{u}\rightarrow 0}.
\end{split}
\label{eq: MBC real space}
\end{equation}
Since the MBC appears in real space as an effective gauge field, a finite value requires the electronic system to break time-reversal symmetry in real space. 

To facilitate the calculation of the MBC, we express it in terms of single-particle wave functions obtained by diagonalizing the tight-binding Hamiltonian in Eq.~\eqref{eq: TB in k-space}. The derivation follows Ref.~\cite{Chiral_Phonons_Angular_Momentum}; here we briefly summarize the main steps. We assume that we are dealing with an insulating system at zero temperature. Firstly, we Fourier transform the curvature in Eq.~\eqref{eq: MBC real space} and introduce a complete basis of states denoted by $\left \{ \Phi_n\right \}$ with corresponding energies $\left \{ E_n\right \}$, where $n>0$ corresponds to excited states, to get 
\begin{equation}
\begin{split}
    G_{\kappa' \beta}^{\kappa \alpha} \left( \vec{k} \right) &= 
    \frac{1}{N} \sum_{l,l'}  G_{\kappa' \beta}^{\kappa \alpha} \left( \vec{R}_{l}^0 - \vec{R}_{l'}^0\right) \mathrm{e}^{-\mathrm{i}\left(\vec{R}_{l \kappa}^0 - \vec{R}_{l' \kappa'}^0 \right) \cdot \vec{k}} \\
     &=  
     \frac{\mathrm{i} }{N}
     \sum_{n \neq 0} \left[ \left\langle \Phi_0 \middle| M_{\vec{k} \kappa \alpha} \middle|\Phi_n \right \rangle  \left \langle  \Phi_n \middle| M_{-\vec{k} \kappa'\beta} \middle|\Phi_0 \right \rangle \right. \\
     &\quad - \left.
     \left\langle \Phi_0 \middle| M_{-\vec{k} \kappa'\beta} \middle|\Phi_n \right \rangle  \left \langle  \Phi_n \middle| M_{\vec{k} \kappa \alpha} \middle|\Phi_0 \right \rangle
     \right] \frac{1}{\left(E_n - E_0 \right)^2 },
\end{split}    
    \label{eq: MBC reciprocal space stage 1}
\end{equation} 
where 
\begin{align}
    M_{\vec{k} \kappa \alpha} = \sum_{l} \frac{\partial H_{\text{el}}}{\partial u_{l \kappa \alpha}}\mathrm{e}^{-\mathrm{i} \vec{R}_{l \kappa}^0 \cdot \vec{k}}
\end{align}
and 
\begin{align}
    M_{-\vec{k} \kappa' \beta} = \sum_{l'} \frac{\partial H_{\text{el}}}{\partial u_{l' \kappa' \beta}}\mathrm{e}^{\mathrm{i} \vec{R}_{l' \kappa'}^0 \cdot \vec{k}}.
\end{align} 
To obtain Eq.~\eqref{eq: MBC reciprocal space stage 1} we made use of the identity
\begin{equation}
  \left \langle \frac{\partial \Phi_n}{\partial \vec{u}} \middle| \Phi_0\right \rangle = \frac{1}{E_n - E_0}\left \langle \Phi_n \middle| \frac{\partial H_{\text{el}}}{\partial \vec{u}} \middle| \Phi_0\right \rangle.
\end{equation}  

By Fourier transforming the fermion operators using Eq.~\eqref{eq: FT electron operators} we obtain
\begin{equation}
\begin{split}
   M_{\vec{k} \kappa \alpha} 
     & = 
     \sum_{\vec{q}} \vec{\Psi}^{\dagger}_{\vec{q}} \mathcal{M}_{\vec{k} \kappa \alpha} \vec{\Psi}_{{\vec{q}+\vec{k}}}, \\ 
     M_{-\vec{k} \kappa' \beta} &= \sum_{\vec{q}} \vec{\Psi}^{\dagger}_{\vec{q} + \vec{k}} \mathcal{M}_{-\vec{k} \kappa' \beta} \vec{\Psi}_{{\vec{q}}},
\end{split}
\label{eq: phonon-electron operators}
\end{equation}
where $\vec{\Psi}_{\vec{q}+ \vec{k}} = \left(c_{\vec{q} + \vec{k} \kappa_1}, ...., c_{\vec{q} + \vec{k} \kappa_n}\right ) ^T$, with $n$ the being the total number of sublattices and $\mathcal{M}_{\vec{k} \kappa \alpha}$ and $\mathcal{M}_{-\vec{k} \kappa' \beta}$ are (non-Hermitian) electron-phonon coupling kernels. 

To compute the analytical form of $\mathcal{M}_{-\vec{k} \kappa' \beta}$ and $\mathcal{M}_{\vec{k} \kappa \alpha}$ for $H_\text{el}$ in Eq.~\eqref{eq: TB-H real space}, we assume that the spatial dependence of the tight-binding Hamiltonian enters through the nearest-neighbor hopping amplitude $t$. Accordingly, the electron-phonon coupling parameter $t'$
is identified with the spatial modulation of $t$. Its numerical value is taken to be $t'= t'_0 = 1 \text{ eV} / \text{\AA}$ (see also Tab.~\ref{tab:numerical-values}), corresponding to a material-realizable strength of the electron-phonon interaction as can be seen in Ref.~\cite{PhysRevB.104.195201, PhysRevB.85.205436}. With the technical details delegated to App.~\ref{sec:MBC}, we arrive at
\begin{equation}
    \mathcal{M}_{\vec{k} \kappa \alpha}  = 
    \begin{pmatrix}
        0 & \left[\mathcal{M}_{\vec{k} \kappa \alpha}\right]_{12} \\
        \left[\mathcal{M} _{\vec{k} \kappa \alpha}\right]_{21} & 0  
    \end{pmatrix}, \\
\end{equation}
with the correspoding elements for $\kappa= \text{A}, \text{B}$ and $\alpha=x,y$ being given by
\begin{equation}
    \begin{aligned}
        \left[\mathcal{M}_{\vec{k} \text{A} x}\right]_{12} 
     & = -\frac{t'}{\sqrt{2}}
     \left(
         \mathrm{e}^{-\mathrm{i} \phi} 
         + \mathrm{e}^{\mathrm{i} \phi} \mathrm{e}^{-\mathrm{i} 
         \left(\vec{q} + \vec{k} \right)\cdot \vec{a}_2} 
         -
          \mathrm{e}^{\mathrm{i} \phi} \mathrm{e}^{-\mathrm{i} 
         \left(\vec{q} + \vec{k} \right)\cdot \vec{a}_1 } 
    \right.
    \\
    & 
    \quad 
    \left.
         - \mathrm{e}^{-\mathrm{i} \phi} \mathrm{e}^{-\mathrm{i} 
         \left(\vec{q} + \vec{k} \right)\cdot \left(\vec{a}_2 + \vec{a}_1\right)}
    \right),
    \\
    \left[\mathcal{M} _{\vec{k} \text{A} x}\right]_{21} 
    &=- 
    \frac{t'}{\sqrt{2}}
    \left(
          \mathrm{e}^{\mathrm{i} \phi} - 
         \mathrm{e}^{-\mathrm{i} \phi} \mathrm{e}^{\mathrm{i} \vec{q} \cdot \vec{a}_1} 
         -  \mathrm{e}^{\mathrm{i} \phi} \mathrm{e}^{\mathrm{i} \vec{q} \cdot \left(\vec{a}_2 +\vec{a}_1 \right)} + 
          \mathrm{e}^{-\mathrm{i} \phi}  \mathrm{e}^{\mathrm{i} \vec{q} \cdot \vec{a}_2}
    \right),  
    \\ 
    \left[\mathcal{M} _{\vec{k} \text{A} y}\right]_{12} 
     & = -\frac{t'}{\sqrt{2}}
     \left(
         \mathrm{e}^{-\mathrm{i} \phi} 
         - \mathrm{e}^{\mathrm{i} \phi} \mathrm{e}^{-\mathrm{i} 
         \left(\vec{q} + \vec{k} \right)\cdot \vec{a}_2} 
    \right.
         \\
         & 
    \left.
    \quad
     +
    \mathrm{e}^{\mathrm{i} \phi} \mathrm{e}^{-\mathrm{i} 
    \left(\vec{q} + \vec{k} \right)\cdot \vec{a}_1 }
    - \mathrm{e}^{-\mathrm{i} \phi} \mathrm{e}^{-\mathrm{i} 
    \left(\vec{q} + \vec{k} \right)\cdot \left(\vec{a}_2 + \vec{a}_1\right)}
    \right),
    \\
    \left[\mathcal{M}_{\vec{k} \text{A} y}\right]_{21} 
    &= -
    \frac{t'}{\sqrt{2}}
    \left(
          \mathrm{e}^{\mathrm{i} \phi} +
         \mathrm{e}^{-\mathrm{i} \phi} \mathrm{e}^{\mathrm{i} \vec{q} \cdot \vec{a}_1} 
         -  \mathrm{e}^{\mathrm{i} \phi} \mathrm{e}^{\mathrm{i} \vec{q} \cdot \left(\vec{a}_2 +\vec{a}_1 \right)} - 
          \mathrm{e}^{-\mathrm{i} \phi}  \mathrm{e}^{\mathrm{i} \vec{q} \cdot \vec{a}_2}
          \right), 
    \end{aligned}
\label{eq: coupling elements A}
\end{equation}
and 
\begin{equation}
\begin{aligned}
    \left[\mathcal{M}_{\vec{k} \text{B} x} \right]_{12}
    &= -
    \frac{t'}{\sqrt{2}}
    \left(
    -\mathrm{e}^{-\mathrm{i} \phi} -   \mathrm{e}^{\mathrm{i} \phi}
    \mathrm{e}^{-\mathrm{i} \vec{q}\cdot \vec{a}_2} 
    + \mathrm{e}^{\mathrm{i} \phi} \mathrm{e}^{-\mathrm{i} \vec{q}\cdot \vec{a}_1} 
     \right.
    \\
    & 
    \left.
    \quad
    +
    \mathrm{e}^{-\mathrm{i} \phi} \mathrm{e}^{-\mathrm{i} \vec{q}\cdot \left(\vec{a}_2 +\vec{a}_1 \right)} 
    \right)  \mathrm{e}^{-\mathrm{i} \vec{k} \cdot \vec{\tau}_{\text{B}}}
     , 
     \\
    \left[\mathcal{M}_{\vec{k} \text{B} x} \right]_{21} 
    &= -
    \frac{t'}{\sqrt{2}}
    \left(
    -\mathrm{e}^{\mathrm{i} \phi} 
     -  
    \mathrm{e}^{-\mathrm{i} \phi}
    \mathrm{e}^{\mathrm{i} \left(\vec{q} + \vec{k} \right)\cdot \vec{a}_2} 
    + \mathrm{e}^{-\mathrm{i} \phi} \mathrm{e}^{\mathrm{i} \left(\vec{q}+\vec{k}\right)\cdot \vec{a}_1} 
    \right.
    \\
    & 
    \quad
    \left.
    +
    \mathrm{e}^{\mathrm{i} \phi} \mathrm{e}^{\mathrm{i} \left(\vec{q} + \vec{k}\right)\cdot \left(\vec{a}_2 +\vec{a}_1 \right)}
    \right)  \mathrm{e}^{-\mathrm{i} \vec{k} \cdot \vec{\tau}_{\text{B}}}, 
    \\ 
    \left[\mathcal{M}_{\vec{k} \text{B} y}  \right]_{12}
    &= -
    \frac{t'}{\sqrt{2}}
    \left(
    -
    \mathrm{e}^{-\mathrm{i} \phi} 
    +
    \mathrm{e}^{\mathrm{i} \phi}
    \mathrm{e}^{-\mathrm{i} \vec{q}\cdot \vec{a}_2}  
    -  
    \mathrm{e}^{\mathrm{i} \phi} \mathrm{e}^{-\mathrm{i} \vec{q}\cdot \vec{a}_1} 
    \right.
    \\
    & 
    \left.
    \quad
    +
    \mathrm{e}^{-\mathrm{i} \phi} \mathrm{e}^{-\mathrm{i} \vec{q}\cdot \left(\vec{a}_2 +\vec{a}_1 \right)} 
    \right)  \mathrm{e}^{-\mathrm{i} \vec{k} \cdot \vec{\tau}_{\text{B}}}
        , 
    \\
    \left[\mathcal{M}_{\vec{k} \text{B} y} \right]_{21} 
     &= -
     \frac{t'}{\sqrt{2}}
    \left(
    -
    \mathrm{e}^{\mathrm{i} \phi} 
    +  
    \mathrm{e}^{-\mathrm{i} \phi}
     \mathrm{e}^{\mathrm{i} \left(\vec{q} + \vec{k} \right)\cdot \vec{a}_2} 
     -
    \mathrm{e}^{-\mathrm{i} \phi} \mathrm{e}^{\mathrm{i} \left(\vec{q}+\vec{k}\right)\cdot \vec{a}_1} 
    +
    \right. 
    \\
    & 
    \quad 
    \left.
    \mathrm{e}^{\mathrm{i} \phi} \mathrm{e}^{\mathrm{i} \left(\vec{q} + \vec{k}\right)\cdot \left(\vec{a}_2 +\vec{a}_1 \right)}
    \right)  \mathrm{e}^{-\mathrm{i} \vec{k} \cdot \vec{\tau}_{\text{B}}}.
\end{aligned}
\label{eq: coupling elements B}
\end{equation}
The elements of $\mathcal{M}_{-\vec{k} \kappa'\beta}$ are obtained by exchanging $\vec{k} + \vec{q} \leftrightarrow \vec{q}$ in the expression above. 

We continue by further simplifying the expression for $G_{\kappa'\beta}^{\kappa \alpha}\left( \vec{k}\right)$ in Eq.~\eqref{eq: MBC reciprocal space stage 1}.
We do so by taking $\Phi_0$ and $\Phi_n$ to be Slater determinants and by using the Slater-Condon rules to write the matrix element products involving the many-body states in terms of single-particle states. The final expression for computing the MBC is \cite{Chiral_Phonons_Angular_Momentum} (see App.~\ref{sec:MBC} for technical details)
\begin{equation}
\begin{aligned}
     G_{\kappa' \beta}^{\kappa \alpha} \left( \vec{k} \right) &= 
     \frac{\mathrm{i} }{N}
     \sum_{m^-, m^+} \sum_{\vec{q}} \left[
     \frac{u_{\vec{q} m^-} ^{\dagger} \mathcal{M}_{\vec{k} \kappa \alpha} u_{\vec{q} + \vec{k} m^+}  u_{\vec{q} + \vec{k} m^+} ^{\dagger} \mathcal{M}_{-\vec{k} \kappa' \beta} u_{\vec{q} m^-}}{\left(E_{\vec{q} m^-}- E_{\vec{q} + \vec{k} m^+}\right)^2} \right. \\
     &
     \quad - \left.
     \frac{u_{\vec{q} + \vec{k} m^-}^{\dagger} \mathcal{M}_{-\vec{k} \kappa' \beta} u_{\vec{q} m^+}  u_{\vec{q} m^+} ^{\dagger} \mathcal{M}_{\vec{k} \kappa \alpha} u_{\vec{q} + \vec{k} m^-}}{\left(E_{\vec{q} m^+}- E_{\vec{q} + \vec{k} m^-}\right)^2} \right],  
\end{aligned}
\label{eq: Molecular Berry curvature final}
\end{equation}
where $E_{\vec{k} m^{\pm}}$ is the energy of the single-electron state $u_{\vec{k} m^{\pm}}$. We use $m^-$ to label the occupied states, and $m^+$ to label the unoccupied states. We readily see from Eq.~\eqref{eq: Molecular Berry curvature final} that since the matrix elements of Eq.~\eqref{eq: coupling elements A} and  Eq.~\eqref{eq: coupling elements B} are proportional to $t'$, the MBC scales quadratically with the electron-phonon coupling strength. 

On a technical note, to obtain the single-electron states $u_{\vec{k} m^{\pm}}$ appearing in Eq.~\eqref{eq: Molecular Berry curvature final}, the exact choice of the Fourier transform gauge [i.e., including the position of the basis $\tau_{\kappa}$ as done in Eq.~\eqref{eq: gauge for orbital magnetization} or not as in Eq.\eqref{eq: FT electron operators}] of the electronic operators does not play a role. More specifically, one can show that even though the electron-phonon coupling kernels depend on the choice of the Fourier transform gauge, their product elements that appear in $G_{\kappa'\beta}^{\kappa \alpha}$ are gauge invariant. This is in contrast to the band geometric Berry curvature as well as the intrinsic orbital moment in Eq.~\eqref{eq: orbital magnetization}, where one has to exercise caution when it comes to the choice of the Fourier transform gauge.

For a two-band electronic system with particle-hole symmetry, $E_{\vec{k} +} =- E_{\vec{k} -}$, the MBC simplifies considerably. This applies, in particular, to the checkerboard loop-current model in Eq.~\eqref{eq: TB k-space Kernel} for the choice $t_\mathrm{A}=-t_\mathrm{B}$ adopted here 
\begin{equation}
\begin{aligned}
     G_{\kappa' \beta}^{\kappa \alpha} \left( \vec{k} \right) 
     &= \frac{\mathrm{i}}{N} \sum_{\vec{q}}
    \frac{
    \left[\mathcal{M}_{-\vec{k} \kappa'\beta} \right]_{21} \left[\mathcal{M}_{\vec{k} \kappa \alpha} \right]_{12} - 
        \left[\mathcal{M}_{-\vec{k} \kappa'\beta} \right]_{12} \left[\mathcal{M}_{\vec{k} \kappa \alpha} \right]_{21} 
    }{\left(E_{\vec{q} +}- E_{\vec{q} + \vec{k} -}\right)^2}
    \\
    & 
    \quad
    \times 
    \frac{
        \left|\Delta_{\vec{k} + \vec{q}}\right|^2  
        \left|\Delta_{\vec{q}} \right|^2 - 
        \left(\lambda_{\vec{k} + \vec{q}} + E_{\vec{k} + \vec{q} +} \right)^2\left(-\lambda_{\vec{q}} + E_{\vec{q} -}\right)^2}
        {N_{\vec{q}+}^2 N_{\vec{k}+ \vec{q}-}^2}.
\end{aligned}
\label{eq: MBC checkerboard}
\end{equation}
In the following, we will use Eq.~\eqref{eq: MBC checkerboard} when computing the MBC.
\begin{figure*}
    \centering
\includegraphics[width=1\linewidth]{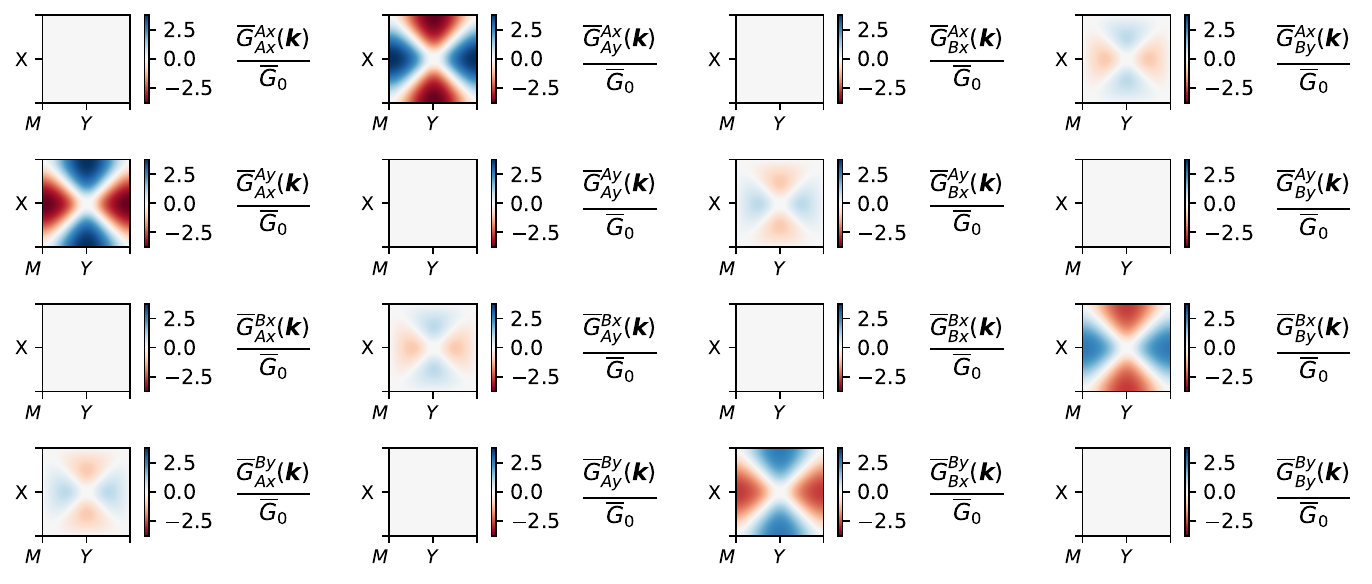}
    \caption{Real part of all elements of the MBC for checkerboard-lattice model in Eq.~\eqref{eq: TB-H real space}. Each element is plotted as a function of $\vec{k}$ within the first Brillouin zone, with $\overline{G}_0 = 10^{-2}g_0 = 10^{-2} \hbar / \left(M_0 a_0^2\right) = 10^{9} \text{Hz}$, where $g_0= \hbar / \left(M_0 a_0^2\right) = 10^{11}$Hz is the baseline scale of our MBC. The ordering of the elements corresponds to the ordering of the MBC matrix in Eq.~\eqref{eq: G-matrix checkerboard elements}.}
    \label{fig:MBC numerics}
\end{figure*}
We continue by defining the rescaled MBC as
\begin{equation}
    \overline{G}_{\kappa' \beta}^{\kappa \alpha} \left( \vec{k} \right) = \frac{\hbar}{2}\frac{1}{\sqrt{M_\kappa M_{\kappa'}}} G_{\kappa' \beta}^{\kappa \alpha} \left( \vec{k} \right), 
\end{equation}
 which has units of frequency and---as it will become clear in Sec.~\ref{main sec: MBC dynamics}---is the relevant quantity entering the phonon dynamics.
The elements of $\overline{G}_{\kappa' \beta}^{\kappa \alpha} \left( \vec{k} \right)$ derived using Eq.~\eqref{eq: MBC checkerboard} for $\kappa, \kappa'= \text{A}, \text{B}$ and $\alpha, \beta = x,y$ can be  grouped up in a matrix reading

\begin{equation}
    \overline{G}_{\vec{k}} = 
    \begin{pmatrix}
        \overline{G}_{\text{A}x}^{\text{A}x}\left( \vec{k} \right)  & \overline{G}_{\text{A}y}^{\text{A}x} \left( \vec{k} \right)& \overline{G}_{\text{B}x}^{\text{A}x}\left( \vec{k} \right)    & \overline{G}_{\text{B}y}^{\text{A}x}\left( \vec{k} \right)  \\
        \overline{G}_{\text{A}x}^{\text{A}y}\left( \vec{k} \right)  & \overline{G}_{\text{A}y}^{\text{A}y}\left( \vec{k} \right)  & \overline{G}_{\text{B}x}^{\text{A}y}\left( \vec{k} \right)  & \overline{G}_{\text{B}y}^{\text{A}y}\left(\vec{k} \right)  \\
        \overline{G}_{\text{A}x}^{\text{B}x}\left( \vec{k} \right) & \overline{G}^{\text{B}x}_{\text{A}y} \left( \vec{k} \right) & \overline{G}^{\text{B}x}_{\text{B}x}\left( \vec{k} \right)    & \overline{G}^{\text{B}y}_{\text{B}x}\left( \vec{k} \right)  \\
        \overline{G}^{\text{B}y}_{\text{A}x} \left( \vec{k} \right) & \overline{G}^{\text{B}y}_{\text{A}y}\left( \vec{k} \right) & \overline{G}_{\text{B}x}^{\text{B}y}\left( \vec{k} \right)  & \overline{G}_{\text{B}y}^{\text{B}y}\left( \vec{k} \right) 
    \end{pmatrix}. 
    \label{eq: G-matrix checkerboard elements}
\end{equation}

The MBC matrix in Eq.~\eqref{eq: G-matrix checkerboard elements} is anti-Hermitian and is constrained by the symmetries of the electronic system. Here, we highlight how inversion symmetry $\mathcal{P}$, time-reversal with rotation $\mathcal{C}_{4z,+}\mathcal{T}$, and the mirror $\mathcal{M}_{xy}$ constrain $\overline{G}_{\vec{k}}$.
For the checkerboard lattice where every site is a center of inversion,  $\mathcal{P}$ enforces 
\begin{equation}
    \overline{G}_{\kappa \alpha}^{\kappa'\beta}\left(\vec{k}\right) = \left(\overline{G}_{\kappa \alpha}^{\kappa'\beta}\left(\vec{k}\right)\right)^*,  
    \label{eq: inversion MBC constrain}
\end{equation}
rendering the MBC matrix real and antisymmetric. Moreover, $\mathcal{T} \mathcal{C}_{4z,+}$ leads to the constraint 
\begin{equation}
      \overline{G}_{\kappa'\beta}^{\kappa \alpha}\left(\vec{k} \right) = - \sum_{\overline{\alpha} \overline{\beta}} \left[\text{C}_{4z, +}\right]_{\overline{\alpha} \alpha}  \left[\text{C}_{4z, +}\right]_{\overline{\beta} \beta} \overline{G}_{\kappa'\overline{\beta}}^{\kappa \overline{\alpha}}\left(- \text{C}_{4z,+}\vec{k} \right), 
      \label{eq: C4T constrain}
\end{equation}
with $\left[\text{C}_{4z, +}\right]_{\alpha \overline{\alpha}}$ being the elements of the $90^{\circ}$ clockwise rotation matrix 
\begin{equation}
      \text{C}_{4z, +} = 
    \begin{pmatrix}
        0 & 1 \\
        - 1 & 0 
    \end{pmatrix},
\end{equation}
and $\text{C}_{4z,+}\vec{k}$ being the rotated $\vec{k}$ vector. Finally, the reflection $\mathcal{M}_{xy}$ enforces  
\begin{equation}
      \overline{G}_{\kappa'\beta}^{\kappa \alpha}\left( \vec{k}\right) = \sum_{\overline{\alpha} \overline{\beta}} \left[\text{M}_{xy}\right]_{\overline{\alpha} \alpha}
    \left[\text{M}_{xy}\right]_{\overline{\beta} \beta} \overline{G}_{\kappa'\overline{\beta}}^{\kappa \overline{\alpha}}\left( \text{M}_{xy}\vec{k}\right),
    \label{eq: Mxy constraint}
\end{equation}
where $\left[\text{M}_{xy}\right]_{\overline{\alpha} \alpha}$ is an element of the reflection matrix 
\begin{equation}
        \text{M}_{xy}= \begin{pmatrix}
        0 & 1 \\
        1 & 0 
    \end{pmatrix}.
\end{equation} 
We provide a detailed proof of these symmetry constrains in App.~\ref{AP: subsec: MBC symmetries}. The simple form of the above relations relies on the definition of the MBC Fourier transform in Eq.~\eqref{eq: MBC reciprocal space stage 1}, which includes the atomic basis positions.

Considering the aforementioned symmetry relations, we identify only two independent elements of $\overline{G}_{\vec{k}}$, namely $\overline{G}^{Ax}_{Ay}\left(\vec{k} \right)$ and $\overline{G}^{Ax}_{By}\left(\vec{k}\right)$. 
More specifically, we identify the following relations for the independent elements of $G_{\vec{k}}$
\begin{subequations}
\begin{align} 
    \overline{G}_{Ax}^{Ax}\left(\vec{k}\right) &=  \overline{G}_{Ay}^{Ay}\left(\vec{k}\right) =  \overline{G}_{Bx}^{Bx}\left(\vec{k}\right) =  \overline{G}_{By}^{By}\left(\vec{k}\right) = 0, 
    \label{eq:line1} 
    \\
    \overline{G}_{By}^{Ax}\left(\vec{k}\right) &= \overline{G}_{Bx}^{Ay} \left(-\text{C}_{4z, +}\vec{k}\right) = \overline{G}_{Bx}^{Ay} \left(\text{M}_{xy}\vec{k}\right),
    \label{eq:line2} 
    \\
    \overline{G}_{By}^{Ay}\left(\vec{k} \right) &= \overline{G}_{Bx}^{Ax}\left( \vec{k} \right) = 0, 
    \label{eq:line3}
    \\
    \overline{G}_{Ay}^{Ax}\left(\vec{k}\right) &= 
         \frac{M_\text{B}}{M_{\text{A}}}\overline{G}_{By}^{Bx}\left(\vec{k}\right).
         \label{eq:line4}
\end{align}
\end{subequations}
The missing elements corresponding to the lower triangular part of $G_{\vec{k}}$ can be obtained from the ones above by means of anti-Hermiticity.
Equation~\eqref{eq:line1} follows from inversion symmetry forcing the matrix to be real and thus antisymmetric, leading to vanishing diagonal elements. The symmetry constrains in Eq.~\eqref{eq: C4T constrain} and Eq.~\eqref{eq: Mxy constraint} lead to Eq.~\eqref{eq:line2}. However, Eqs.~\eqref{eq:line3} and \eqref{eq:line4} are not a product of symmetries. They rather derive from the simple nature of the two-band electronic system that we consider and the approximations made in the construction of the $\mathcal{M}_{\vec{k} \kappa \alpha}$ and $\mathcal{M}_{-\vec{k} \kappa' \beta}$ kernels in Eqs.~\eqref{eq: coupling elements A} and \eqref{eq: coupling elements B}, where only the spatial modulation of the nearest-neighbor hopping is taken into account. Figure~\ref{fig:MBC numerics} shows all elements of $\overline{G}_{\vec{k}}$ across the Brillouin zone, making the symmetry constraints manifest.

\section{\texorpdfstring{$D$}{D}-wave axial phonons from molecular Berry curvature}

\label{main sec: MBC dynamics}
The effects of electron-phonon coupling can be incorporated in the kinetic part of the phonon Hamiltonian by means of minimal coupling \cite{7f8x-fkx9} \cite{PhysRevX.15.011036, Chiral_Phonons_Angular_Momentum, MBP_Phonon_THE_Nagaosa, li2025phonondichroismsrevealingunusual, PhysRevLett.119.075301, PhysRevLett.134.206701, das2026antiferrochiralphononsmathcalpmathcaltsymmetricantiferromagnets,  dhakal2025theoryintrinsicphononthermal}
\begin{equation}
    \frac{1}{2}\sum_{l, \kappa, \alpha}\frac{1}{ M_\kappa} p^2_{l \kappa \alpha} \rightarrow 
      \frac{1}{2}\sum_{l, \kappa, \alpha}\frac{1}{ M_\kappa} \left(p_{l \kappa \alpha}- \hbar A_{l \kappa \alpha}\right)^2, 
\end{equation}
where $M_{\kappa}$ is the mass of the ion  at the basis site $\kappa = \text{A}, \text{B}$.
By employing the harmonic approximation for the lattice potential, the full phonon Hamiltonian is given by
\begin{equation}
\begin{aligned}
     H_{\text{ph}}
     &  
     = \frac{1}{2} \sum_{l, \kappa, \alpha} \frac{1}{ M_\kappa} \left(p_{l \kappa \alpha}- \hbar A_{l \kappa \alpha}\right)^2 \\
     & 
     \quad + \frac{1} {2} \sum_{l,l'} \sum_{\kappa \kappa'} \sum_{\alpha \beta} u_{l \kappa \alpha} \Phi_{\kappa'\beta}^{\kappa \alpha} \left(\vec{R}_l^0 -\vec{R}_{l'}^0 \right)u_{l' \kappa' \beta}, 
\end{aligned}
\end{equation}
where  $\Phi_{\kappa'\beta}^{\kappa \alpha} \left(\vec{R}_l^0 -\vec{R}_{l'}^0 \right)$ is the element of the force constant matrix (FCM) generated by the displacements $u_{l \kappa \alpha}$ and  $u_{l' \kappa' \beta}$.
The molecular Berry connection can be expressed in terms of the MBC using an antisymmetric gauge that is valid in equilibrium as \cite{bfll-sdrb, MBP_Phonon_THE_Nagaosa, Chiral_Phonons_Angular_Momentum}
\begin{equation}
    A_{l \kappa \alpha} = - 
    \frac{1}{2}\sum_{\beta, l', \kappa'} G_{\kappa'\beta}^{\kappa \alpha}\left( \vec{R}_l^0 - \vec{R}_{l'}^0\right) u_{l' \kappa'\beta}. 
    \label{eq: gauge for molecular connection}
\end{equation}
By Fourier transforming the momenta and the displacements,
\begin{equation}
\begin{aligned}
    & p_{l \kappa \alpha} = \frac{1}{\sqrt{N}} \sum_{\vec{k}} p_{\vec{k} \kappa \alpha} \mathrm{e}^{\mathrm{i} \vec{k} \cdot \vec{R}_{l \kappa}^0}, \\
    & 
    u_{l \kappa \alpha} = \frac{1}{\sqrt{N}} \sum_{\vec{k}} u_{\vec{k} \kappa \alpha} \mathrm{e}^{\mathrm{i} \vec{k} \cdot \vec{R}_{l \kappa}^0 },
\label{eq: FT p-u}
\end{aligned}  
\end{equation}
one can write $H_{\text{ph}}$ in compact form
\begin{equation}
    H_{\text{ph}} = \frac{1}{2}\sum_{\vec{k}}
    \begin{pmatrix}
        \vec{p}_{\vec{k}} \\
       \vec{u}_{\vec{k}} 
    \end{pmatrix}^{\dagger}
    \begin{pmatrix}
        I &  \overline{G}_{\vec{k}} \\
        \overline{G}^{\dagger}_{\vec{k}} & \overline{D}_{\vec{k}} + \overline{G}^{\dagger}_{\vec{k}}\overline{G}_{\vec{k}}
    \end{pmatrix}
    \begin{pmatrix}
        \vec{p}_{k} \\
       \vec{u}_{k} 
    \end{pmatrix},
    \label{eq: phonon Hamiltonian with MBC}
\end{equation}
where $\overline{G}_{\vec{k}}$ is the rescaled MBC matrix defined in Eq.~\eqref{eq: MBC checkerboard}, $\overline{D}_{\vec{k}}$ is the phonon dynamical matrix with elements
\begin{equation}
     \overline{D}_{\kappa' \beta}^{\kappa \alpha}\left(\vec{k} \right) = \frac{1}{N} \frac{1}{\sqrt{M_\kappa M_\kappa'}}\sum_{l l'} \mathrm{e}^{-\mathrm{i} \vec{k} \cdot \left(\vec{R}_{l \kappa}^0  - \vec{R}_{l' \kappa'}^0 \right)} \Phi_{\kappa'\beta}^{\kappa \alpha}\left(\vec{R}_l^0  - \vec{R}_{l'}^0\right),
\label{eq: dynamical matrix}
\end{equation}
 and $I$ is an identity matrix of the same dimension as $\overline{G}_{\vec{k}}$ and $\overline{D}_{\vec{k}}$. For the checkerboard lattice, the vectors $\vec{\vec{p}}_{\vec{k}}$ and $\vec{\vec{u}}_{\vec{k}}$ are given by
\begin{equation}
    \begin{aligned}
        \vec{p}_{\vec{k}} &= 
        \begin{pmatrix}
            p_{\vec{k} \text{A} x} /\sqrt{M_\text{A}} & p_{\vec{k} \text{A} y} /\sqrt{M_\text{A}} & p_{\vec{k} \text{B} x} /\sqrt{M_\text{B}} & p_{\vec{k} \text{B} y} /\sqrt{M_\text{B}} 
        \end{pmatrix}, \\
        \vec{u}_{\vec{k}} &= 
        \begin{pmatrix}
            u_{\vec{k} \text{A} x} \sqrt{M_\text{A}} & u_{\vec{k} \text{A} y} \sqrt{M_\text{A}} & u_{\vec{k} \text{B} x}\sqrt{M_\text{B}} & u_{\vec{k} \text{B} y} \sqrt{M_\text{B}} 
        \end{pmatrix}.
    \end{aligned}
\end{equation}
With $\overline{G}_{\vec{k}}$ from Eq.~\eqref{eq: MBC checkerboard}, we now construct the dynamical matrix for the checkerboard lattice to solve the harmonic phonon problem coupled to the electrons.

We consider up to second-neighbor couplings in the FCM, whose elements $\Phi_{\kappa'\beta}^{\kappa \alpha} \left(\vec{R}_l^0 -\vec{R}_{l'}^0 \right)$ have units of spring constants. To deduce the number of independent elements of FCM we use the crystallographic symmetries of the checkerboard lattice, as shown in detail in App.~\ref{AP:Derivation of Phonon Dynamical Matrix}. For simplicity we assume that the blocks $\Phi_{\text{A}}^{\text{A}}$ and $\Phi_{\text{B}}^{\text{B}}$ of the FCM (corresponding to the second-neighbor couplings that describe intra-sublattice force constants) are identical. 
Within this approximation, a total of four independent spring constants are required; two spring constants, namely $n_{11}$, and $n_{12}$, corresponding to longitudinal and transverse nearest-neighbor coupling, and two more, $\gamma_{11}$, and $\gamma_{22}$, for longitudinal and transverse second-neighbor interactions. Their numerical values are given in Tab.~\ref{tab:numerical-values}.
After a Fourier transformation [see Eq.~\eqref{eq: dynamical matrix}], we obtain the dynamical matrix as 
\begin{equation}
    \overline{D}_{\vec{k}} = 
    \begin{pmatrix}
        \overline{D}_{\text{A}x}^{\text{A}x}\left( \vec{k} \right) &  \overline{D}_{\text{A}y}^{\text{A}x}\left( \vec{k} \right) & \overline{D}_{\text{B}x}^{\text{A}x}\left( \vec{k} \right)  & \overline{D}_{\text{B}y}^{\text{A}x}\left( \vec{k} \right) \\
        \overline{D}_{\text{A}x}^{\text{A}y}\left( \vec{k} \right) & \overline{D}_{\text{A}y}^{\text{A}y}\left( \vec{k} \right)  & \overline{D}_{\text{B}x}^{\text{A}y}\left( \vec{k} \right)  & \overline{D}_{\text{B}y}^{\text{A}y}\left( \vec{k} \right)  \\
       \overline{D}_{\text{A}x}^{\text{B}x}\left(\vec{k} \right)  & \overline{D}_{\text{A}y}^{\text{B}x}\left( \vec{k} \right)  & \overline{D}_{\text{B}x}^{\text{B}x}\left( \vec{k} \right) & \overline{D}_{\text{B}y}^{\text{B}x}\left( \vec{k} \right) \\
        \overline{D}_{\text{A}x}^{\text{B}y}\left( \vec{k} \right)  & \overline{D}_{\text{A}y}^{\text{B}y}\left( \vec{k} \right) & \overline{D}_{\text{B}x}^{\text{B}y}\left( \vec{k} \right) & \overline{D}_{\text{B}y}^{\text{B}y}\left( \vec{k} \right) 
    \end{pmatrix}, 
\end{equation}
with the corresponding elements being 
\begin{equation}
\begin{aligned}
     \overline{D}_{\text{B}x}^{\text{A}x}\left(\vec{k}\right) &= \frac{n_{11}}{\sqrt{M_\text{A} M_\text{B}}}
     \left(
     \mathrm{e}^{-\mathrm{i}\vec{k} \cdot \vec{a}_1} + \mathrm{e}^{-\mathrm{i}\vec{k} \cdot \vec{a}_2} 
     \right. 
     \\
     & 
     \left.
     \quad 
     + \mathrm{e}^{-\mathrm{i}\vec{k} \cdot \left( \vec{a}_1 + \vec{a}_2\right)} + 1
     \right)
     \mathrm{e}^{\mathrm{i}\left( k_x + k_y\right) /2},  \\
       \overline{D}_{\text{B} y}^{\text{A}x}\left(\vec{k}\right) &= \frac{n_{12}}{\sqrt{M_A M_B}} \left(-\mathrm{e}^{-\mathrm{i}\vec{k} \cdot \vec{a}_1} - \mathrm{e}^{-\mathrm{i}\vec{k} \cdot \vec{a}_2} 
       \right. 
       \\
       & 
       \left. 
       \quad
       + \mathrm{e}^{-\mathrm{i}\vec{k} \cdot \left( \vec{a}_1 + \vec{a}_2\right)} + 1 \right)\mathrm{e}^{\mathrm{i}\left( k_x + k_y\right) /2}, \\
      \overline{D}_{\text{B}x}^{\text{A}y}\left(\vec{k}\right) &=  \overline{D}_{\text{B}y}^{\text{A}x}\left(\vec{k}\right), \\
       \overline{D}_{\text{B}y}^{\text{A}y}\left(\vec{k}\right) &= \overline{D}_{\text{B}x}^{\text{A}x}\left(\vec{k}\right), 
\end{aligned}
\label{eq: D elements FN}
\end{equation}
for the nearest-neighbor couplings, and 
\begin{equation}
    \begin{aligned}
        \overline{D}_{\text{A}x}^{\text{A}x}\left(\vec{k} \right) &= \frac{1}{M_\text{A}} 
        \left(2 \left( \gamma_{11} \cos k_x + \gamma_{22} \cos k_y\right) 
        \right. \\
        & 
        \left. 
        \quad
        - 2 \left( \gamma_{22} + \gamma_{11} \right)- 4 n_{11}\right), \\
        \overline{D}_{\text{A}y}^{\text{A}y}\left(\vec{k} \right) &= \frac{1}{M_\text{A}} \left(2 
        \left( \gamma_{22}\cos k_x + \gamma_{11} \cos k_y\right) 
        \right. 
        \\ 
        & 
        \quad
        \left.
        - 2 \left( \gamma_{22} + \gamma_{11} \right)- 4 n_{11}\right), \\
        \overline{D}_{\text{B}x}^{\text{B}x}\left(\vec{k} \right) &= \frac{1}{M_\text{B}} \left(2 
        \left( \gamma_{11} \cos k_x + \gamma_{22}\cos k_y\right) 
        \right. 
        \\
        & 
        \left. 
        \quad
        - 2 \left( \gamma_{22} + \gamma_{11} \right)- 4 n_{11}\right), \\
        \overline{D}_{\text{B}y}^{\text{B}y}\left(\vec{k} \right) &= \frac{1}{M_\text{B}} \left(2 
        \left( \gamma_{22} \cos k_x + \gamma_{11} \cos k_y\right) 
        \right. 
        \\
        & 
        \left. 
        \quad
        - 2 \left( \gamma_{22} + \gamma_{11} \right)- 4 n_{11}
        \right), \\
        \overline{D}_{\text{A}y}^{\text{A}x}\left( \vec{k}\right) &=  \overline{D}_{\text{A}x}^{\text{A}y}\left(\vec{k} \right)=  \overline{D}_{\text{B}x}^{\text{B}y}\left(\vec{k} \right) = \overline{D}_{\text{B}y}^{\text{B}x}\left(\vec{k} \right) = 0,  
    \end{aligned}
\label{eq: second neighbors D elements}
\end{equation}
for the second-neighbor coupling. The remaining elements of $\overline{D}_{\vec{k}}$ in Eq.~\eqref{eq: D elements FN} are given by Hermitian conjugation. We note that, in general, the reduced symmetries of a magnetic system compared to the non-magnetic case would also affect the symmetries of the dynamical matrix by inducing correction terms to the FCM. These are, however, subleading and do not break time-reversal symmetry, which is why we have neglected them. 

\begin{figure*}
    \centering
    \includegraphics[width=1\linewidth]{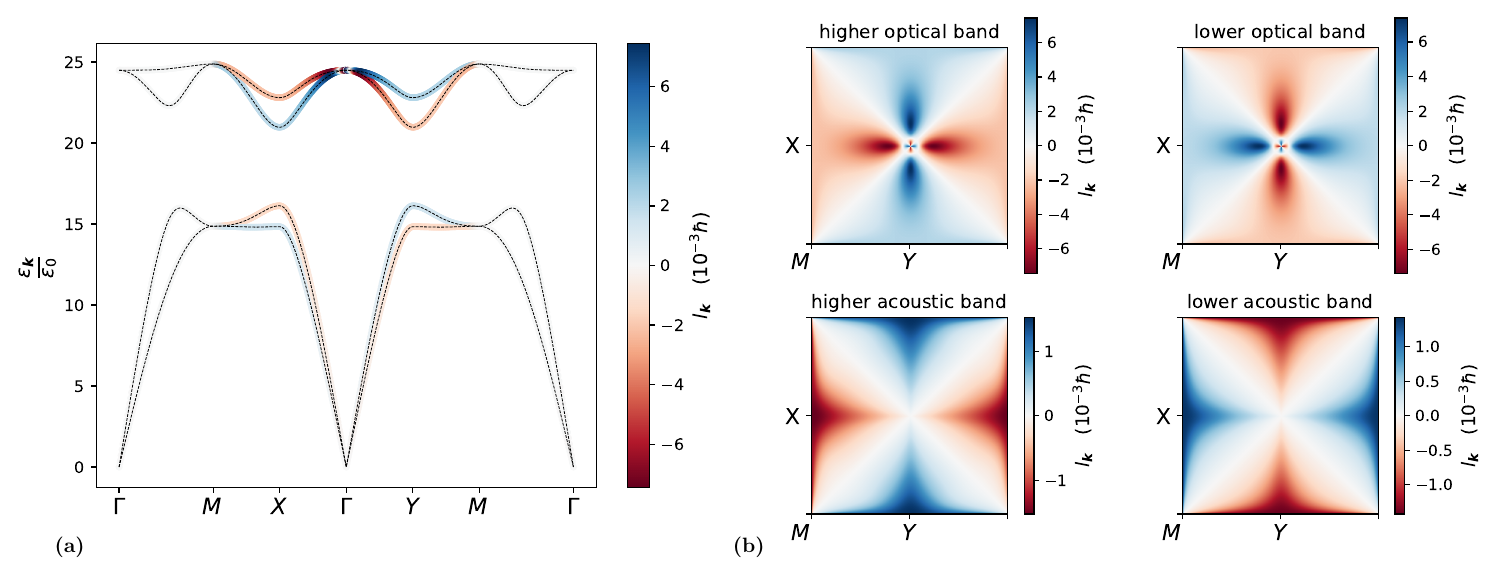}
    \caption{(a) Phonon band structure of the in-plane phonons on the checkerboard lattice, including the effects of the MBC in the phonon Hamiltonian, with $\varepsilon_0 = \hbar \omega_0=1\text{ meV}$ being the characteristic phonon energy scale. Dashed lines indicate the harmonic phonon energies and color shows the out-of-plane phonon angular momentum $l_{\vec{k} \sigma}$. (b) $d$-wave texture of $l_{\vec{k} \sigma}$ for all four phonon bands.}
   \label{fig: POAM}
\end{figure*}

With $H_{\text{ph}}$ fully constructed, we continue by promoting $\vec{p}_{\vec{k}}$ and $\vec{u}_{\vec{k}}$ to operators and determining their time evolution. We do so by assuming plane-wave solutions of the form $\vec{p}_{\vec{k}} = \vec{\mu}_{\vec{k} \sigma} \mathrm{e}^{-\mathrm{i} \omega_{\vec{k} \sigma} t}$ and $\vec{u}_{\vec{k}} = \vec{\epsilon}_{\vec{k} \sigma} \mathrm{e}^{-\mathrm{i} \omega_{\vec{k} \sigma} t}$ and derive the eigenvalue problem 
\begin{equation}
   \omega_{\vec{k} \sigma} 
    \begin{pmatrix}
       \tilde{\vec{\mu}}_{\vec{k} \sigma} \\
    \tilde{\vec{\epsilon}}_{\vec{k} \sigma}
    \end{pmatrix}= 
H_{\vec{k}}^{\text{eff}} \begin{pmatrix}
      \tilde{\vec{\mu}}_{\vec{k} \sigma} \\
    \tilde{\vec{\epsilon}}_{\vec{k} \sigma}
    \end{pmatrix},
\end{equation}
with 
\begin{equation}
    H_{\vec{k}}^{\text{eff}} = 
    \mathrm{i}
    \begin{pmatrix}
           \overline{G}\left(\vec{k}\right) & -\overline{D}\left(\vec{k}\right) - \overline{G}^{\dagger}\left(\vec{k}\right)\overline{G}\left(\vec{k}\right) \\
           I & \overline{G}\left(\vec{k}\right)
       \end{pmatrix}.
    \label{eq: effective H_ph}
\end{equation}
Here, $\sigma$ is the phonon band index, $\omega_{\vec{k} \sigma}$ is the phonon frequency,  $\vec{\mu}_{\vec{k} \sigma}$ and $\vec{\epsilon}_{\vec{k} \sigma}$ are the momentum and displacement polarization vectors, respectively, while $\tilde{\vec{\mu}}_{\vec{k} \sigma}$, and $\tilde{\vec{\epsilon}}_{\vec{k} \sigma}$ are their normalized counterparts. The technical steps to derive and diagonalize $H_{\vec{k}}^{\text{eff}}$, and to proceed with the second quantization of the $\vec{u}_{\vec{k}}$ and $\vec{p}_{\vec{k}}$ are described in detail in Ref.~\cite{THE_Lifa_Zhang} and reviewed also for completeness in App.~\ref{AP:Phonon Dynamics with the Molecular Berry Curvature}. The positive phonon energies $\omega_{\vec{k} \sigma}$ resulting from the diagonalization of Eq.~\eqref{eq: effective H_ph} [see also Eq.~\eqref{eq:diagonalphonons}] are depicted in Fig.~\ref{fig: POAM}(a).

Eventually, we can proceed as in Ref.~\cite{Lifa_phonon_De_Hass_effect} and compute the angular momentum of the phonons. Starting from the semiclassical definition of the kinetic angular momentum
\begin{equation}
    \boldsymbol{J} = \sum_{l, \kappa} \vec{u}_{l \kappa} \times \dot{\vec{u}}_{l \kappa}, 
    \label{eq: semiclassical OAM}
\end{equation}
and using the canonical momentum for the derivative of the displacements, we obtain the following relation for the thermal average of the out-of-plane component of $\vec{J}$,
\begin{equation}
    \langle  J_{z} \rangle = \sum_{\vec{k}, \sigma}  l_{\vec{k} \sigma} \left( f_{\vec{k} \sigma} + \frac{1}{2}\right).
    \label{eq: OAM thermal average}
\end{equation}
In accordance with Ref.~\cite{Lifa_phonon_De_Hass_effect}, we have defined the angular momentum of a phonon in band $\sigma$ with momentum $\vec{k}$ as
\begin{equation}
  l_{\vec{k} \sigma} = \hbar \tilde{\vec{\epsilon}}_{\vec{k} \sigma}^{\dagger} M \tilde{\vec{\epsilon}}_{\vec{k}\sigma}, \quad \text{with} \quad M = I \otimes \begin{pmatrix}
         0 & - \mathrm{i} \\
        \mathrm{i} & 0 
    \end{pmatrix}.
    \label{eq: OAM phonons texture}
\end{equation}
Note that Eq.~\eqref{eq: OAM phonons texture} is fully consistent with Ref.~\cite{Chiral_Phonons_Angular_Momentum}, even though the technical route that we have followed is different.

For the orbital altermagnet on the checkerboard lattice, we show $l_{\vec{k} \sigma}$ in Fig.~\ref{fig: POAM}(a) along high-symmetry directions, and in Fig.~\ref{fig: POAM}(b) across the full Brillouin zone for all four phonon bands. The resulting $d$-wave phonon angular momentum texture caused by the electron-phonon coupling is clearly visible. 

Interestingly, the acoustic modes also exhibit finite angular momentum, albeit of smaller magnitude compared to the optical modes. This is in contrast to the $s$-wave axial phonons in the Haldane model studied in Ref.~\cite{Chiral_Phonons_Angular_Momentum}, where the obtained acoustic modes, compared to the optical ones,  have a negligibly small angular momentum over the full Brillouin zone. Moreover, the $s$-wave axial phonons in Ref.~\cite{Chiral_Phonons_Angular_Momentum} come with lifted degeneracy of the optical phonon modes at the $\Gamma$ point. In contrast, the MBC of our $d$-wave system does not permit a splitting of the optical phonon modes. Specifically, the combined constraints of $\mathcal{M}_{xy}$ and $\mathcal{C}_{4z,+}\mathcal{T}$ [Eqs.~\eqref{eq: C4T constrain} and \eqref{eq: Mxy constraint}] restrict the form of $G_{\vec{k}}$ such that the optical modes remain degenerate. As shown in Appendix~\ref{AP: subsubsection: Gamma splitting}, this result follows solely from the underlying symmetries and is independent of the complexity of the toy model. We therefore conclude that the magnetic point group $4'/mm'm$ of the orbital altermagnet introduced in Sec.~\ref{main sec: electrons}, which is incompatible with ferromagnetism, forbids a finite time-reversal-odd axial moment at the $\Gamma$ point.

The magnitude of $l_{\vec{k} \sigma}$ is directly determined by the strength of the MBC, which in turn is governed by two key factors. The first factor is the microscopic origin of the MBC, namely its relation to electron-phonon coupling. Since the MBC depends quadratically on $t'$, its characteristic scale is set by
$
g_0 = \hbar / \left(M_0 a_0^2 \right)$.
For a representative mass scale of $M_0 = 10^{-25}$ kg and a lattice constant of $a_0 = 1 \text{ \AA}$, we obtain $g_0 = 10^{11}$ Hz. While electron-phonon coupling determines this fundamental scale, the actual magnitude of the MBC is highly sensitive to the underlying electronic band structure. In particular, Eq.~\eqref{eq: MBC checkerboard} shows that the MBC at the $\Gamma$ point ($\vec{k}=0$) scales inversely with the square of the \emph{direct} electronic band gap $E_{\vec{q}+}- E_{\vec{q}-}$. Therefore, systems with small direct electronic gaps are expected to exhibit strongly enhanced MBC effects.
In our model, however, the MBC vanishes at the $\Gamma$ point due to the numerator in Eq.~\eqref{eq: MBC checkerboard}. Instead, it is finite only at $\vec{k} \ne 0$ (and largest at the $X$ and $Y$ points of the Brillouin zone), where it grows with the inverse square of an \emph{indirect} gap $E_{\vec{q}+}- E_{\vec{q}+\vec{k}-}$. We emphasize that the applicability of the MBC requires that the phonon band width is much smaller than the electronic gap, as otherwise non-adiabatic terms beyond the MBC must be considered. For the parameters we have chosen, we obtain a direct electronic gap of $E_g = 2V = 0.6 \text{ eV}$, and the phonon bandwidth has been set to $\Delta\varepsilon = 25\text{ meV}$. Thus, we find $\Delta\varepsilon / E_g = 1/24 \ll 1 $ and the adiabatic approximation is justified.

With the above discussion taken into account and for $t'/t'_0 = 1$, the magnitude of the MBC is restricted to $G_0 = 10^{-2}g_0 = 10^9\text{ Hz}$.
For a fixed phonon bandwidth of $\Delta\varepsilon = 25\text{ meV}$, obtained by choosing the dominant force constant as $n_{11} = -100\text{ N/m}$, the corresponding MBC energy scale is four orders of magnitude smaller than $\Delta\varepsilon$. This hierarchy of energy scales is directly reflected in the relatively small values of $l_{\vec{k}\sigma}$, which remain in the range of $\sim 10^{-3} \hbar$. We expect, however, that the magnitude of $d$-wave angular momentum $l_{\vec{k}\sigma}$ can vary substantially across materials, as it is controlled by the underlying electronic band structure, in particular by the relevant indirect electronic gaps.

\section{Phonon Angular Momentum Transport}
\label{main sec: splitter}

\begin{figure*}
\includegraphics[width=1\linewidth]{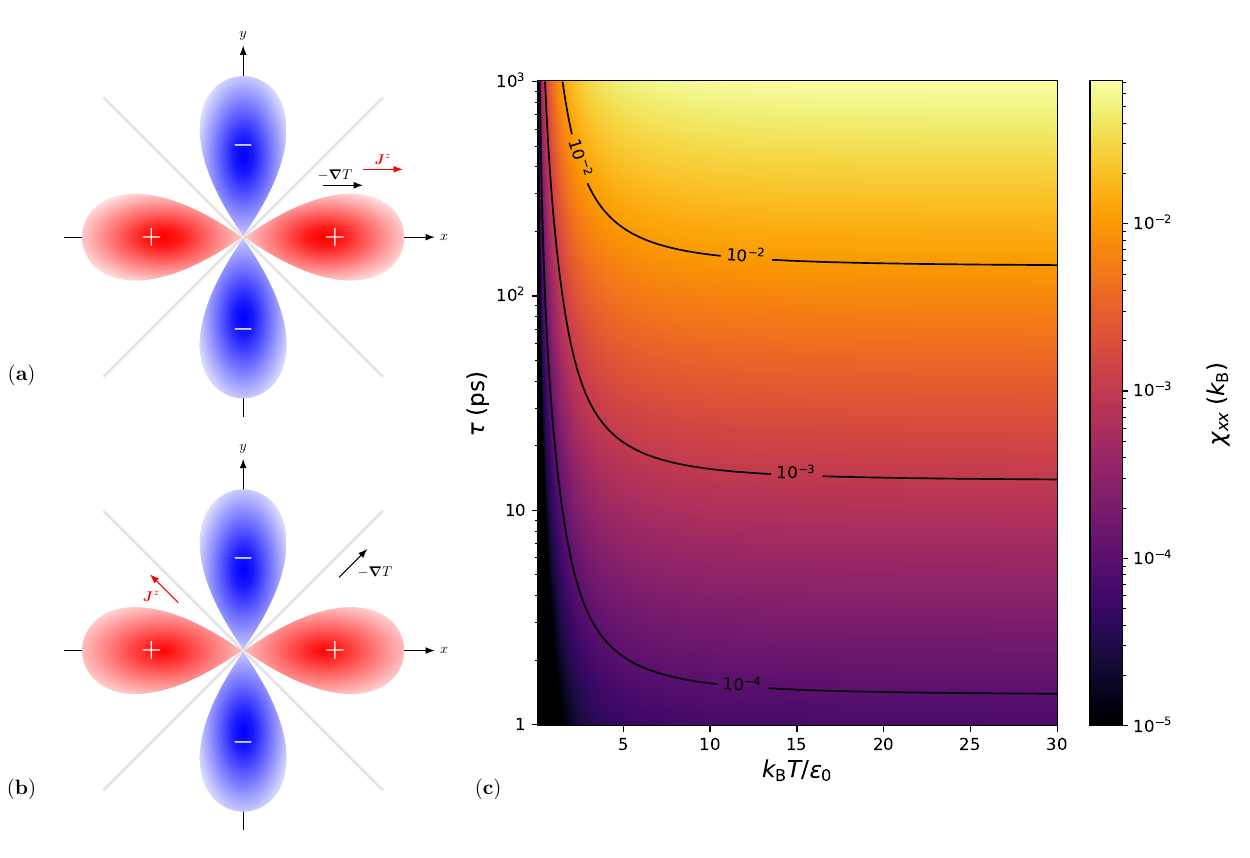}
\caption{
(a,b) Schematic of the angular-momentum splitter effect, with the red and blue lobes representing the $d$-wave phonon angular momentum texture in reciprocal space. The red color denotes the positive part of the texture and the blue the negative one. The nodal lines of the texture along the diagonals are colored in gray. We distinguish the cases where $-\vec{\nabla}T$ is applied (a) along the $x$-axis, resulting in a longitudinal angular momentum current response, and (b)
along the nodal lines of the texture, thus generating a transverse angular momentum current response.
(c) Response tensor element $\chi_{xx}$ (angular momentum conductivity) as a function of temperature $T$ and constant phonon relaxation time $\tau$. 
}
\label{fig:orbital_splitter}
\end{figure*}

Axial phonons with a $d$-wave angular momentum texture can exhibit characteristic thermal transport responses such as an angular momentum splitter effect, as pointed out in Ref.~\cite{bendin2026dwavephononangularmomentum}. To see so, consider the constitutive equation
\begin{equation}
    j_{\mu}^z = - \chi_{\mu \nu} \partial_{\nu}T, 
    \label{eq:constitutive}
\end{equation}
where $j_{\mu}^z$ is the $\mu$th component of a non-equilibrium angular momentum current (polarized along the $z$ direction), $\partial_{\nu}T$ is a temperature gradient along the $\nu$th direction, and the response tensor $\chi_{\mu \nu}$ is the thermal angular momentum conductivity. 
To compute $\chi_{\mu \nu}$ from microscopics, we define the angular momentum polarized current as 
 \begin{equation}
     \vec{j}^z = \frac{1}{V}\sum_{\vec{k}, \sigma} l_{\vec{k} \sigma} \vec{v}_{\vec{k} \sigma} f_{\vec{k} \sigma},
\label{eq: orbital current}
 \end{equation}
where $l_{\vec{k} \sigma} $ is the phonon angular momentum given in Eq.~\eqref{eq: OAM phonons texture}, $\vec{v}_{\vec{k} \sigma}= \frac{1}{\hbar}\frac{\partial \varepsilon_{\vec{k} \sigma}}{\partial \vec{k}}$ is the phonon group velocity, and $f_{\vec{k} \sigma}$ is the steady-state distribution function of the phonons induced by the temperature gradient. 
We assume that the temperature gradient is applied in the $\nu$th direction and utilize Boltzmann's semiclassical kinetic equation which upon linearization in the temperature gradient reduces to

\begin{equation}
    \frac{\partial f^0_{\vec{k} \sigma}}{\partial T} v_{\vec{k} \sigma}^{\nu} \partial_{\nu} T = - \frac{f_{\vec{k} \sigma} - f_{\vec{k} \sigma}^0}{\tau}, 
    \label{eq: Boltzmann equation}
\end{equation}
where $ f^0_{\vec{k} \sigma} = \left(\mathrm{e}^{\beta \varepsilon_{\vec{k} \sigma} }-1 \right)^{-1}$ is the equilibrium Bose-Einstein distribution function, and $\tau$ is a constant phonon relaxation time. We solve Eq.~\eqref{eq: Boltzmann equation} for $f_{\vec{k} \sigma}$ and compare with Eq.~\eqref{eq:constitutive}, to extract the response tensor
\begin{equation}
     \chi_{\mu \nu} = \frac{\tau}{k_{\text{B}} T^2 V}\sum_{\sigma \vec{k}} l_{\vec{k} \sigma} v_{\vec{k} \sigma}^{\mu} v_{\vec{k} \sigma}^{\nu} \varepsilon_{\vec{k} \sigma} f_{\vec{k} \sigma}^0 \left(f_{\vec{k} \sigma}^0 + 1\right). 
    \label{eq: response tensor}
\end{equation}
The symmetries of the $d$-wave restrict the shape of the response tensor as follows:
\begin{equation}
\begin{aligned}
    \chi = 
    \begin{pmatrix}
        \chi_{xx} & \chi_{xy} \\
       \chi_{yx} &   \chi_{yy} 
    \end{pmatrix}
    = 
    \begin{pmatrix}
        \chi_{xx} & 0 \\
        0 &   -\chi_{xx}
    \end{pmatrix}
\end{aligned}.
\label{eq: orbital Seeback}
\end{equation}
The off-diagonal elements are zero because the product $v_{\vec{k} \sigma}^{x} v_{\vec{k} \sigma}^{y} l_{\vec{k} \sigma}$ is an odd function in $\vec{k}$ and thus vanishes upon integration. The diagonal elements are finite and related by a factor of $-1$ due to $\mathcal{C}_4 \mathcal{T}$ symmetry.

Since, according to Eq.~\eqref{eq: orbital Seeback}, a temperature gradient generates a longitudinal angular-momentum current, we refer to this response as the phonon angular-momentum Seebeck effect. This effect is depicted in Fig.~\ref{fig:orbital_splitter}(a). Its key signature is the pronounced directional dependence: rotating the temperature gradient by $90^\circ$ reverses the sign of the response, reflecting the $d$-wave character. We note that this effect is necessarily accompanied by a phonon heat current parallel to the temperature gradient, in accordance with Fourier’s law.

To produce a pure angular momentum response not accompanied by a heat current, we consider  \textit{transverse} responses and apply $-\vec{\nabla}T$ along the nodal lines of $l_{\vec{k}\sigma}$ that are located along the two diagonals $k_x=k_y$ and $k_x=-k_y$, 
as seen in Fig.~\ref{fig:orbital_splitter}(b).
Rotating the diagonal response tensor $\chi$ by $45^\circ$ (with $R$ indicating the corresponding rotation matrix) we obtain 
\begin{equation}
    \widetilde{\chi} = R^\text{T} \chi R 
    =
    \begin{pmatrix}
        0 & - \chi_{xx} \\
        -\chi_{xx} & 0
    \end{pmatrix}. 
\label{eq: Nernst response}
\end{equation}
This purely transverse but symmetric response is reminiscent of the spin splitter effect in $d$-wave spin altermagnets \cite{PhysRevLett.126.127701} and can be referred to as phonon angular momentum splitter effect. 

This splitter effect (1) is an intraband response proportional to the relaxation time $\tau$, (2) enters the symmetric part of $\chi$, and (3) requires the force to be applied along the nodal lines of the angular momentum texture. It is therefore sharply different from the phonon angular momentum Nernst effect discussed in Ref.~\cite{doi:10.1021/acs.nanolett.0c03220}. The latter (1) is an interband response that is not linear in $\tau$ but instead converges to a finite value as $\tau \to 0$, (2) enters the \emph{antisymmetric} part of $\chi$, and (3) appears for any direction of the force $\nabla T$. 

The angular momentum splitter effect, or likewise the sign change of the corresponding angular momentum Seebeck effect upon rotating the temperature gradient by $90$ degrees, is a characteristic hallmark of $d$-wave axial phonons. Using Eq.~\eqref{eq: response tensor}, the quantitative value of $\chi_{xx}$ as a function of temperature and relaxation time (between $\tau = 1$ ps to $\tau = 1$ ns) is depicted in Fig.~\ref{fig:orbital_splitter}(c). Relaxation times of the order of ns are realistic for long-lived acoustic phonons in crystalline solids at sufficiently low or intermediate temperatures, where Umklapp scattering due to lattice anharmonicity is suppressed \cite{zhao2008fullspectrum}.
Assuming $\tau = 1\text{ ns}$ and the parameters for which we have calculated the phonon angular momentum in Fig.~\ref{fig: POAM} of Sec.~\ref{main sec: MBC dynamics},
we obtain a peak phonon angular momentum conductivity of $\chi_{xx} \sim 10^{-4}\; k_\text{B}$.  

\section{Discussion and Conclusion}
\label{sec:Conclusion}
We have shown that orbital altermagnets can host axial phonons with a $d$-wave angular-momentum texture in reciprocal space. This texture originates from the ubiquitous non-relativistic electron-phonon coupling present in all solids. The resulting phonon bands exhibit a pronounced $d$-wave angular-momentum structure in both acoustic and optical branches.  
In addition, we predict a phonon angular-momentum splitter effect, which provides a promising route for experimental detection.

Regarding candidate materials, orbital altermagnets---systems with $d$-wave loop-current order as considered here---offer a natural platform. Besides cuprates \cite{PhysRevB.55.14554}, recent theoretical work identifies kagome metals of the family AV$_3$Sb$_5$ \cite{chakraborty2025orbitalaltermagnetismkagomelattice} as promising candidates, where $A$ denotes alkali metals such as K, Rb, or Cs. Other studies suggest a possible coexistence of orbital altermagnetism with (conventional or unconventional) spin magnetism in several materials, such as CuBr$_2$, VS$_2$, MoO, and CrO \cite{pan2026orbitalaltermagnetism}. Orbital altermagnetism may also be engineered via adatoms, as demonstrated in Ref.~\cite{PhysRevLett.134.146001}. Beyond orbital systems, analogous physics should also arise in spin altermagnets when spin-orbit coupling is included \cite{bendin2026dwavephononangularmomentum}. Our results therefore establish a general principle: both orbital and spin altermagnets can host $d$-wave axial phonons as a robust consequence of electron-phonon coupling.

The phonon angular-momentum splitter effect leads to an accumulation of angular momentum at sample boundaries, which may be detected via an inverse spin Hall voltage, similar to experiments on quartz \cite{PhysRevLett.132.056302, nabei2026orbital}. While our analysis demonstrates that the effect is symmetry-allowed and generically expected, a quantitative description calls for further refinement. In particular, a more complete treatment of the angular-momentum current in Eq.~\eqref{eq: orbital current} should account for the non-commutativity of velocity and angular-momentum operators, and for the non-conservation of phonon angular momentum. With this in mind, our definition of the angular momentum polarized current $\vec{j}^z$ stems from a semi-classical, Boltzmann-like treatment of the angular momentum current density operator, where potential interband coherent contributions have been neglected.
In addition, a realistic description of phonon transport requires incorporating momentum- and temperature-dependent phonon lifetimes arising from anharmonic scattering. These additions will refine the temperature dependence of $\chi_{xx}$, 
which in the present work is governed by Bose-Einstein statistics alone.

The peak value of the estimated response tensor element corresponds to $\chi_{xx} \sim 10^{-4} k_{\text{B}}$ for the simple toy model considered here. To benchmark the magnitude of our result, we compare it with previously reported transverse and longitudinal angular momentum current responses. Reference~\cite{go2023intrinsicmagnonorbitalhall} predicts an orbital angular momentum Nernst conductivity of magnons of the order of $10^{-2} k_{\text{B}}$ in collinear antiferromagnets, while the magnon spin Nernst conductivity for the same type of system has been computed to be of the order of $10^{-4} k_{\text{B}}$ in Ref.~\cite{PhysRevLett.117.217202}. Regarding spin altermagnets, Ref.~\cite{cui2023efficientspinseebeckspin} obtained a value of $\sim 0.1 k_{\text{B}}$ for the magnon spin Seebeck and splitter coefficients, albeit for an exceptionally long magnon relaxation time, corresponding to systems of high magnetic quality. Consequently, the magnitude of our phonon angular momentum splitter conductivity 
is comparable to established spincaloritronic responses, and may be further enhanced in materials with smaller indirect electronic band gaps, as discussed at the end of Sec.~\ref{main sec: MBC dynamics}. We have thus demonstrated that $d$-wave phonon angular momentum transport can generate sizable longitudinal and transverse signals in orbital altermagnets. 

More broadly, our work opens several directions for future research. Extensions to $g$- and $i$-wave axial phonons, as recently proposed in Ref.~\cite{wang2026alteraxialphononscollinearmagnets}, in (orbital) altermagnets are natural next steps, as is the exploration of their dynamical and transport signatures. It will be particularly interesting to investigate the interplay between odd-partiy wave angular-momentum textures, such as $p$-wave textures \cite{li2026pwaveorbitalmagnetism}, and axial phonon responses. Together, these developments point toward a rich landscape of symmetry-controlled phonon phenomena in quantum materials with unconventional magnetic order.

\section{acknowledgments}
We would like to thank Yafei Ren, Verena Brehm, Dominik Juraschek and Robin R.~Neumann for helpful discussions.
This work was funded by the German Research Foundation (DFG) as part of Project No.~504261060 (Emmy Noether Programme). We acknowledge support by the Dynamics and Topology Center (TopDyn) funded by the State of Rhineland-Palatinate.

\bibliography{chiral_phonons}

@article{Juraschek2025,
  title = {Chiral phonons},
  volume = {21},
  ISSN = {1745-2481},
  url = {http://dx.doi.org/10.1038/s41567-025-03001-9},
  DOI = {10.1038/s41567-025-03001-9},
  number = {10},
  journal = {Nature Physics},
  publisher = {Springer Science and Business Media LLC},
  author = {Juraschek,  Dominik M. and Geilhufe,  R. Matthias and Zhu,  Hanyu and Basini,  Martina and Baum,  Peter and Baydin,  Andrey and Chaudhary,  Swati and Fechner,  Michael and Flebus,  Benedetta and Grissonnanche,  Gael and Kirilyuk,  Andrei I. and Lemeshko,  Mikhail and Maehrlein,  Sebastian F. and Mignolet,  Maxime and Murakami,  Shuichi and Niu,  Qian and Nowak,  Ulrich and Romao,  Carl P. and Rostami,  Habib and Satoh,  Takuya and Spaldin,  Nicola A. and Ueda,  Hiroki and Zhang,  Lifa},
  year = {2025},
  month = Sep,
  pages = {1532–1540}
}

@misc{wang2026alteraxialphononscollinearmagnets,
      title={Alteraxial Phonons in Collinear Magnets}, 
      author={Fuyi Wang and Junqi Xu and Xinqi Liu and Huaiqiang Wang and Lifa Zhang and Haijun Zhang},
      year={2026},
      eprint={2512.07518},
      archivePrefix={arXiv},
      primaryClass={cond-mat.mtrl-sci},
      url={https://arxiv.org/abs/2512.07518}, 
}

@article{PhysRevLett.127.125901,
  title = {Phonon Helicity Induced by Electronic Berry Curvature in Dirac Materials},
  author = {Hu, Lun-Hui and Yu, Jiabin and Garate, Ion and Liu, Chao-Xing},
  journal = {Phys. Rev. Lett.},
  volume = {127},
  issue = {12},
  pages = {125901},
  numpages = {6},
  year = {2021},
  month = {Sep},
  publisher = {American Physical Society},
  doi = {10.1103/PhysRevLett.127.125901},
  url = {https://link.aps.org/doi/10.1103/PhysRevLett.127.125901}
}

@article{doi:10.1021/acs.nanolett.4c00606,
author = {Wang, Tingting and Sun, Hong and Li, Xiaozhe and Zhang, Lifa},
title = {Chiral Phonons: Prediction, Verification, and Application},
journal = {Nano Letters},
volume = {24},
number = {15},
pages = {4311-4318},
year = {2024},
doi = {10.1021/acs.nanolett.4c00606},
note ={PMID: 38587210},
URL = {https://doi.org/10.1021/acs.nanolett.4c00606},
eprint = {https://doi.org/10.1021/acs.nanolett.4c00606}
}

@misc{chen2026topologicalphononics,
      title={Topological phononics}, 
      author={Zeguo Chen and Tiantian Zhang and Xulong Wang and Jiangxu Li and Zhi-Kang Lin and Feng Gao and Li-Wei Wang and Yizhou Liu and Qi Wang and Xiujuan Zhang and Guancong Ma and Xingqiu Chen and Minghui Lu and Yanfeng Chen and Jian-Hua Jiang},
      year={2026},
      eprint={2605.20900},
      archivePrefix={arXiv},
      primaryClass={cond-mat.mtrl-sci},
      url={https://arxiv.org/abs/2605.20900}, 
}

@article{Zhang2025,
  title = {Weyl phonons: the connection of topology and chirality},
  volume = {16},
  ISSN = {2041-1723},
  url = {http://dx.doi.org/10.1038/s41467-025-58913-0},
  DOI = {10.1038/s41467-025-58913-0},
  number = {1},
  journal = {Nature Communications},
  publisher = {Springer Science and Business Media LLC},
  author = {Zhang, Tiantian and Murakami, Shuichi and Miao, Hu},
  year = {2025},
  month = Apr 
}

@misc{zhang2025advancesphononsbandtopology,
      title={Advances in Phonons: From Band Topology to Phonon Chirality}, 
      author={Tiantian Zhang and Yizhou Liu and Hu Miao and Shuichi Murakami},
      year={2025},
      eprint={2505.06179},
      archivePrefix={arXiv},
      primaryClass={cond-mat.mtrl-sci},
      url={https://arxiv.org/abs/2505.06179}, 
}

@article{THE_Lifa_Zhang,
  title = {The phonon Hall effect: theory and application},
  author = { Lifa Zhang, Jie Ren, Jian-Sheng Wang and Baowen Li },
  journal = { Journal of Physics: Condensed Matter, Volume 23, Number 30},
  volume = {23},
  year = {2011},
  month = {Jul},
  publisher = {American Physical Society},
  doi = {10.1088/0953-8984/23/30/305402},
  url = {https://iopscience.iop.org/article/10.1088/0953-8984/23/30/305402}
}

@misc{bendin2026dwavephononangularmomentum,
      title={D-Wave Phonon Angular Momentum Texture in Altermagnets by Magnon-Phonon-Hybridization}, 
      author={Hannah Bendin and Alexander Mook and Ingrid Mertig and Robin R. Neumann},
      year={2026},
      eprint={2511.08357},
      archivePrefix={arXiv},
      primaryClass={cond-mat.mes-hall},
      url={https://arxiv.org/abs/2511.08357}, 
}

@article{Lifa_phonon_De_Hass_effect,
  title = {Angular Momentum of Phonons and the Einstein--de Haas Effect},
  author = {Zhang, Lifa and Niu, Qian},
  journal = {Phys. Rev. Lett.},
  volume = {112},
  issue = {8},
  pages = {085503},
  numpages = {5},
  year = {2014},
  month = {Feb},
  publisher = {American Physical Society},
  doi = {10.1103/PhysRevLett.112.085503},
  url = {https://link.aps.org/doi/10.1103/PhysRevLett.112.085503}
}

@article{PhysRevB.108.134307,
  title = {Classification of materials with phonon angular momentum and microscopic origin of angular momentum},
  author = {Coh, Sinisa},
  journal = {Phys. Rev. B},
  volume = {108},
  issue = {13},
  pages = {134307},
  numpages = {8},
  year = {2023},
  month = {Oct},
  publisher = {American Physical Society},
  doi = {10.1103/PhysRevB.108.134307},
  url = {https://link.aps.org/doi/10.1103/PhysRevB.108.134307}
}

@article{
doi:10.1126/science.aar2711,
author = {Hanyu Zhu  and Jun Yi  and Ming-Yang Li  and Jun Xiao  and Lifa Zhang  and Chih-Wen Yang  and Robert A. Kaindl  and Lain-Jong Li  and Yuan Wang  and Xiang Zhang },
title = {Observation of chiral phonons},
journal = {Science},
volume = {359},
number = {6375},
pages = {579-582},
year = {2018},
doi = {10.1126/science.aar2711},
URL = {https://www.science.org/doi/abs/10.1126/science.aar2711}
}

@article{PhysRevLett.115.115502,
  title = {Chiral Phonons at High-Symmetry Points in Monolayer Hexagonal Lattices},
  author = {Zhang, Lifa and Niu, Qian},
  journal = {Phys. Rev. Lett.},
  volume = {115},
  issue = {11},
  pages = {115502},
  numpages = {5},
  year = {2015},
  month = {Sep},
  publisher = {American Physical Society},
  doi = {10.1103/PhysRevLett.115.115502},
  url = {https://link.aps.org/doi/10.1103/PhysRevLett.115.115502}
}

@article{doi:10.1021/acs.nanolett.0c03220,
author = {Park, Sungjoon and Yang, Bohm-Jung},
title = {Phonon Angular Momentum Hall Effect},
journal = {Nano Letters},
volume = {20},
number = {10},
pages = {7694-7699},
year = {2020},
doi = {10.1021/acs.nanolett.0c03220},
note ={PMID: 32955897},
URL = { 
        https://doi.org/10.1021/acs.nanolett.0c03220
},
eprint = { 
        https://doi.org/10.1021/acs.nanolett.0c03220
}
}

@article{PhysRevB.100.094303,
  title = {Chiral phonons in kagome lattices},
  author = {Chen, Hao and Wu, Weikang and Yang, Shengyuan A. and Li, Xiao and Zhang, Lifa},
  journal = {Phys. Rev. B},
  volume = {100},
  issue = {9},
  pages = {094303},
  numpages = {6},
  year = {2019},
  month = {Sep},
  publisher = {American Physical Society},
  doi = {10.1103/PhysRevB.100.094303},
  url = {https://link.aps.org/doi/10.1103/PhysRevB.100.094303}
}

@misc{yao2026dynamicalorbitalangularmomentum,
      title={Dynamical Orbital Angular Momentum Induced by Circularly Polarized Phonons}, 
      author={Dapeng Yao and Dongwook Go and Yuriy Mokrousov and Shuichi Murakami},
      year={2026},
      eprint={2511.09271},
      archivePrefix={arXiv},
      primaryClass={cond-mat.mes-hall},
      url={https://arxiv.org/abs/2511.09271}, 
}

@article{Mrudul_2025,
   title={Generation of phonons with angular momentum during ultrafast demagnetization},
   volume={112},
   ISSN={2469-9969},
   url={http://dx.doi.org/10.1103/nt8w-47hb},
   DOI={10.1103/nt8w-47hb},
   number={18},
   journal={Physical Review B},
   publisher={American Physical Society (APS)},
   author={Mrudul, M. S. and Weißenhofer, Markus and Oppeneer, Peter M.},
   year={2025},
   month=Nov }

@article{Tauchert2022,
  title = {Polarized phonons carry angular momentum in ultrafast demagnetization},
  volume = {602},
  ISSN = {1476-4687},
  url = {http://dx.doi.org/10.1038/s41586-021-04306-4},
  DOI = {10.1038/s41586-021-04306-4},
  number = {7895},
  journal = {Nature},
  publisher = {Springer Science and Business Media LLC},
  author = {Tauchert,  S. R. and Volkov,  M. and Ehberger,  D. and Kazenwadel,  D. and Evers,  M. and Lange,  H. and Donges,  A. and Book,  A. and Kreuzpaintner,  W. and Nowak,  U. and Baum,  P.},
  year = {2022},
  month = Feb,
  pages = {73–77}
}

@article{PhysRevLett.132.056302,
  title = {Chirality-Induced Selectivity of Phonon Angular Momenta in Chiral Quartz Crystals},
  author = {Ohe, Kazuki and Shishido, Hiroaki and Kato, Masaki and Utsumi, Shoyo and Matsuura, Hiroyasu and Togawa, Yoshihiko},
  journal = {Phys. Rev. Lett.},
  volume = {132},
  issue = {5},
  pages = {056302},
  numpages = {7},
  year = {2024},
  month = {Jan},
  publisher = {American Physical Society},
  doi = {10.1103/PhysRevLett.132.056302},
  url = {https://link.aps.org/doi/10.1103/PhysRevLett.132.056302}
}

@article{Davies2024,
  title = {Phononic switching of magnetization by the ultrafast Barnett effect},
  volume = {628},
  ISSN = {1476-4687},
  url = {http://dx.doi.org/10.1038/s41586-024-07200-x},
  DOI = {10.1038/s41586-024-07200-x},
  number = {8008},
  journal = {Nature},
  publisher = {Springer Science and Business Media LLC},
  author = {Davies,  C. S. and Fennema,  F. G. N. and Tsukamoto,  A. and Razdolski,  I. and Kimel,  A. V. and Kirilyuk,  A.},
  year = {2024},
  month = Apr,
  pages = {540–544}
}

@article{Luo2023,
  title = {Large effective magnetic fields from chiral phonons in rare-earth halides},
  volume = {382},
  ISSN = {1095-9203},
  url = {http://dx.doi.org/10.1126/science.adi9601},
  DOI = {10.1126/science.adi9601},
  number = {6671},
  journal = {Science},
  publisher = {American Association for the Advancement of Science (AAAS)},
  author = {Luo,  Jiaming and Lin,  Tong and Zhang,  Junjie and Chen,  Xiaotong and Blackert,  Elizabeth R. and Xu,  Rui and Yakobson,  Boris I. and Zhu,  Hanyu},
  year = {2023},
  month = Nov,
  pages = {698–702}
}

@article{PhysRevMaterials.3.064405,
  title = {Orbital magnetic moments of phonons},
  author = {Juraschek, Dominik M. and Spaldin, Nicola A.},
  journal = {Phys. Rev. Mater.},
  volume = {3},
  issue = {6},
  pages = {064405},
  numpages = {8},
  year = {2019},
  month = {Jun},
  publisher = {American Physical Society},
  doi = {10.1103/PhysRevMaterials.3.064405},
  url = {https://link.aps.org/doi/10.1103/PhysRevMaterials.3.064405}
}

@article{PhysRevB.110.094401,
  title = {Giant effective magnetic moments of chiral phonons from orbit-lattice coupling},
  author = {Chaudhary, Swati and Juraschek, Dominik M. and Rodriguez-Vega, Martin and Fiete, Gregory A.},
  journal = {Phys. Rev. B},
  volume = {110},
  issue = {9},
  pages = {094401},
  numpages = {24},
  year = {2024},
  month = {Sep},
  publisher = {American Physical Society},
  doi = {10.1103/PhysRevB.110.094401},
  url = {https://link.aps.org/doi/10.1103/PhysRevB.110.094401}
}

@article{nabei2026orbital,
  title = {Orbital Seebeck effect induced by chiral phonons},
  author = {Yoji Nabei and Cong Yang and Hong Sun and Hana Jones and Thuc Mai and Tian Wang and Rikard Bodin and Binod Pandey and Ziqi Wang and Yuzan Xiong and Andrew H. Comstock and Benjamin Ewing and John Bingen and Rui Sun and Dmitry Smirnov and Wei Zhang and Axel Hoffmann and Rahul Rao and Ming Hu and Z. Valy Vardeny and Binghai Yan and Xiaosong Li and Jun Zhou and Jun Liu and Dali Sun},
  year = {2026},
  publisher = {Springer Science and Business Media LLC},
  journal = {Nature Physics},
  volume = {22},
  pages = {245-251},
  doi = {10.1038/s41567-025-03134-x},
  url = {https://doi.org/10.1038/s41567-025-03134-x}
}

@article{Grissonnanche2020,
  title = {Chiral phonons in the pseudogap phase of cuprates},
  volume = {16},
  ISSN = {1745-2481},
  url = {http://dx.doi.org/10.1038/s41567-020-0965-y},
  DOI = {10.1038/s41567-020-0965-y},
  number = {11},
  journal = {Nature Physics},
  publisher = {Springer Science and Business Media LLC},
  author = {Grissonnanche,  G. and Thériault,  S. and Gourgout,  A. and Boulanger,  M.-E. and Lefran\c{c}ois,  E. and Ataei,  A. and Laliberté,  F. and Dion,  M. and Zhou,  J.-S. and Pyon,  S. and Takayama,  T. and Takagi,  H. and Doiron-Leyraud,  N. and Taillefer,  L.},
  year = {2020},
  month = Jul,
  pages = {1108–1111}
}

@article{tyf9-thv8,
  title = {Thermal Hall conductivity of electron-doped cuprates: Electrons and phonons},
  author = {Boulanger, Marie-Eve and Chen, Lu and Oliviero, Vincent and Vignolles, David and Grissonnanche, Ga\"el and Xu, Ke-Jun and Shen, Zhi-Xun and Proust, Cyril and Baglo, Jordan and Taillefer, Louis},
  journal = {Phys. Rev. B},
  volume = {113},
  issue = {9},
  pages = {094511},
  numpages = {8},
  year = {2026},
  month = {Mar},
  publisher = {American Physical Society},
  doi = {10.1103/tyf9-thv8},
  url = {https://link.aps.org/doi/10.1103/tyf9-thv8}
}

@article{PhysRevB.111.134414,
  title = {Theory of spin magnetization driven by chiral phonons},
  author = {Yao, Dapeng and Murakami, Shuichi},
  journal = {Phys. Rev. B},
  volume = {111},
  issue = {13},
  pages = {134414},
  numpages = {9},
  year = {2025},
  month = {Apr},
  publisher = {American Physical Society},
  doi = {10.1103/PhysRevB.111.134414},
  url = {https://link.aps.org/doi/10.1103/PhysRevB.111.134414}
}

@article{PhysRevB.103.214302,
  title = {Universal features of canonical phonon angular momentum without time-reversal symmetry},
  author = {Komiyama, Hisayoshi and Murakami, Shuichi},
  journal = {Phys. Rev. B},
  volume = {103},
  issue = {21},
  pages = {214302},
  numpages = {10},
  year = {2021},
  month = {Jun},
  publisher = {American Physical Society},
  doi = {10.1103/PhysRevB.103.214302},
  url = {https://link.aps.org/doi/10.1103/PhysRevB.103.214302}
}

@article{PhysRevB.110.094427,
  title = {Atomistic spin dynamics simulations of magnonic spin Seebeck and spin Nernst effects in altermagnets},
  author = {Wei\ss{}enhofer, Markus and Marmodoro, Alberto},
  journal = {Phys. Rev. B},
  volume = {110},
  issue = {9},
  pages = {094427},
  numpages = {14},
  year = {2024},
  month = {Sep},
  publisher = {American Physical Society},
  doi = {10.1103/PhysRevB.110.094427},
  url = {https://link.aps.org/doi/10.1103/PhysRevB.110.094427}
}

@article{Romao2024,
  title = {Phonon-Induced Geometric Chirality},
  volume = {18},
  ISSN = {1936-086X},
  url = {http://dx.doi.org/10.1021/acsnano.4c05978},
  DOI = {10.1021/acsnano.4c05978},
  number = {43},
  journal = {ACS Nano},
  publisher = {American Chemical Society (ACS)},
  author = {Romao,  Carl P. and Juraschek,  Dominik M.},
  year = {2024},
  month = Oct,
  pages = {29550–29557}
}

@article{RevModPhys.64.51,
  title = {The geometric phase in molecular systems},
  author = {Mead, C. Alden},
  journal = {Rev. Mod. Phys.},
  volume = {64},
  issue = {1},
  pages = {51--85},
  numpages = {0},
  year = {1992},
  month = {Jan},
  publisher = {American Physical Society},
  doi = {10.1103/RevModPhys.64.51},
  url = {https://link.aps.org/doi/10.1103/RevModPhys.64.51}
}

@article{PhysRevLett.113.263004,
  title = {Is the Molecular Berry Phase an Artifact of the Born-Oppenheimer Approximation?},
  author = {Min, Seung Kyu and Abedi, Ali and Kim, Kwang S. and Gross, E. K. U.},
  journal = {Phys. Rev. Lett.},
  volume = {113},
  issue = {26},
  pages = {263004},
  numpages = {5},
  year = {2014},
  month = {Dec},
  publisher = {American Physical Society},
  doi = {10.1103/PhysRevLett.113.263004},
  url = {https://link.aps.org/doi/10.1103/PhysRevLett.113.263004}
}

@article{PhysRevLett.126.225703,
  title = {Intrinsic Vibrational Angular Momentum from Nonadiabatic Effects in Noncollinear Magnetic Molecules},
  author = {Bistoni, Oliviero and Mauri, Francesco and Calandra, Matteo},
  journal = {Phys. Rev. Lett.},
  volume = {126},
  issue = {22},
  pages = {225703},
  numpages = {5},
  year = {2021},
  month = {Jun},
  publisher = {American Physical Society},
  doi = {10.1103/PhysRevLett.126.225703},
  url = {https://link.aps.org/doi/10.1103/PhysRevLett.126.225703}
}

@article{7f8x-fkx9,
  title = {Inherited Berry curvature of phonons in Dirac materials with time-reversal symmetry},
  author = {Ghosh, Sayandip and Parlak, Sel\ifmmode \mbox{\c{c}}\else \c{c}\fi{}uk and Garate, Ion},
  journal = {Phys. Rev. B},
  volume = {112},
  issue = {17},
  pages = {174305},
  numpages = {17},
  year = {2025},
  month = {Nov},
  publisher = {American Physical Society},
  doi = {10.1103/7f8x-fkx9},
  url = {https://link.aps.org/doi/10.1103/7f8x-fkx9}
}

@misc{das2026antiferrochiralphononsmathcalpmathcaltsymmetricantiferromagnets,
      title={Antiferro-Chiral Phonons in $\mathcal{P}\mathcal{T}$-Symmetric Antiferromagnets}, 
      author={Sanjib Kumar Das and Randy Yeh and Yafei Ren},
      year={2026},
      eprint={2605.08490},
      archivePrefix={arXiv},
      primaryClass={cond-mat.mes-hall},
      url={https://arxiv.org/abs/2605.08490}, 
}

@article{PhysRevLett.134.206701,
  title = {Nonreciprocal Phonons in $\mathcal{P}\mathcal{T}$-Symmetric Antiferromagnets},
  author = {Ren, Yafei and Saparov, Daniyar and Niu, Qian},
  journal = {Phys. Rev. Lett.},
  volume = {134},
  issue = {20},
  pages = {206701},
  numpages = {7},
  year = {2025},
  month = {May},
  publisher = {American Physical Society},
  doi = {10.1103/PhysRevLett.134.206701},
  url = {https://link.aps.org/doi/10.1103/PhysRevLett.134.206701}
}

@article{Chiral_Phonons_Angular_Momentum,
  title = {Lattice dynamics with molecular Berry curvature: Chiral optical phonons},
  author = {Saparov, Daniyar and Xiong, Bangguo and Ren, Yafei and Niu, Qian},
  journal = {Phys. Rev. B},
  volume = {105},
  issue = {6},
  pages = {064303},
  numpages = {11},
  year = {2022},
  month = {Feb},
  publisher = {American Physical Society},
  doi = {10.1103/PhysRevB.105.064303},
  url = {https://link.aps.org/doi/10.1103/PhysRevB.105.064303}
}

@article{PhysRevLett.119.075301,
  title = {Circular Phonon Dichroism in Weyl Semimetals},
  author = {Liu, Donghao and Shi, Junren},
  journal = {Phys. Rev. Lett.},
  volume = {119},
  issue = {7},
  pages = {075301},
  numpages = {5},
  year = {2017},
  month = {Aug},
  publisher = {American Physical Society},
  doi = {10.1103/PhysRevLett.119.075301},
  url = {https://link.aps.org/doi/10.1103/PhysRevLett.119.075301}
}

@misc{li2025phonondichroismsrevealingunusual,
      title={Phonon Dichroisms Revealing Unusual Electronic Quantum Geometry}, 
      author={Ding Li and Guoao Yang and Tao Qin and Jianhui Zhou and Yugui Yao},
      year={2025},
      eprint={2511.16141},
      archivePrefix={arXiv},
      primaryClass={cond-mat.mes-hall},
      url={https://arxiv.org/abs/2511.16141}, 
}

@phdthesis{Saparov_thesis,
  author       = {Daniyar Saparov},
  title        = {Effect of Molecular Berry Curvature on the Dynamics of Phonons},
  school       = {The University of Texas at Austin},
  year         = {2022},
  address      = {Austin, Texas, USA},
  url          = {dx.doi.org/10.26153/
tsw/42807}
}

@article{PhysRevX.15.011036,
  title = {Phonon Thermal Hall Effect in Mott Insulators via Skew Scattering by the Scalar Spin Chirality},
  author = {Oh, Taekoo and Nagaosa, Naoto},
  journal = {Phys. Rev. X},
  volume = {15},
  issue = {1},
  pages = {011036},
  numpages = {13},
  year = {2025},
  month = {Feb},
  publisher = {American Physical Society},
  doi = {10.1103/PhysRevX.15.011036},
  url = {https://link.aps.org/doi/10.1103/PhysRevX.15.011036}
}

@article{bfll-sdrb,
  title = {Emergence of Chiral Phonons in Two-Dimensional Kagome Lattices Harboring Electronic Chirality},
  author = {Chen, Yanru and Qin, Wei and Zhang, Shunhong and Cui, Ping and Niu, Qian and Zhang, Zhenyu},
  journal = {Phys. Rev. Lett.},
  volume = {135},
  issue = {12},
  pages = {126608},
  numpages = {9},
  year = {2025},
  month = {Sep},
  publisher = {American Physical Society},
  doi = {10.1103/bfll-sdrb},
  url = {https://link.aps.org/doi/10.1103/bfll-sdrb}
}

@article{MBP_Phonon_THE_Nagaosa,
  title = {Berry Phase of Phonons and Thermal Hall Effect in Nonmagnetic Insulators},
  author = {Saito, Takuma and Misaki, Kou and Ishizuka, Hiroaki and Nagaosa, Naoto},
  journal = {Phys. Rev. Lett.},
  volume = {123},
  issue = {25},
  pages = {255901},
  numpages = {5},
  year = {2019},
  month = {Dec},
  publisher = {American Physical Society},
  doi = {10.1103/PhysRevLett.123.255901},
  url = {https://link.aps.org/doi/10.1103/PhysRevLett.123.255901}
}

@article{tpjd-dh1m,
  title = {Ab Initio Theory of Phonon Magnetic Moment Induced by Electron-Phonon Coupling in Magnetic Materials},
  author = {Wang, Fuyi and Liu, Xinqi and Sun, Hong and Wang, Huaiqiang and Murakami, Shuichi and Zhang, Lifa and Zhang, Haijun and Xing, Dingyu},
  journal = {Phys. Rev. Lett.},
  volume = {135},
  issue = {25},
  pages = {256701},
  numpages = {8},
  year = {2025},
  month = {Dec},
  publisher = {American Physical Society},
  doi = {10.1103/tpjd-dh1m},
  url = {https://link.aps.org/doi/10.1103/tpjd-dh1m}
}

@misc{dhakal2025theoryintrinsicphononthermal,
      title={Theory of Intrinsic Phonon Thermal Hall Effect in $\alpha$-RuCl$_3$}, 
      author={Ramesh Dhakal and David A. S. Kaib and Kate Choi and Sananda Biswas and Roser Valenti and Stephen M. Winter},
      year={2025},
      eprint={2407.00660},
      archivePrefix={arXiv},
      primaryClass={cond-mat.str-el},
      url={https://arxiv.org/abs/2407.00660}, 
}

@article{PhysRevLett.130.086701,
  title = {Frequency Splitting of Chiral Phonons from Broken Time-Reversal Symmetry in ${\mathrm{CrI}}_{3}$},
  author = {Bonini, John and Ren, Shang and Vanderbilt, David and Stengel, Massimiliano and Dreyer, Cyrus E. and Coh, Sinisa},
  journal = {Phys. Rev. Lett.},
  volume = {130},
  issue = {8},
  pages = {086701},
  numpages = {6},
  year = {2023},
  month = {Feb},
  publisher = {American Physical Society},
  doi = {10.1103/PhysRevLett.130.086701},
  url = {https://link.aps.org/doi/10.1103/PhysRevLett.130.086701}
}

@article{PhysRevX.14.011041,
  title = {Adiabatic Dynamics of Coupled Spins and Phonons in Magnetic Insulators},
  author = {Ren, Shang and Bonini, John and Stengel, Massimiliano and Dreyer, Cyrus E. and Vanderbilt, David},
  journal = {Phys. Rev. X},
  volume = {14},
  issue = {1},
  pages = {011041},
  numpages = {30},
  year = {2024},
  month = {Mar},
  publisher = {American Physical Society},
  doi = {10.1103/PhysRevX.14.011041},
  url = {https://link.aps.org/doi/10.1103/PhysRevX.14.011041}
}

@misc{zhang2025generalabinitioframework,
      title={General ab initio framework for chiral phonons induced by electronic order}, 
      author={Shuai Zhang and Mengqi Wang and Tiantian Zhang},
      year={2025},
      eprint={2509.09253},
      archivePrefix={arXiv},
      primaryClass={cond-mat.mtrl-sci},
      url={https://arxiv.org/abs/2509.09253}, 
}

@article{mmdm-hrj4,
  title = {High-throughput quantification of altermagnetic band splitting},
  author = {Sufyan, Ali and Marfoua, Brahim and Larsson, J. Andreas and Loon, Erik van and Armiento, Rickard},
  journal = {Phys. Rev. Mater.},
  pages = {--},
  year = {2026},
  month = {Mar},
  publisher = {American Physical Society},
  doi = {10.1103/mmdm-hrj4},
  url = {https://link.aps.org/doi/10.1103/mmdm-hrj4}
}

@article{PhysRevX.12.031042,
  title = {Beyond Conventional Ferromagnetism and Antiferromagnetism: A Phase with Nonrelativistic Spin and Crystal Rotation Symmetry},
  author = {\ifmmode \check{S}\else \v{S}\fi{}mejkal, Libor and Sinova, Jairo and Jungwirth, Tomas},
  journal = {Phys. Rev. X},
  volume = {12},
  issue = {3},
  pages = {031042},
  numpages = {16},
  year = {2022},
  month = {Sep},
  publisher = {American Physical Society},
  doi = {10.1103/PhysRevX.12.031042},
  url = {https://link.aps.org/doi/10.1103/PhysRevX.12.031042}
}

@article{PhysRevX.12.040501,
  title = {Emerging Research Landscape of Altermagnetism},
  author = {\ifmmode \check{S}\else \v{S}\fi{}mejkal, Libor and Sinova, Jairo and Jungwirth, Tomas},
  journal = {Phys. Rev. X},
  volume = {12},
  issue = {4},
  pages = {040501},
  numpages = {27},
  year = {2022},
  month = {Dec},
  publisher = {American Physical Society},
  doi = {10.1103/PhysRevX.12.040501},
  url = {https://link.aps.org/doi/10.1103/PhysRevX.12.040501}
}

@article{fgc1-5blp,
  title = {Altermagnetic splitting of magnons in hematite $\ensuremath{\alpha}\text{\ensuremath{-}}{\mathrm{Fe}}_{2}{\mathrm{O}}_{3}$},
  author = {Hoyer, Rhea and Stavropoulos, P. Peter and Razpopov, Aleksandar and Valent\'{\i}, Roser and \ifmmode \check{S}\else \v{S}\fi{}mejkal, Libor and Mook, Alexander},
  journal = {Phys. Rev. B},
  volume = {112},
  issue = {6},
  pages = {064425},
  numpages = {22},
  year = {2025},
  month = {Aug},
  publisher = {American Physical Society},
  doi = {10.1103/fgc1-5blp},
  url = {https://link.aps.org/doi/10.1103/fgc1-5blp}
}

@article{PhysRevLett.131.256703,
  title = {Chiral Magnons in Altermagnetic ${\mathrm{RuO}}_{2}$},
  author = {\ifmmode \check{S}\else \v{S}\fi{}mejkal, Libor and Marmodoro, Alberto and Ahn, Kyo-Hoon and Gonz\'alez-Hern\'andez, Rafael and Turek, Ilja and Mankovsky, Sergiy and Ebert, Hubert and D'Souza, Sunil W. and \ifmmode \check{S}\else \v{S}\fi{}ipr, Ond\ifmmode \check{r}\else \v{r}\fi{}ej and Sinova, Jairo and Jungwirth, Tom\'a\ifmmode \check{s}\else \v{s}\fi{}},
  journal = {Phys. Rev. Lett.},
  volume = {131},
  issue = {25},
  pages = {256703},
  numpages = {6},
  year = {2023},
  month = {Dec},
  publisher = {American Physical Society},
  doi = {10.1103/PhysRevLett.131.256703},
  url = {https://link.aps.org/doi/10.1103/PhysRevLett.131.256703}
}

@article{PhysRevLett.133.156702,
  title = {Chiral Split Magnon in Altermagnetic MnTe},
  author = {Liu, Zheyuan and Ozeki, Makoto and Asai, Shinichiro and Itoh, Shinichi and Masuda, Takatsugu},
  journal = {Phys. Rev. Lett.},
  volume = {133},
  issue = {15},
  pages = {156702},
  numpages = {6},
  year = {2024},
  month = {Oct},
  publisher = {American Physical Society},
  doi = {10.1103/PhysRevLett.133.156702},
  url = {https://link.aps.org/doi/10.1103/PhysRevLett.133.156702}
}

@misc{pan2026orbitalaltermagnetism,
      title={Orbital Altermagnetism}, 
      author={Mingxiang Pan and Feng Liu and Huaqing Huang},
      year={2026},
      eprint={2510.00509},
      archivePrefix={arXiv},
      primaryClass={cond-mat.mtrl-sci},
      url={https://arxiv.org/abs/2510.00509}, 
}

@misc{chakraborty2025orbitalaltermagnetismkagomelattice,
      title={Orbital altermagnetism on the kagome lattice and possible application to $A$V$_3$Sb$_5$}, 
      author={Anzumaan R. Chakraborty and Fan Yang and Turan Birol and Rafael M. Fernandes},
      year={2025},
      eprint={2509.26596},
      archivePrefix={arXiv},
      primaryClass={cond-mat.str-el},
      url={https://arxiv.org/abs/2509.26596}, 
}

@article{Yu2025,
  title = {Altermagnetism from coincident Van Hove singularities: application to $\kappa$-Cl},
  volume = {16},
  ISSN = {2041-1723},
  url = {http://dx.doi.org/10.1038/s41467-025-57970-9},
  DOI = {10.1038/s41467-025-57970-9},
  number = {1},
  journal = {Nature Communications},
  publisher = {Springer Science and Business Media LLC},
  author = {Yu,  Yue and Suh,  Han Gyeol and Roig,  Mercè and Agterberg,  Daniel F.},
  year = {2025},
  month = Mar 
}

@article{ydjj-6r9l,
  title = {Collinear $p$-Wave Magnetism and Hidden Orbital Ferrimagnetism},
  author = {Leeb, Valentin and Knolle, Johannes},
  journal = {Phys. Rev. Lett.},
  volume = {137},
  issue = {10},
  pages = {106701},
  numpages = {10},
  year = {2026},
  month = {Aug},
  publisher = {American Physical Society},
  doi = {10.1103/ydjj-6r9l},
  url = {https://link.aps.org/doi/10.1103/ydjj-6r9l}
}

@article{PhysRevLett.134.146001,
  title = {Adatom Engineering Magnetic Order in Superconductors: Applications to Altermagnetic Superconductivity},
  author = {Pupim, Lucas V. and Scheurer, Mathias S.},
  journal = {Phys. Rev. Lett.},
  volume = {134},
  issue = {14},
  pages = {146001},
  numpages = {9},
  year = {2025},
  month = {Apr},
  publisher = {American Physical Society},
  doi = {10.1103/PhysRevLett.134.146001},
  url = {https://link.aps.org/doi/10.1103/PhysRevLett.134.146001}
}

@article{g4dl-1ff2,
  title = {Phonon-mediated spin-polarized superconductivity in altermagnets},
  author = {Leraand, Kristoffer and M\ae{}land, Kristian and Sudb\o{}, Asle},
  journal = {Phys. Rev. B},
  volume = {112},
  issue = {10},
  pages = {104510},
  numpages = {12},
  year = {2025},
  month = {Sep},
  publisher = {American Physical Society},
  doi = {10.1103/g4dl-1ff2},
  url = {https://link.aps.org/doi/10.1103/g4dl-1ff2}
}

@article{THONHAUSER2011,
  title = {THEORY OF ORBITAL MAGNETIZATION IN SOLIDS},
  volume = {25},
  ISSN = {1793-6578},
  url = {http://dx.doi.org/10.1142/S0217979211058912},
  DOI = {10.1142/s0217979211058912},
  number = {11},
  journal = {International Journal of Modern Physics B},
  publisher = {World Scientific Pub Co Pte Lt},
  author = {THONHAUSER,  T.},
  year = {2011},
  month = Apr,
  pages = {1429–1458}
}

@article{PhysRevLett.95.137204,
  title = {Berry Phase Correction to Electron Density of States in Solids},
  author = {Xiao, Di and Shi, Junren and Niu, Qian},
  journal = {Phys. Rev. Lett.},
  volume = {95},
  issue = {13},
  pages = {137204},
  numpages = {4},
  year = {2005},
  month = {Sep},
  publisher = {American Physical Society},
  doi = {10.1103/PhysRevLett.95.137204},
  url = {https://link.aps.org/doi/10.1103/PhysRevLett.95.137204}
}

@article{PhysRevLett.95.137205,
  title = {Orbital Magnetization in Periodic Insulators},
  author = {Thonhauser, T. and Ceresoli, Davide and Vanderbilt, David and Resta, R.},
  journal = {Phys. Rev. Lett.},
  volume = {95},
  issue = {13},
  pages = {137205},
  numpages = {4},
  year = {2005},
  month = {Sep},
  publisher = {American Physical Society},
  doi = {10.1103/PhysRevLett.95.137205},
  url = {https://link.aps.org/doi/10.1103/PhysRevLett.95.137205}
}

@misc{li2026pwaveorbitalmagnetism,
      title={$P$-wave Orbital Magnetism}, 
      author={Yantao Li and Pavlo Sukhachov},
      year={2026},
      eprint={2604.18695},
      archivePrefix={arXiv},
      primaryClass={cond-mat.mes-hall},
      url={https://arxiv.org/abs/2604.18695}, 
}

@book{Phonon_Book,
  author    = {A. A. Maradudin and E. W. Montroll and G. H. Weiss and I. P. Ipatova},
  title     = {Theory of Lattice Dynamics in the Harmonic Approximation},
  publisher = {Academic Press},
  year      = {1971},
  address   = {New York},
  edition   = {2nd}
}

@article{PhysRevLett.131.186702,
  title = {Spurious Symmetry Enhancement in Linear Spin Wave Theory and Interaction-Induced Topology in Magnons},
  author = {Gohlke, Matthias and Corticelli, Alberto and Moessner, Roderich and McClarty, Paul A. and Mook, Alexander},
  journal = {Phys. Rev. Lett.},
  volume = {131},
  issue = {18},
  pages = {186702},
  numpages = {7},
  year = {2023},
  month = {Oct},
  publisher = {American Physical Society},
  doi = {10.1103/PhysRevLett.131.186702},
  url = {https://link.aps.org/doi/10.1103/PhysRevLett.131.186702}
}

@article{PhysRevB.55.14554,
  title = {Non-Fermi-liquid states and pairing instability of a general model of copper oxide metals},
  author = {Varma, C. M.},
  journal = {Phys. Rev. B},
  volume = {55},
  issue = {21},
  pages = {14554--14580},
  numpages = {0},
  year = {1997},
  month = {Jun},
  publisher = {American Physical Society},
  doi = {10.1103/PhysRevB.55.14554},
  url = {https://link.aps.org/doi/10.1103/PhysRevB.55.14554}
}

@article{Yu2024,
  title = {Non-trivial quantum geometry and the strength of electron–phonon coupling},
  volume = {20},
  ISSN = {1745-2481},
  url = {http://dx.doi.org/10.1038/s41567-024-02486-0},
  DOI = {10.1038/s41567-024-02486-0},
  number = {8},
  journal = {Nature Physics},
  publisher = {Springer Science and Business Media LLC},
  author = {Yu,  Jiabin and Ciccarino,  Christopher J. and Bianco,  Raffaello and Errea,  Ion and Narang,  Prineha and Bernevig,  B. Andrei},
  year = {2024},
  month = May,
  pages = {1262–1268}
}

@article{zhao2008fullspectrum,
  title = {Full-spectrum phonon relaxation times in crystalline Si from molecular dynamics simulations},
  author = {Hong Zhao and Jonathan B. Freund},
  year = {2008},
  publisher = {AIP Publishing},
  journal = {Journal of Applied Physics},
  volume = {104},
  doi = {10.1063/1.2963721},
  url = {https://doi.org/10.1063/1.2963721}
}

@misc{go2023intrinsicmagnonorbitalhall,
      title={Intrinsic Magnon Orbital Hall Effect in Honeycomb Antiferromagnets}, 
      author={Gyungchoon Go and Daehyeon An and Hyun-Woo Lee and Se Kwon Kim},
      year={2023},
      eprint={2303.11687},
      archivePrefix={arXiv},
      primaryClass={cond-mat.mes-hall},
      url={https://arxiv.org/abs/2303.11687}, 
}

@article{PhysRevLett.117.217202,
  title = {Spin Nernst Effect of Magnons in Collinear Antiferromagnets},
  author = {Cheng, Ran and Okamoto, Satoshi and Xiao, Di},
  journal = {Phys. Rev. Lett.},
  volume = {117},
  issue = {21},
  pages = {217202},
  numpages = {5},
  year = {2016},
  month = {Nov},
  publisher = {American Physical Society},
  doi = {10.1103/PhysRevLett.117.217202},
  url = {https://link.aps.org/doi/10.1103/PhysRevLett.117.217202}
}

@misc{cui2023efficientspinseebeckspin,
      title={Efficient Spin Seebeck and Spin Nernst Effects of Magnons in Altermagnets}, 
      author={Qirui Cui and Bowen Zeng and Tao Yu and Hongxin Yang and Ping Cui},
      year={2023},
      eprint={2306.08976},
      archivePrefix={arXiv},
      primaryClass={cond-mat.mes-hall},
      url={https://arxiv.org/abs/2306.08976}, 
}

@article{PhysRevLett.126.127701,
  title = {Efficient Electrical Spin Splitter Based on Nonrelativistic Collinear Antiferromagnetism},
  author = {Gonz\'alez-Hern\'andez, Rafael and \ifmmode \check{S}\else \v{S}\fi{}mejkal, Libor and V\'yborn\'y, Karel and Yahagi, Yuta and Sinova, Jairo and Jungwirth, Tom\'a\ifmmode \check{s}\else \v{s}\fi{} and \ifmmode \check{Z}\else \v{Z}\fi{}elezn\'y, Jakub},
  journal = {Phys. Rev. Lett.},
  volume = {126},
  issue = {12},
  pages = {127701},
  numpages = {6},
  year = {2021},
  month = {Mar},
  publisher = {American Physical Society},
  doi = {10.1103/PhysRevLett.126.127701},
  url = {https://link.aps.org/doi/10.1103/PhysRevLett.126.127701}
}

@article{PhysRevB.85.205436,
  title = {Hopping-resolved electron-phonon coupling in bilayer graphene},
  author = {Cappelluti, E. and Profeta, G.},
  journal = {Phys. Rev. B},
  volume = {85},
  issue = {20},
  pages = {205436},
  numpages = {9},
  year = {2012},
  month = {May},
  publisher = {American Physical Society},
  doi = {10.1103/PhysRevB.85.205436},
  url = {https://link.aps.org/doi/10.1103/PhysRevB.85.205436}
}

@article{PhysRevB.104.195201,
  title = {Deformation potential extraction and computationally efficient mobility calculations in silicon from first principles},
  author = {Li, Zhen and Graziosi, Patrizio and Neophytou, Neophytos},
  journal = {Phys. Rev. B},
  volume = {104},
  issue = {19},
  pages = {195201},
  numpages = {18},
  year = {2021},
  month = {Nov},
  publisher = {American Physical Society},
  doi = {10.1103/PhysRevB.104.195201},
  url = {https://link.aps.org/doi/10.1103/PhysRevB.104.195201}
}

\widetext
\appendix

\begin{section}{Molecular Berry Curvature}
\label{sec:MBC}
    This section reviews the molecular Berry curvature (MBC). We first follow Ref.~\cite{Chiral_Phonons_Angular_Momentum} to derive an expression for the MBC in terms of single-particle eigenstates and the microscopic electron-phonon coupling. We then examine how the symmetries of the checkerboard model constrain the MBC and prove that the MBC matrix is anti-Hermitian. Finally, for the checkerboard model, we derive the explicit electron-phonon coupling matrices and prove that the magnetic point group of our electronic system does not allow for a finite phonon angular momentum at the $\Gamma$ point. 

\begin{subsection}{Molecular Berry Curvature as a Gauge Field: General Derivation}
Our starting point is the MBC in Eq.~\eqref{eq: MBC reciprocal space stage 1} reading
\begin{equation}
\begin{aligned}
    G_{\kappa' \beta}^{\kappa \alpha} \left( \vec{k} \right) 
     &= 
     \frac{\mathrm{i} }{N}
     \sum_{n \neq 0} \left[ \left\langle \Phi_0 \middle| M_{\vec{k} \kappa \alpha} \middle|\Phi_n \right \rangle  \left \langle  \Phi_n \middle| M_{-\vec{k} \kappa'\beta} \middle|\Phi_0 \right \rangle -
     \left\langle \Phi_0 \middle| M_{-\vec{k} \kappa'\beta} \middle|\Phi_n \right \rangle  \left \langle  \Phi_n \middle| M_{\vec{k} \kappa \alpha} \middle|\Phi_0 \right \rangle
     \right] \frac{1}{\left(E_n - E_0 \right)^2}, 
\end{aligned}
\label{AP: eq: MBC reciprocal space stage 2}
\end{equation}
where we recall that $M_{\vec{k} \kappa \alpha} = \sum_{l} \frac{\partial H_{\text{el}}}{\partial u_{l  \kappa \alpha}}\mathrm{e}^{-\mathrm{i} \vec{R}_{l \kappa}^0  \cdot \vec{k}}$ and $M_{-\vec{k} \kappa' \beta} = \sum_{l'} \frac{\partial H_{\text{el}}}{\partial u_{l'  \kappa' \beta}}\mathrm{e}^{\mathrm{i} \vec{R}_{l' \kappa'}^0 \cdot \vec{k}}$ are the Fourier transform of the electron-phonon coupling operators. To make analytical progress we assume that the electronic Hamiltonian is a non-interacting tight-binding Hamiltonian containing hopping terms of electrons between different sites, which generally assumes the form
\begin{equation}
    H_{\text{el}} = -\sum_{l,l'} \sum_{\kappa \kappa'} t_{ll'} c^{\dagger}_{l \kappa} c_{l' \kappa'}.
    \label{AP: eq: tight binding H general}
\end{equation}
Here, $t_{ll'}$ are, in general complex, hopping amplitudes between the $l$-th and $l'$-th unit cell, and $\kappa, \kappa'$ are internal indices which, without loss of generality, are assumed to be sublattice indices. We note that $t_{ll'}$ are chosen such so that $H_{\text{el}}= H_{\text{el}}^{\dagger}$ applies. Later on, we furthermore assume the $t_{ll'}$ to carry distance dependence that ultimately gives rise to electron-phonon coupling, as shown in Sec.~\ref{AP:subsection: MBC checkerboard} for the case of the checkerboard lattice. By Fourier transforming the fermion operators as
\begin{equation}
    c_{l  \kappa} = \frac{1}{\sqrt{N}} \sum_{\vec{q}} c_{\vec{q} \kappa} \mathrm{e}^{\mathrm{i} \vec{q} \cdot \vec{R}_l ^0},
    \label{AP: eq: FT fermions}
\end{equation}
one can write $M_{\vec{k} \kappa \alpha}$ and $M_{-\vec{k} \kappa' \beta}$ as
\begin{equation}
\begin{aligned}
    & M_{\vec{k} \kappa \alpha} =\sum_{\vec{q}} \vec{\Psi}^{\dagger}_{\vec{q}} \mathcal{M}_{\vec{k} \kappa \alpha} \vec{\Psi}_{{\vec{q}+\vec{k}}}, \\
    &  M_{-\vec{k} \kappa' \beta} = \sum_{\vec{q}} \vec{\Psi}^{\dagger}_{\vec{q} + \vec{k}} \mathcal{M}_{-\vec{k} \kappa' \beta} \vec{\Psi}_{{\vec{q}}},
\end{aligned}
\label{AP:eq: e-ph coupling matrices}
\end{equation}
where $\vec{\Psi}_{\vec{q}+ \vec{k}} = \left(c_{\vec{q} + \vec{k} \kappa_1}, ...., c_{\vec{q} + \vec{k} \kappa_N}  \right )^T$, and $\mathcal{M}_{\vec{k} \kappa \alpha}$ and $\mathcal{M}_{-\vec{k} \kappa' \beta}$  are the kernels of the electron-phonon coupling. 

In reciprocal space, the Hamiltonian in Eq.~\eqref{AP: eq: tight binding H general} can be written as
\begin{equation}
    H_{\text{el}} = \sum_{\vec{k}} \vec{\Psi}_{\vec{k}}^{\dagger} H_{\vec{k}} \vec{\Psi}_{\vec{k}}.
\end{equation}
Denoting the normalized column eigenvectors of $H_{\vec{k}}$ by $u_{\vec{k} m}$ (with $m$ the band index) we can construct a unitary matrix 
\begin{equation}
    U_{\vec{k}} = 
    \begin{pmatrix}
        u_{\vec{k} m_1} & u_{\vec{k} m_2} & \ldots & u_{\vec{k} m_N}
    \end{pmatrix},
\end{equation}
whose action on the vector $\vec{\Psi}_{\vec{k}}$ is 
\begin{equation}
    U_{\vec{k}}^{\dagger} \vec{\Psi}_{\vec{k}} = \tilde{\vec{\Psi}}_{\vec{k}}, 
\end{equation}
where now  $\tilde{\vec{\Psi}}_{\vec{k}} = \left(c_{\vec{k} m_1}, ...., c_{\vec{k} m_N}  \right )^T$ contains the eigenmodes of the Hamiltonian. 

To compute matrix elements such as $\left\langle \Phi_0 \middle| M_{\vec{k} \kappa \alpha} \middle|\Phi_n \right \rangle$ appearing in Eq.~\eqref{AP: eq: MBC reciprocal space stage 2} we assume that the many-body states $\Phi_n, \Phi_0 $ are Slater determinants \cite{Saparov_thesis}. This allows us to apply the Slater-Condon rules, according to which the matrix element of a one-body operator between two Slater determinants is nonzero only if the determinants differ by a single occupied single-particle state. In that case, the many-body matrix element reduces to the matrix element of the one-body operator between the two single-particle states in which the Slater determinants differ, rather than requiring an evaluation between the many-body states themselves.

Assuming that our electronic model is insulating with a well-defined gap separating the occupied from the unoccupied states at $T=0$, we can write \cite{Saparov_thesis}
\begin{equation}
    \ket{\Phi_0} = \mathcal{A} \ket{u_{\vec{k}_1 m_1^-}} \ket{u_{\vec{k}_2 m_2^-}} \ldots \ket{u_{\vec{k}_r m_r^-}} \ldots, 
    \label{AP:eq: slater ground state}
\end{equation}
where the minus sign in the superscript of the band index denotes that all the states in Eq.~\eqref{AP:eq: slater ground state} are occupied states below the Fermi level of the system at $T=0$, and $\mathcal{A}$ is an operator anti-symmetrizing the state product to the right of it. An excited state $\Phi_n$ is then defined as
\begin{equation}
     \ket{\Phi_n} = \mathcal{A} \ket{u_{\vec{k}_1 m_1^-}} \ket{u_{\vec{k}_2 m_2^-}} \ldots \ket{u_{\vec{k}_n m_n^+}} \ldots, 
\end{equation}
where in contrast to $\Phi_0$ the electron originally in the single-particle state $\ket{u_{\vec{k}_rm_r^-}}$ has been excited to the state $\ket{u_{\vec{k}_n m_n^+}}$, whose energy lies above the Fermi level. Thus, we find for the product of the elements $ P \equiv \left\langle \Phi_0 \middle| \mathcal{M}_{\vec{k}, \kappa \alpha} \middle|\Phi_n \right \rangle  \left\langle \Phi_n \middle| \mathcal{M}_{-\vec{k}, \kappa' \beta} \middle|\Phi_0 \right \rangle $ that 
\begin{equation}
\begin{aligned}
    P &= \left\langle 
    \Phi_0 \middle| M_{\vec{k} \kappa \alpha} \middle|\Phi_n \right \rangle  \left\langle \Phi_n \middle| M_{-\vec{k} \kappa' \beta} \middle|\Phi_0 \right \rangle
    \\
    &= 
    \left\langle u_{\vec{k}_r m_r^-} \middle| M_{\vec{k}, \kappa \alpha} \middle|u_{\vec{k}_n m_n^+} \right \rangle  \left\langle u_{\vec{k}_n m_n^+} \middle| M_{-\vec{k} \kappa' \beta} \middle|u_{\vec{k}_r m_r^-} \right \rangle
    \\
    &= \sum_{\vec{q} \vec{q}'} \left\langle u_{\vec{k}_r m_r^-} \middle| \vec{\Psi}^{\dagger}_{\vec{q}} \mathcal{M}_{\vec{k} \kappa \alpha} \vec{\Psi}_{{\vec{q}+\vec{k}}} \middle|u_{\vec{k}_n m_n^+} \right \rangle  \left\langle u_{\vec{k}_n m_n^+} \middle| \vec{\Psi}^{\dagger}_{\vec{q}' + \vec{k}} \mathcal{M}_{-\vec{k}, \kappa' \beta} \vec{\Psi}_{{\vec{q}'}} \middle|u_{\vec{k}_r m_r^-} \right \rangle \\
    &= \sum_{\vec{q} \vec{q}'} \left\langle u_{\vec{k}_r m_r^-} \middle| \tilde{\vec{\Psi}}^{\dagger}_{\vec{q}} U_{\vec{q}}^{\dagger}\mathcal{M}_{\vec{k} \kappa \alpha} U_{\vec{q} + \vec{k}}  \tilde{\vec{\Psi}}_{{\vec{q}+\vec{k}}} \middle|u_{\vec{k}_n m_n^+} 
    \right \rangle 
    \left\langle u_{\vec{k}_n m_n^+} \middle| \tilde{\vec{\Psi}}^{\dagger}_{\vec{q}' + \vec{k}} U_{\vec{q}'+ \vec{k}}^{\dagger} \mathcal{M}_{-\vec{k} \kappa' \beta} 
     U_{\vec{q}'} \tilde{\vec{\Psi}}_{{\vec{q}'}} \middle|u_{\vec{k}_r m_r^-} \right \rangle \\
     & = 
     \sum_{\vec{q}} u_{\vec{q} m_r^-}^{\dagger} \mathcal{M}_{\vec{k} \kappa \alpha} u_{\vec{q} + \vec{k} m_n^+}  u_{\vec{q} + \vec{k} m_n^+}^{\dagger} \mathcal{M}_{-\vec{k} \kappa' \beta} u_{\vec{q} m_r^-}.
\end{aligned}
\label{AP:eq: result for P}
\end{equation}
To go from the fourth to the fifth line we noticed that one of the two terms in the expression of $P$, namely
$P_1 = \left\langle u_{\vec{k}_r m_r^-} \middle| \tilde{\vec{\Psi}}^{\dagger}_{\vec{q}} U_{\vec{q}}^{\dagger}\mathcal{M}_{\vec{k} \kappa \alpha} U_{\vec{q} + \vec{k}}  \tilde{\vec{\Psi}}_{{\vec{q}+\vec{k}}} \middle|u_{\vec{k}_n m_n^+} 
    \right \rangle$,  can be written as
\begin{equation}
\begin{aligned}
     P_1 &= \left\langle u_{\vec{k}_r m_r^-} \middle| \tilde{\vec{\Psi}}^{\dagger}_{\vec{q}} U_{\vec{q}}^{\dagger}\mathcal{M}_{\vec{k} \kappa \alpha} U_{\vec{q} + \vec{k}}  \tilde{\vec{\Psi}}_{{\vec{q}+\vec{k}}} \middle|u_{\vec{k}_n m_n^+} 
    \right \rangle \\
    &=
    \sum_{m,m'}  \left\langle u_{\vec{k}_r m_r^-} \middle| \left[\tilde{\vec{\Psi}}^{\dagger}_{\vec{q}}\right]_m \left[U_{\vec{q}}^{\dagger}\mathcal{M}_{\vec{k} \kappa \alpha} U_{\vec{q} + \vec{k}}\right]_{m m'}  \left[\tilde{\vec{\Psi}}_{{\vec{q}+\vec{k}}}\right]_{m'} \middle|u_{\vec{k}_n m_n^+} 
    \right \rangle \\
    &=
    \sum_{m,m'} \left[U_{\vec{q}}^{\dagger}\mathcal{M}_{\vec{k} \kappa \alpha} U_{\vec{q} + \vec{k}}\right]_{m m'}  \left\langle u_{\vec{k}_r m_r^-} \middle| \left[\tilde{\vec{\Psi}}^{\dagger}_{\vec{q}}\right]_m  \left[\tilde{\vec{\Psi}}_{{\vec{q}+\vec{k}}}\right]_{m'} \middle|u_{\vec{k}_n m_n^+} 
    \right \rangle \\
    & = 
    \sum_{m,m'}
    \left[U_{\vec{q}}^{\dagger}\mathcal{M}_{\vec{k} \kappa \alpha} U_{\vec{q} + \vec{k}}\right]_{m m'}
    \left\langle u_{\vec{k}_r m_r^-} \middle| c^{\dagger}_{\vec{q}m} c_{{\vec{q}+\vec{k}}{m'}} \middle|u_{\vec{k}_n m_n^+} 
    \right \rangle \\
    & = 
    \sum_{m,m'}
    \left[U_{\vec{q}}^{\dagger}\mathcal{M}_{\vec{k} \kappa \alpha} U_{\vec{q} + \vec{k}}\right]_{m m'}
    \left\langle 0 \middle| c_{\vec{k}_r m_r^-}c^{\dagger}_{\vec{q}m} c_{{\vec{q}+\vec{k}}{m'}}c^{\dagger}_{\vec{k}_n m_n^+} \middle|0 
    \right \rangle \\
    &= 
    \sum_{m,m'}
    \left[U_{\vec{q}}^{\dagger}\mathcal{M}_{\vec{k} \kappa \alpha} U_{\vec{q} + \vec{k}}\right]_{m m'}
    \delta_{\vec{k}_r, \vec{q}} \delta_{m, m_r^-}
     \delta_{\vec{q} + \vec{k}, \vec{k}_n} \delta_{m, m_n^+} \\
     & = 
    u_{\vec{q} m_r^-}^{\dagger}\mathcal{M}_{\vec{k} \kappa \alpha} u_{\vec{q} + \vec{k} m_n^+}
    \delta_{\vec{k}_r, \vec{q}} 
     \delta_{\vec{q} + \vec{k}, \vec{k}_n}, 
\end{aligned} 
\end{equation}
where we expanded the product of the four operators as 
\begin{equation}
\begin{aligned}
    \left\langle 0 \middle| c_{\vec{k}_r m_r^-}c^{\dagger}_{\vec{q}m} c_{{\vec{q}+\vec{k}}{m'}}c^{\dagger}_{\vec{k}_n m_n^+} \middle|0 
    \right \rangle
    &= \left\langle 0 \middle| c_{\vec{k}_r m_r^-}c^{\dagger}_{\vec{q}m} \middle| 0 
     \right \rangle
     \left \langle 0 \middle|
    c_{{\vec{q}+\vec{k}}{m'}}c^{\dagger}_{\vec{k}_n m_n^+} \middle|0 
    \right \rangle  
    -
    \left\langle 0 \middle| c_{\vec{k}_r m_r^-}c^{\dagger}_{\vec{k}_n m_n^+} \middle| 0 
    \right \rangle
    \left \langle 0 \middle|
    c^{\dagger}_{\vec{q}m}
    c_{{\vec{q}+\vec{k}}{m'}} \middle|0 
    \right \rangle \\
    &= \left\langle 0 \middle| c_{\vec{k}_r m_r^-}c^{\dagger}_{\vec{q}m} \middle| 0 
     \right \rangle
     \left \langle 0 \middle|
    c_{{\vec{q}+\vec{k}}{m'}}c^{\dagger}_{\vec{k}_n m_n^+} \middle|0 
    \right \rangle, 
\end{aligned}
\end{equation}
using Wick's theorem. Using the result for $P_1$ one can simplify in a similar same manner the expression for $P_2 = \left\langle u_{\vec{k}_n m_n^+} \middle| \tilde{\vec{\Psi}}^{\dagger}_{\vec{q}' + \vec{k}} U_{\vec{q}'+ \vec{k}}^{\dagger} \mathcal{M}_{-\vec{k} \kappa' \beta} 
     U_{\vec{q}'} \tilde{\vec{\Psi}}_{{\vec{q}'}} \middle|u_{\vec{k}_r m_r^-} \right \rangle$ and reach the final line in Eq.~\eqref{AP:eq: result for P}. 
 Finally, following the same logic for the second term in Eq.~\eqref{AP: eq: MBC reciprocal space stage 2}, the MBC can be obtained as
\begin{equation}
     G_{\kappa' \beta}^{\kappa \alpha} \left( \vec{k} \right) = 
     \frac{\mathrm{i} }{N}
     \sum_{m^-, m^+} \sum_{\vec{q}} \left[
     \frac{u_{\vec{q} m^-} ^{\dagger} \mathcal{M}_{\vec{k} \kappa \alpha} u_{\vec{q} + \vec{k} m^+}  u_{\vec{q} + \vec{k} m^+} ^{\dagger} \mathcal{M}_{-\vec{k} \kappa' \beta} u_{\vec{q} m^-}}{\left(E_{\vec{q} m^-}- E_{\vec{q} + \vec{k} m^+}\right)^2} 
     -
     \frac{u_{\vec{q} + \vec{k} m^-}^{\dagger} \mathcal{M}_{-\vec{k} \kappa' \beta} u_{\vec{q} m^+}  u_{\vec{q} m^+} ^{\dagger} \mathcal{M}_{\vec{k} \kappa \alpha} u_{\vec{q} + \vec{k} m^-}}{\left(E_{\vec{q} m^+}- E_{\vec{q} + \vec{k} m^-}\right)^2} \right], 
\label{AP:eq: Molecular Berry curvature final}
\end{equation}
where $E_{\vec{k} m^{\pm}}$ is the energy of the single-particle state $u_{\vec{k} m^{\pm}}$. Moreover, $m^{-}$ labels the occupied states and $m^{+}$ the unoccupied ones.

\end{subsection}

\begin{subsection}{Properties and Symmetries of the Molecular Berry Curvature}
\label{AP: subsec: MBC symmetries}

In this section, we discuss how the symmetries of the crystal constrain the MBC and subsequently show that the MBC is anti-Hermitian. Specifically, we consider the following symmetry operations:
\begin{enumerate}
    \item time-reversal ($\mathcal{T}$),
    \item inversion ($\mathcal{P}$), 
    \item  clockwise $90^{\circ}$ rotation around the $z$-axis followed by time-reversal ($\mathcal{C}_{4z, +} \mathcal{T}$),  
    \item reflection about the $x=y$-line ($\mathcal{M}_{xy}$).
\end{enumerate}
In the following, we consider the checkerboard lattice introduced in the main text and analyze the action of all crystal symmetry operations. Among the symmetries discussed above, only (b)-(d) are symmetries of the altermagnetic loop-current model defined in Eq.~\eqref{eq: TB-H real space}. They are of particular relevance because the MBC originates from the coupling between the electronic degrees of freedom and the crystal lattice. By contrast, time-reversal symmetry $\mathcal{T}$ by itself is not a symmetry of the electronic model considered here. Nevertheless, analyzing its action provides valuable insight into the physical origin and symmetry properties of the MBC.

First, recall that the MBC enters the phonon Hamiltonian as a magnetic Raman-like coupling, such that the relevant contribution in reciprocal space is
\begin{equation}
    H_{\text{RM}} = \frac{1}{2} \sum_{\vec{k}} \sum_{\alpha \beta} \sum_{\kappa \kappa'} \left[\frac{1}{M_{\kappa}} p_{\vec{k} \kappa \alpha}^{\dagger} G_{\kappa'\beta}^{\kappa \alpha}\left( \vec{k} \right) u_{\vec{k} \kappa'\beta} + \frac{1}{M_{\kappa'}}u^{\dagger}_{\vec{k} \kappa \alpha} 
\left(G_{\kappa \alpha}^{\kappa'\beta}\left(\vec{k} \right)\right)^*p_{\vec{k} \kappa'\beta} \right]. 
\label{AP:eq:Raman coupling H_phonon}
\end{equation}
We first determine how the relevant symmetry operations act on the momentum and displacement operators entering $ H_{\text{RM}}$. We specify their action on the corresponding real-space operators $p_{l \kappa \alpha}$ and $u_{l \kappa \alpha}$, where $l$ labels the unit cells. The transformation properties of the reciprocal-space operators, $p_{\vec{k} \kappa \alpha}$ and $u_{\vec{k} \kappa \alpha}$, then follow directly from the Fourier transformations
\begin{equation}
    \begin{aligned}
        & u_{\vec{k} \kappa \alpha} = \frac{1}{N} \sum_{l} \mathrm{e}^{-\mathrm{i} \vec{k} \cdot \left(\vec{R}_l^0 + \vec{\tau}_{\kappa}\right)} u_{l \kappa \alpha}, \\
        & p_{\vec{k} \kappa \alpha} = \frac{1}{N} \sum_{l} \mathrm{e}^{-\mathrm{i} \vec{k} \cdot \left(\vec{R}_l^0 + \vec{\tau}_{\kappa}\right)} p_{l\kappa \alpha}. 
    \end{aligned}
\label{APP: FT atomic gauge}
\end{equation}
We recall that, for the checkerboard lattice considered in the main text, the basis vectors are chosen as $\vec{\tau}_{\text{A}} =0$ and
$\vec{\tau}_{\text{B}} = \left(\hat{\vec{x}} +  \hat{\vec{y}}\right)/2$, while the primitive lattice vectors are $\vec{a}_1 = \hat{\vec{x}}$ and $\vec{a}_2 = \hat{\vec{y}}$. Throughout, the lattice constant is set to unity.

Before analyzing the symmetry constraints, we briefly comment on the choice of Fourier convention. The atomic gauge introduced in Eq.~\eqref{APP: FT atomic gauge} is particularly convenient for lattices with a basis, as it leads to considerably simpler symmetry transformation laws than the convention in which the basis vectors are omitted from the Fourier transform.  

\begin{enumerate}
\item \textbf{Time-reversal Symmetry} ($\mathcal{T}$)

The action of the time-reversal operator $\mathcal{T}$ on the momentum and displacement operators in real space is defined as \cite{Yu2024}
\begin{equation}
    \begin{aligned}
        & \mathcal{T} p_{l \kappa \alpha} \mathcal{T}^{-1} = - p_{l \kappa \alpha}, \\
        & \mathcal{T} u_{l \kappa \alpha} \mathcal{T}^{-1} = u_{l \kappa \alpha}. 
    \end{aligned}
\end{equation}
Since $\mathcal{T}$ is anti-unitary, its action on a scalar $c$ is complex conjugation, $\mathcal{T}c \mathcal{T}^{-1} = c^*$. It then follows for the displacement that
\begin{equation}
    \begin{aligned}
        \mathcal{T} u_{\vec{k} \kappa \alpha} \mathcal{T}^{-1} 
        & = \frac{1}{\sqrt{N}} \sum_{l}  \mathcal{T} \mathrm{e}^{-\mathrm{i} \vec{k} \cdot \left(\vec{R}_l +\vec{\tau}_\kappa \right)} u_{l \kappa \alpha}  \mathcal{T}^{-1} \\
        & = 
        \frac{1}{\sqrt{N}} \sum_{l}  \mathcal{T} \mathrm{e}^{-\mathrm{i} \vec{k} \cdot \left(\vec{R}_l +\vec{\tau}_\kappa \right)} \mathcal{T}^{-1} \mathcal{T} u_{l \kappa \alpha}  \mathcal{T}^{-1} \\
        & = 
         \frac{1}{\sqrt{N}} \sum_{l}  \mathrm{e}^{\mathrm{i} \vec{k} \cdot \left(\vec{R}_l +\vec{\tau}_\kappa \right)}  u_{l \kappa \alpha}
         \\
      & = 
      u_{-\vec{k} \kappa \alpha}, 
    \end{aligned}
\label{APP: eq: TR on u k-space}
\end{equation}
and similarly for the momentum that
\begin{equation}
    \mathcal{T} p_{\vec{k} \kappa \alpha} \mathcal{T}^{-1} = -p_{-\vec{k} \kappa \alpha}.
\label{APP: eq: TR on momentum k-space}
\end{equation}
Using Eqs.~\eqref{APP: eq: TR on u k-space} and \eqref{APP: eq: TR on momentum k-space} in Eq.~\eqref{AP:eq:Raman coupling H_phonon}, one can show that
\begin{equation}
\begin{aligned}
   \mathcal{T} \left[\sum_{\vec{k}} \sum_{ \kappa \kappa'} \sum_{\beta 
\alpha}\frac{1}{M_{\kappa}}p_{\vec{k} \kappa \alpha}^{\dagger} G_{\kappa'\beta}^{\kappa \alpha}\left( \vec{k} \right) u_{\vec{k} \kappa'\beta} \right]\mathcal{T}^{-1} 
   & = 
    \frac{1}{2} \sum_{\vec{k}} \sum_{\alpha \beta} \sum_{\kappa \kappa'} \frac{1}{M_{\kappa}}\mathcal{T} p_{\vec{k} \kappa \alpha}^{\dagger} \mathcal{T}^{-1} \mathcal{T} G_{\kappa'\beta}^{\kappa \alpha}\left( \vec{k} \right)\mathcal{T}^{-1} \mathcal{T} u_{\vec{k} \kappa'\beta}  \mathcal{T}^{-1} \\
    & = 
    -
    \frac{1}{2} \sum_{\vec{k}} \sum_{\alpha \beta} \sum_{\kappa \kappa'} \frac{1}{M_{\kappa}}p_{-\vec{k} \kappa \alpha}^{\dagger} \left[G_{\kappa'\beta}^{\kappa \alpha}\left( \vec{k} \right)\right]^* u_{-\vec{k} \kappa'\beta} \\
    & = 
    -
    \frac{1}{2} \sum_{\vec{k}} \sum_{\alpha \beta} \sum_{\kappa \kappa'}\frac{1}{M_{\kappa}} p_{\vec{k} \kappa \alpha}^{\dagger} \left[G_{\kappa'\beta}^{\kappa \alpha}\left( -\vec{k} \right)\right]^* u_{\vec{k} \kappa'\beta} \\
    & 
    = 
    - 
     \frac{1}{2} \sum_{\vec{k}} \sum_{\alpha \beta} \sum_{\kappa \kappa'} \frac{1}{M_{\kappa}} p_{\vec{k} \kappa \alpha}^{\dagger} G_{\kappa'\beta}^{\kappa \alpha}\left(\vec{k} \right) u_{\vec{k} \kappa'\beta}, 
\end{aligned}
\end{equation}
and similarly 
\begin{equation}
\mathcal{T} \left[\sum_{\vec{k}} \sum_{\kappa \kappa'} \sum_{\beta 
\alpha}\frac{1}{M_{\kappa'}} u_{\vec{k} \kappa \alpha}^{\dagger} G_{\kappa \alpha}^{\kappa' \beta}\left( \vec{k} \right) p_{\vec{k} \kappa'\beta} \right]\mathcal{T}^{-1} = - \sum_{\vec{k}}\sum_{\kappa \kappa'} \frac{1}{M_{\kappa'}}\sum_{\beta 
\alpha}u_{\vec{k} \kappa \alpha}^{\dagger} G_{\kappa \alpha}^{\kappa' \beta}\left( \vec{k} \right) p_{\vec{k} \kappa'\beta}. 
\end{equation}
Therefore, if $\mathcal{T}$ is a symmetry of the system, that is, if $\mathcal{T} H_{\text{RM}} \mathcal{T}^{-1} = H_{\text{RM}}$, all elements of the MBC matrix must vanish identically:
\begin{equation}
    G_{\kappa'\beta}^{\kappa \alpha}\left( \vec{k} \right) = 0,  \quad \text{if $\mathcal{T}$ is a symmetry}. 
\end{equation} 
\item \textbf{Inversion Symmetry} ($\mathcal{P}$) 

Inversion symmetry acts on displacement and momentum operators in real space as
\begin{equation}
    \begin{aligned}
        \mathcal{P} u_{l \kappa \alpha} \mathcal{P}^{-1} 
        &= -u_{l' \kappa \alpha}, \\
       \mathcal{P} p_{l \kappa \alpha} \mathcal{P}^{-1} 
        & = -p_{l' \kappa \alpha}, 
    \end{aligned}
    \label{APP: eq:invllprime}
\end{equation}
where $l'$ is the index of the inverted unit cell. 
Unlike time-reversal symmetry, inversion symmetry is unitary. Since the checkerboard lattice is non-Bravais, the lattice vectors associated with unit cells $l$ and $l'$ in Eq.~\eqref{APP: eq:invllprime} are not simply related by inversion ($\vec{R}_{l}^0 \ne - \vec{R}_{l'}^0$). Instead, they satisfy 
\begin{equation}
    \vec{R}_{l'}^0 = - \vec{R}_{l}^0 - 2 \vec{\tau}_{\kappa}.
    \label{APP:eq:position-relation}
\end{equation} 
We emphasize that, since every lattice site of the checkerboard lattice is an inversion center, the basis index $\kappa$ remains unchanged under inversion. This situation may differ for other lattices, such as the honeycomb lattice, where inversion symmetry interchanges the two sublattices.

Using Eq.~\eqref{APP:eq:position-relation} as well as the Fourier transformation in Eq.~\eqref{APP: FT atomic gauge}, we determine the action of $\mathcal{P}$ on $u_{\vec{k} \kappa \alpha}$ and $p_{\vec{k} \kappa \alpha}$ as 
\begin{equation}
\begin{aligned}
            \mathcal{P}u_{\vec{k} \kappa \alpha}  \mathcal{P}^{-1} &= 
            \frac{1}{\sqrt{N}} \sum_{l} \mathcal{P}u_{l \kappa \alpha}  \mathcal{P}^{-1} \mathrm{e}^{-\mathrm{i} \vec{k} \cdot \vec{R}_l^0} \mathrm{e}^{-\mathrm{i} \vec{k} \cdot \vec{\tau}_{\kappa}} \\
            & 
            = 
            -\frac{1}{\sqrt{N}} \sum_{l} u_{l' \kappa \alpha}   \mathrm{e}^{-\mathrm{i} \vec{k} \cdot \vec{R}_l^0}\mathrm{e}^{-\mathrm{i} \vec{k} \cdot \vec{\tau}_{\kappa}} \\
            & = 
            -\frac{1}{\sqrt{N}} \sum_{l} u_{l' \kappa \alpha}   \mathrm{e}^{-\mathrm{i} \vec{k} \cdot \left(-\vec{R}_{l'}^0 - 2\vec{\tau}_{\kappa} \right)}\mathrm{e}^{-\mathrm{i} \vec{k} \cdot \vec{\tau}_{\kappa}} \\
            & =
            -\frac{1}{\sqrt{N}} \sum_{l} u_{l \kappa \alpha}   \mathrm{e}^{\mathrm{i} \vec{k} \cdot \vec{R}_{l}^0}
             \mathrm{e}^{\mathrm{i} \vec{k} \cdot \vec{\tau}_{\kappa}}
            \\
            & = 
            - u_{-\vec{k} \kappa \alpha}, 
        \end{aligned}, 
        \label{eq:u-inv-trafo}
\end{equation}
and
\begin{equation}
    \mathcal{P}p_{\vec{k} \kappa \alpha} \mathcal{P}^{-1} = - p_{-\vec{k} \kappa \alpha}.
    \label{eq:p-inv-trafo}
\end{equation} 

Using Eqs.~\eqref{eq:u-inv-trafo} and \eqref{eq:p-inv-trafo} in Eq.~\eqref{AP:eq:Raman coupling H_phonon}, the two terms comprising $H_{\text{RM}}$ can be seen to transform as follows
\begin{equation}
\begin{aligned}
      \mathcal{P}\left[\sum_{\vec{k}} \sum_{\kappa \kappa'} \frac{1}{M_{\kappa}} \sum_{\beta 
\alpha} p_{\vec{k} \kappa \alpha}^{\dagger} G_{\kappa'\beta}^{\kappa \alpha}\left( \vec{k} \right) u_{\vec{k} \kappa'\beta} \right]\mathcal{P}^{-1} & 
= \sum_{\vec{k}} \sum_{ \kappa \kappa'}\frac{1}{M_{\kappa}} \sum_{\beta 
\alpha}\mathcal{P} p_{\vec{k} \kappa \alpha}^{\dagger} G_{\kappa'\beta}^{\kappa \alpha}\left( \vec{k} \right) u_{\vec{k} \kappa'\beta} \mathcal{P}^{-1} \\
& = 
\sum_{\vec{k}} \sum_{ \kappa \kappa'} \frac{1}{M_{\kappa}}\sum_{\beta 
\alpha}\mathcal{P} p_{\vec{k} \kappa \alpha}^{\dagger} \mathcal{P}^{-1} \mathcal{P}G_{\kappa'\beta}^{\kappa \alpha}\left( \vec{k} \right) \mathcal{P}^{-1} \mathcal{P} u_{\vec{k} \kappa'\beta} \mathcal{P}^{-1} \\
& = 
\sum_{\vec{k}} \sum_{ \kappa \kappa'} \frac{1}{M_{\kappa}}\sum_{\beta 
\alpha}p_{-\vec{k} \kappa \alpha}^{\dagger} G_{\kappa'\beta}^{\kappa \alpha}\left( \vec{k} \right)  u_{-\vec{k} \kappa'\beta} \\
&= 
\sum_{\vec{k}} \sum_{ \kappa \kappa'} \frac{1}{M_{\kappa}}\sum_{\beta 
\alpha}p_{\vec{k} \kappa \alpha}^{\dagger} G_{\kappa'\beta}^{\kappa \alpha}\left(-\vec{k} \right)  u_{\vec{k} \kappa'\beta}\\
& = 
\sum_{\vec{k}} \sum_{ \kappa \kappa'} \frac{1}{M_{\kappa}}\sum_{\beta 
\alpha}p_{\vec{k} \kappa \alpha}^{\dagger} 
\left[G_{\kappa'\beta}^{\kappa \alpha}\left(\vec{k} \right) 
\right]^* u_{\vec{k} \kappa'\beta}, 
\end{aligned}
\end{equation}
and 
\begin{equation}
\mathcal{P} \left[\sum_{\vec{k}} \sum_{ \kappa \kappa'} \frac{1}{M_{\kappa'}}\sum_{\beta 
\alpha} u^{\dagger}_{\vec{k} \kappa \alpha} 
\left(G_{\kappa \alpha}^{\kappa'\beta}\left(\vec{k} \right)\right)^*p_{\vec{k} \kappa'\beta}\right] \mathcal{P}^{-1}= 
 \sum_{\vec{k}} \sum_{ \kappa \kappa'}\frac{1}{M_{\kappa'}}\sum_{\beta 
\alpha} u^{\dagger}_{\vec{k} \kappa \alpha} 
G_{\kappa \alpha}^{\kappa'\beta}\left(\vec{k} \right)p_{\vec{k} \kappa'\beta}. 
\end{equation}
Therefore, in centrosymmetric lattices with $\mathcal{P} H_{\text{RM}} \mathcal{P}^{-1} = H_{\text{RM}}$, where each lattice site is an inversion center, we find that
\begin{equation}
        G_{\kappa \alpha}^{\kappa'\beta}\left(\vec{k} \right) = \left[G_{\kappa \alpha}^{\kappa'\beta}\left(\vec{k} \right)\right]^* 
        \in \text{Real}.
        \label{AP:eq: inversion on MBC}
\end{equation}
We emphasize that the above result applies only within the Fourier gauge defined in Eq.~\eqref{APP: FT atomic gauge}, in analogy to the well-known case of the dynamical phonon matrix \cite{Phonon_Book}. 
\item \textbf{$90^{\circ}$ rotation followed by time-reversal} ($\mathcal{C}_{4z,+} \mathcal{T}$)  

We start off with the effect of the clockwise $90^{\circ}$ rotation in real space:
\begin{equation}
    \begin{aligned}
        & \mathcal{C}_{4z, +} u_{l \kappa \alpha} \mathcal{C}_{4z, +}^{-1} = 
        \sum_{\overline{a}} \left[\text{C}_{4z, +}\right]_{\alpha \overline{\alpha}}
        u_{l' \kappa \overline{\alpha}}, 
        \\
        & 
        \mathcal{C}_{4z, +} p_{l \kappa \alpha} \mathcal{C}_{4z, +}^{-1} = 
        \sum_{\overline{a}} \left[\text{C}_{4z, +}\right]_{\alpha \overline{\alpha}}
        p_{l' \kappa \overline{\alpha}}, 
    \end{aligned}
\end{equation}
where $l'$ is the transformed unit cell index under $\mathcal{C}_{4z, +}$ and $\left[\text{C}_{4z, +}\right]_{\alpha \overline{\alpha}}$ are the elements of the $90^{\circ}$ clockwise rotation matrix 
\begin{equation}
   \text{C}_{4z, +} = 
    \begin{pmatrix}
        0 & 1 \\
        - 1 & 0 
    \end{pmatrix}. 
\end{equation}
We recall that we use calligraphic letters for the symmetry operators, while upright letters denote their representations. 
For the checkerboard lattice, we find
\begin{equation}
    \begin{aligned}
        & \vec{R}_{l'}^0 = \text{C}_{4z, +} \vec{R}_{l}^0, \quad \kappa = \text{A}, \\
        & \vec{R}_{l'}^0 = \text{C}_{4z, +} \vec{R}_{l}^0 - \vec{a}_2, \quad \kappa = \text{B},
    \end{aligned}
    \label{AP: rotated vectors}
\end{equation}
with $\text{C}_{4z, +} \vec{R}_{l}^0$ being the rotated unit cell vector. For displacements of ions belonging to the A sublattice, we obtain
\begin{equation}
    \begin{aligned}
        \mathcal{C}_{4z, +} u_{\vec{k} \text{A} \alpha} \mathcal{C}_{4z, +}^{-1} 
        & = \frac{1}{\sqrt{N}} \sum_{l} \mathcal{C}_{4z, +} u_{l \text{A} \alpha}  \left(\mathcal{C}_{4z, +} \right)^{-1}\mathrm{e}^{-\mathrm{i} \vec{k} \cdot \vec{R}_l^0} \\
        &= 
        \frac{1}{\sqrt{N}} \sum_{\overline{a}} \left[\text{C}_{4z, +}\right]_{a \overline{a}} \sum_{l} u_{l' \text{A} \overline{\alpha}} \mathrm{e}^{-\mathrm{i} \left(\text{C}_{4z, +}^{-1}\vec{R}_{l'}^0 \right)\cdot \vec{k}}  \\
        & = 
         \frac{1}{\sqrt{N}} \sum_{\overline{a}} \left[\text{C}_{4z, +}\right]_{a \overline{a}} \sum_{l} u_{l \text{A} \overline{\alpha}} \mathrm{e}^{-\mathrm{i} \left(\text{C}_{4z, +} \vec{k}\right) \cdot \vec{R}_{l}^0} \\ 
         & = 
         \sum_{\overline{\alpha}} \left[\text{C}_{4z, +}\right]_{\alpha \overline{\alpha}} u_{\vec{k}' \text{A} \overline{\alpha}}, 
    \end{aligned}
    \label{eq:rotuA}
\end{equation}
and, similarly, for displacements of the B sublattice,
\begin{equation}
    \begin{aligned}
        \mathcal{C}_{4z, +} u_{\vec{k} \text{B} \alpha} \mathcal{C}_{4z, +}^{-1} 
        & = \frac{1}{\sqrt{N}} \sum_{l} \mathcal{C}_{4z, +} u_{l \text{B} \alpha}  \left(\mathcal{C}_{4z, +} \right)^{-1}\mathrm{e}^{-\mathrm{i} \vec{k} \cdot \vec{R}_l^0} \mathrm{e}^{-\mathrm{i} \vec{k} \cdot \vec{\tau}_\text{B}}\\
        &= 
        \frac{1}{\sqrt{N}} \sum_{\overline{a}} \left[\text{C}_{4z, +}\right]_{a \overline{a}} \sum_{l} u_{l' \text{B} \overline{\alpha}} \mathrm{e}^{-\mathrm{i} \left(\text{C}_{4z, +}^{-1}\vec{R}_{l'}^0 \right) \cdot \vec{k}} \mathrm{e}^{-\mathrm{i} \vec{k} \cdot \left(R^- \vec{a}_2 \right) }\mathrm{e}^{-\mathrm{i} \vec{k} \cdot \vec{\tau}_\text{B}} \\
        & = 
         \frac{1}{\sqrt{N}} \sum_{\overline{a}} \left[\text{C}_{4z, +}\right]_{a \overline{a}} \sum_{l} u_{l' \text{B} \overline{\alpha}} \mathrm{e}^{-\mathrm{i} \left( \text{C}_{4z, +}^{-1}\vec{R}_{l'}^0 \right)\cdot \vec{k}} \mathrm{e}^{-\mathrm{i} \vec{k} \cdot \left(- \vec{a}_1\right)}\mathrm{e}^{-\mathrm{i} \vec{k} \cdot \vec{\tau}_\text{B}}  \\ 
         & = 
          \frac{1}{\sqrt{N}} \sum_{\overline{a}} \left[\text{C}_{4z, +}\right]_{a \overline{a}} \sum_{l} u_{l' \text{B} \overline{\alpha}} \mathrm{e}^{-\mathrm{i} \left(\text{C}_{4z, +}^{-1}\vec{R}_{l'}^0 \right) \cdot \vec{k}} \mathrm{e}^{-\mathrm{i} \left(\text{C}_{4z, +}^{-1}\vec{\tau}_\text{B}\right) \cdot \vec{k}} \\
          & = 
          \frac{1}{\sqrt{N}} \sum_{\overline{a}} \left[\text{C}_{4z, +}\right]_{a \overline{a}} \sum_{l} u_{l' \text{B} \overline{\alpha}} \mathrm{e}^{-\mathrm{i} \left( \text{C}_{4z, +} \vec{k} \right)\cdot \left(\vec{R}_{l'}^0 + \vec{\tau}_\text{B} \right)} \\
          & = 
        \sum_{\overline{\alpha}} \left[\text{C}_{4z, +}\right]_{\alpha \overline{\alpha}} u_{\vec{k}' \text{B} \overline{\alpha}}. 
    \end{aligned}
    \label{eq:rotuB}
\end{equation}
Here, we have defined the rotated momentum $\vec{k}'= \text{C}_{4z, +} \vec{k}$, used Eq.~\eqref{AP: rotated vectors}, and noticed that $\text{C}_{4z, +}^{-1} \vec{a}_2 = -\vec{a}_1$, and $\vec{\tau}_{\text{B}} -\vec{a}_1 = \text{C}_{4z, +}^{-1}\vec{\tau}_\text{B}$ as well as $\vec{k} \cdot \left(\text{C}_{4z, +}^{-1} \vec{R}_{l}^0 \right) = \left( \text{C}_{4z, +} \vec{k} \right) \cdot \vec{R}_l^0$. 
Equations~\eqref{eq:rotuA} and \eqref{eq:rotuB} can be summarized as 
  \begin{equation}
       \mathcal{C}_{4z, +} u_{\vec{k} \kappa \alpha} \mathcal{C}_{4z, +}^{-1} =  \sum_{\overline{\alpha}} \left[\text{C}_{4z, +}\right]_{\alpha \overline{\alpha}} u_{\vec{k}' \kappa \overline{\alpha}}. 
       \label{AP: u_k rotated}
  \end{equation}
Following the same logic, one can derive the relevant relation for the momentum as 
    \begin{equation}
       \mathcal{C}_{4z, +} p_{\vec{k} \kappa \alpha} \mathcal{C}_{4z, +}^{-1} =  \sum_{\overline{\alpha}} \left[\text{C}_{4z, +}\right]_{\alpha \overline{\alpha}} p_{\vec{k}' \kappa \overline{\alpha}}. 
       \label{AP: p_k rotated}
  \end{equation}
  Using Eqs.~\eqref{AP: u_k rotated} and \eqref{AP: p_k rotated} in Eq.~\eqref{AP:eq:Raman coupling H_phonon}, we get for the Raman-like coupling 
  \begin{equation}
      \begin{aligned}
          \mathcal{C}_{4z, +} 
          \left[
          \frac{1}{2} \sum_{\vec{k}} \sum_{\alpha \beta} \sum_{\kappa \kappa'} \frac{1}{M_{\kappa}}p_{\vec{k} \kappa \alpha}^{\dagger} G_{\kappa'\beta}^{\kappa \alpha}\left( \vec{k} \right) u_{\vec{k} \kappa'\beta}
          \right] \mathcal{C}_{4z, +}^{-1}  
          & = 
          \frac{1}{2} \sum_{\vec{k}} \sum_{\alpha \beta} \sum_{\kappa \kappa'} \frac{1}{M_{\kappa}} \mathcal{C}_{4z, +} p_{\vec{k} \kappa \alpha}^{\dagger} \mathcal{C}_{4z, +}^{-1}G_{\kappa'\beta}^{\kappa \alpha}\left( \vec{k} \right)
          \mathcal{C}_{4z, +}
          u_{\vec{k} \kappa'\beta}
         \mathcal{C}_{4z, +}^{-1} \\
         &
         = 
         \frac{1}{2}
         \sum_{\overline{\alpha}\overline{\beta}} 
         \sum_{\vec{k}}
         \sum_{\alpha \beta}\sum_{\kappa \kappa'} \left[\text{C}_{4z, +}\right]_{\alpha \overline{\alpha}}  \left[\text{C}_{4z, +}\right]_{\beta \overline{\beta}}
         \frac{1}{M_{\kappa}}p_{\vec{k}' \kappa \overline{\alpha}}^{\dagger} G_{\kappa'\beta}^{\kappa \alpha}\left(\vec{k} \right) u_{\vec{k}' \kappa'\overline{\beta}} 
         \\
         & = 
         \frac{1}{2}
         \sum_{\vec{k}}
         \sum_{\alpha \beta} \sum_{\kappa \kappa'} \sum_{\overline{\alpha}\overline{\beta}}
          \left[\text{C}_{4z, +}\right]_{\overline{\alpha} \alpha} \left[\text{C}_{4z, +}\right]_{\overline{\beta} \beta}
         \frac{1}{M_{\kappa}} p_{\vec{k} \kappa \alpha}^{\dagger} G_{\kappa' \overline{\beta}}^{\kappa \overline{\alpha}}\left(-\vec{k}'\right) u_{\vec{k} \kappa' \beta}.  
      \end{aligned}
  \end{equation}
  Combining these results with the time-reversal symmetry relations in Eqs.~\eqref{APP: eq: TR on u k-space} and \eqref{APP: eq: TR on momentum k-space} provides the $\mathcal{C}_{4z,+} \mathcal{T}$ operation. If it is a symmetry, i.e., $\left(\mathcal{C}_{4z,+} \mathcal{T} \right)H_{\text{RM}} \left(\mathcal{C}_{4z,+} \mathcal{T} \right)^{-1} = H_{\text{RM}}$, the elements of the MBC obey 
  \begin{equation}
      G_{\kappa'\beta}^{\kappa \alpha}\left( \vec{k} \right) = - \sum_{\overline{\alpha} \overline{\beta}} \left[\text{C}_{4z, +}\right]_{\overline{\alpha} \alpha}  \left[\text{C}_{4z, +}\right]_{\overline{\beta} \beta}  G_{\kappa'\overline{\beta}}^{\kappa \overline{\alpha}}\left( -\vec{k}' \right). 
      \label{APP: eq: C4T constrain}
  \end{equation}
  Therefore, $\mathcal{C}_{4z, +} \mathcal{T}$ imposes a relation between different elements of the MBC matrix at two different momenta related by a $90^{\circ}$ rotation, rather than constraining one of the elements specifically. 
  \item \textbf{ Reflection about the $x=y$ line} ($\mathcal{M}_{xy}$)
  
  For a reflection about the $x=y$ line, the displacement and momentum operators change in real space as 
  \begin{equation}
      \begin{aligned}
         \mathcal{M}_{xy} u_{l \kappa \alpha} \mathcal{M}_{xy}^{-1} = \sum_{\overline{\alpha}} \left[\text{M}_{xy}\right]_{\alpha \overline{\alpha}} u_{l' \kappa \overline{\alpha}}, \\
         \mathcal{M}_{xy} p_{l \kappa \alpha} \mathcal{M}_{xy}^{-1} = \sum_{\overline{\alpha}} \left[\text{M}_{xy}\right]_{\alpha \overline{\alpha}} p_{l' \kappa \overline{\alpha}}, 
      \end{aligned}
  \end{equation}
  where $l'$ is the index of the reflected unit cell and $\left[\text{M}\right]_{\alpha \overline{\alpha}}$ are elements of the reflection matrix 
  \begin{equation}
      \text{M}_{xy} = \begin{pmatrix}
          0 & 1 \\
          1 & 0 
      \end{pmatrix}. 
  \end{equation}
  For the checkerboard lattice, the lattice vectors transform as $\vec{R}_{l'}^0 = \text{M}_{xy} \vec{R}_{l}^0$ while the basis index remains unchanged ($\kappa = \text{A}, \text{B}$). Proceeding analogously to the previous symmetry operations, one finds that
  \begin{equation}
  \begin{aligned}
       & \mathcal{M}_{xy} u_{\vec{k} \kappa \alpha} \mathcal{M}_{xy}^{-1} = \sum_{\overline{\alpha}} \left[\text{M}_{xy}\right]_{\alpha \overline{\alpha}} u_{\vec{k}' \kappa \overline{\alpha}}, 
       \\
       & 
       \mathcal{M}_{xy} p_{\vec{k} \kappa \alpha} \mathcal{M}_{xy}^{-1} = \sum_{\overline{\alpha}} \left[\text{M}_{xy}\right]_{\alpha \overline{\alpha}} p_{\vec{k}'\kappa \overline{\alpha}}, 
  \end{aligned}
  \end{equation}
  where now $\vec{k}' = \text{M}_{xy} \vec{k}$ is the reflected momentum. Consequently,  we find
  \begin{equation}
      \begin{aligned}
         \mathcal{M}_{xy} \left[
          \frac{1}{2} \sum_{\vec{k}} \sum_{\alpha \beta} \sum_{\kappa \kappa'}\frac{1}{M_{\kappa}} p_{\vec{k} \kappa \alpha}^{\dagger} G_{\kappa'\beta}^{\kappa \alpha}\left( \vec{k} \right) u_{\vec{k} \kappa'\beta}
          \right]\mathcal{M}_{xy}^{-1}
          & = 
          \frac{1}{2} \sum_{\vec{k}} \sum_{\alpha \beta} 
          \sum_{\kappa \kappa'} \frac{1}{M_{\kappa}} \mathcal{M}_{xy} p_{\vec{k} \kappa \alpha}^{\dagger} \mathcal{M}_{xy}^{-1} G_{\kappa'\beta}^{\kappa \alpha}\left( \vec{k} \right) \mathcal{M}_{xy} u_{\vec{k} \kappa'\beta} \mathcal{M}_{xy}^{-1} \\
          & 
          = \frac{1}{2} \sum_{\vec{k}} \sum_{\alpha \beta}  \sum_{\kappa \kappa'} 
          \sum_{\overline{\alpha} \overline{\beta}}
          \left[\text{M}_{xy}\right]_{\alpha \overline{\alpha}}
          \left[\text{M}_{xy}\right]_{\beta \overline{\beta}}
          \frac{1}{M_{\kappa}} \ p_{\vec{k}' \kappa \overline{\alpha}}^{\dagger} G_{\kappa'\beta}^{\kappa \alpha}\left( \vec{k} \right)  u_{\vec{k}' \kappa'\overline{\beta}} \\
          & 
          = 
          \frac{1}{2} \sum_{\vec{k}} \sum_{\alpha \beta} \sum_{\kappa \kappa'}  
          \sum_{\overline{\alpha} \overline{\beta}}
          \left[\text{M}_{xy}\right]_{\overline{\alpha} \alpha}
          \left[\text{M}_{xy}\right]_{\overline{\beta} \beta}
          \frac{1}{M_{\kappa}} \ p_{\vec{k} \kappa \alpha}^{\dagger} G_{\kappa'\overline{\beta}}^{\kappa \overline{\alpha}}\left( \vec{k}' \right)  u_{\vec{k} \kappa'\overline{\beta}}, 
      \end{aligned}
  \end{equation}
  which for a system obeying $\mathcal{M}_{xy} H_{\text{RM}} \mathcal{M}_{xy}^{-1} = H_{\text{RM}}$, constraints the elements of the MBC matrix as follows: 
  \begin{equation}
      G_{\kappa'\beta}^{\kappa \alpha}\left( \vec{k} \right) = \sum_{\overline{\alpha} \overline{\beta}} \left[\text{M}_{xy}\right]_{\overline{\alpha} \alpha}
          \left[\text{M}_{xy}\right]_{\overline{\beta} \beta} G_{\kappa'\overline{\beta}}^{\kappa \overline{\alpha}}\left( \vec{k}' \right). 
  \end{equation}
\end{enumerate}

Finally, we note that the MBC matrix is anti-Hermitian:
\begin{equation}
\begin{aligned}
    G_{\kappa \alpha}^{\kappa' \beta} \left( \vec{k} \right) 
    &=  \frac{\mathrm{i} }{N} \sum_{l,l'} \left[ \left\langle \frac{\partial \Phi_0}{\partial u_{l  \kappa' \beta}} \middle|\frac{\partial \Phi_0}{\partial u_{l'  \kappa \alpha}} \right \rangle - 
     \left\langle \frac{\partial \Phi_0}{\partial u_{l'  \kappa \alpha}} \middle|\frac{\partial \Phi_0}{\partial u_{l  \kappa' \beta}} \right \rangle \right] \mathrm{e}^{-\mathrm{i}\left(\vec{R}_{l \kappa}^0 - \vec{R}_{l'\kappa'}^0 \right) \cdot \vec{k}} \\
     & = 
     \frac{\mathrm{i} }{N} \sum_{l,l'} \left[ \left\langle \frac{\partial \Phi_0}{\partial u_{l'  \kappa' \beta}} \middle|\frac{\partial \Phi_0}{\partial u_{l  \kappa \alpha}} \right \rangle - 
     \left\langle \frac{\partial \Phi_0}{\partial u_{l  \kappa \alpha}} \middle|\frac{\partial \Phi_0}{\partial u_{l'  \kappa' \beta}} \right \rangle \right] \mathrm{e}^{-\mathrm{i}\left(\vec{R}_{l' \kappa'}^0 - \vec{R}_{l \kappa}^0 \right) \cdot \vec{k}} \\
     & =
      -\frac{1}{N} \sum_{l,l'}  G_{\kappa' \beta}^{\kappa \alpha} \left( \vec{R}_l^0 - \vec{R}_{l'}^0\right) \mathrm{e}^{-\mathrm{i}\left(\vec{R}_{l'\kappa'}^0 - \vec{R}_{l \kappa}^0 \right) \cdot \vec{k}} \\
      & = 
      -  \left(G_{\kappa' \beta}^{\kappa \alpha} \left( \vec{k} \right) \right)^*. 
\end{aligned}
\label{AP:eq: anti-hermiticity}
\end{equation}
\end{subsection}

\begin{subsection}{Molecular Berry Curvature: Analytical Results for the Checkerboard Lattice}
\label{AP:subsection: MBC checkerboard}
Here, we present results regarding the MBC for the checkerboard lattice considered in the main text. More specifically, we (a) derive expressions for the electron-phonon coupling kernels defined in Eq.~\eqref{AP:eq: e-ph coupling matrices} and (b) show based on the altermagnetic symmetries of the system that the MBC cannot lead to a splitting of the optical phonon frequencies at the $\Gamma$ point. 

\subsubsection{Derivation of Coupling Matrices}
\label{AP: coupling matrices}
The electron-phonon coupling operators entering Eq.~\eqref{eq: MBC reciprocal space stage 1} are given by
\begin{equation}
    M_{\vec{k} \kappa \alpha} = \sum_{l} \partial_{u_{l \kappa \alpha}} H_{\text{el}} \mathrm{e}^{-\mathrm{i} \vec{k} \cdot \vec{R}_{l \kappa}^0 }, 
\end{equation}
where $\alpha = x, y$ is the Cartesian coordinate, and $\kappa = A, B$ is the sublattice index. Below, we first derive these couplings for both the A and B sublattices for a general $\alpha$. We then specify the expressions for the two different directions $x$ and $y$, assuming for simplicity that the spatial dependence of the electronic Hamiltonian $H_{\text{el}}$ in Eq.~\eqref{eq: TB-H real space} arises solely from nearest-neighbor hopping terms $t$.
Ignoring all other interactions, we restate $H_{\text{el}}$ by explicitly writing all nearest-neighbor hoppings:
\begin{equation}
    \begin{aligned}
        H_{\text{el}}
        = & 
        -t \sum_{l }\left( 
        c^{\dagger}_{l  \text{A}} c_{l  \text{B}} \mathrm{e}^{-\mathrm{i} \phi} + c^{\dagger}_{l + l_y \text{A}} c_{l \text{B}} \mathrm{e}^{\mathrm{i} \phi} 
        + c^{\dagger}_{l + l_x \text{A}} c_{l \text{B}} \mathrm{e}^{\mathrm{i} \phi} + c^{\dagger}_{l + l_y + l_x \text{A}} c_{l \text{B}} \mathrm{e}^{-\mathrm{i} \phi} 
        \right.\\
        & +
        \left.
         c^{\dagger}_{l \text{B}} c_{l \text{A}} \mathrm{e}^{\mathrm{i} \phi} + 
         c^{\dagger}_{l -l_x \text{B}} c_{l \text{A}} \mathrm{e}^{-\mathrm{i} \phi} + c^{\dagger}_{l -l_x - l_y \text{B}} c_{l \text{A}} \mathrm{e}^{\mathrm{i} \phi} + 
         c^{\dagger}_{l -l_y \text{B}} c_{l \text{A}} \mathrm{e}^{-\mathrm{i} \phi}
        \right).
    \end{aligned}
\end{equation}
We have defined $l_x = \left(1,0\right)$ and $l_y = \left(0,1\right)$. The index $l$ corresponds to a unit cell denoted by the real lattice vector  $\vec{R}_l^0$, while the indices $l \pm l_x$, $l \pm l_y$, $l \pm \left(l_x + l_y\right)$ correspond to unit cells with lattice vectors $\vec{R}_{l \pm l_{x}}^0 = \vec{R}_l^0 \pm \vec{a}_1$, $\vec{R}_{l \pm l_{y}}^0 = \vec{R}_l^0 \pm \vec{a}_2$, and $ \vec{R}_{l \pm \left(l_{x} + l_{y}\right)}^0 = \vec{R}_l^0 \pm \left(\vec{a}_1 + \vec{a}_2 \right)$, respectively.
It follows that
\begin{equation}
    \begin{aligned}
         M_{\vec{k} \text{A} \alpha} 
         &= \sum_{l} \mathrm{e}^{-\mathrm{i} \vec{k} \cdot \vec{R}_{l \text{A}}^0 } \partial_{u_{l  \text{A} \alpha}} H_{\text{el}} \\
        & = -
        \sum_{l}  \mathrm{e}^{-\mathrm{i} \vec{k} \cdot \vec{R}_{l \text{A}}^0 }
        \left[
        \left(
        \frac{\partial t_{\vec{R}_l^0  \text{A}}^{\vec{R}_l^0  \text{B}}}{\partial u_{l  \text{A} \alpha}}  c^{\dagger}_{l \text{A}} c_{l \text{B}} \mathrm{e}^{-\mathrm{i} \phi} 
         + \frac{\partial t_{\vec{R}_l^0  \text{A}}^{\vec{R}_l^0  - \vec{a}_2   \text{B}}}{\partial u_{l \text{A} \alpha}} c^{\dagger}_{l \text{A}} c_{l - l_y \text{B}} \mathrm{e}^{\mathrm{i} \phi}
          + 
          \frac{\partial t_{\vec{R}_l^0  \text{A}}^{\vec{R}_l^0  - \vec{a}_1   \text{B}}}{\partial u_{l  \text{A} \alpha}} c^{\dagger}_{l \text{A}} c_{l - l_x \text{B}} \mathrm{e}^{\mathrm{i} \phi} 
         + \frac{\partial t_{\vec{R}_l^0 \text{A}}^{\vec{R}_l^0  - \vec{a}_1 -\vec{a}_2  \text{B}}}{\partial u_{l  \text{A} \alpha}} c^{\dagger}_{l \text{A}} c_{l - l_x -l_y \text{B}} \mathrm{e}^{-\mathrm{i} \phi}
        \right)
        \right. \\
        & 
         \qquad \qquad \left. 
         + 
         \left(
         \frac{\partial t_{\vec{R}_l^0  \text{B}}^{\vec{R}_l^0  \text{A}}}{\partial u_{l  \text{A} \alpha}}
         c^{\dagger}_{l \text{B}} c_{l \text{A}} \mathrm{e}^{\mathrm{i} \phi} + 
         \frac{\partial t_{\vec{R}_l^0 -\vec{a}_1 \text{B}}^{\vec{R}_l^0  \text{A}}}{\partial u_{l  \text{A} \alpha}}
         c^{\dagger}_{l -l_x \text{B}} c_{l \text{A}} \mathrm{e}^{-\mathrm{i} \phi}  
         + \frac{\partial t_{\vec{R}_l^0 -\vec{a}_1 -\vec{a}_2 \text{B}}^{\vec{R}_l^0  \text{A}}}{\partial u_{l \text{A} \alpha}}c^{\dagger}_{l -l_x - l_y \text{B}} c_{l \text{A}} \mathrm{e}^{\mathrm{i} \phi} + 
          \frac{\partial t_{\vec{R}_l^0 -\vec{a}_2 \text{B}}^{\vec{R}_l^0  \text{A}}}{\partial u_{l \text{A} \alpha}}
         c^{\dagger}_{l -l_y \text{B}} c_{l \text{A}} \mathrm{e}^{-\mathrm{i} \phi}
         \right)
        \right]. 
    \end{aligned}
\end{equation}
In each term, the subscripts and superscripts of the hopping $t$ explicitly isolate the hopping process that is modified by the displacement of the A-site atom in the $l$-th unit cell. After a Fourier transformation of the fermionic operators, 
\begin{equation}
    c_{l \kappa} = \frac{1}{\sqrt{N}} \sum_{\vec{q}} \mathrm{e}^{\mathrm{i} \vec{q} \cdot  \vec{R}_l^0 } c_{\vec{q} \kappa}, 
\end{equation}
we obtain
\begin{equation}
    \begin{aligned}
         M_{\vec{k} \text{A} \alpha}
         &= -\frac{1}{N} \mathrm{e}^{-\mathrm{i} \vec{k} \cdot \vec{\tau}_{\text{A}}}\sum_{\vec{q}, \vec{q}'}
        \sum_{l}  
        \left[ \mathrm{e}^{\mathrm{i} \left(\vec{q} -\vec{k}   -\vec{q}' \right)\cdot\vec{R}_l^0 }
        \left(
        \frac{\partial t_{\vec{R}_l^0 \text{A}}^{\vec{R}_l^0  \text{B}}}{\partial u_{l \text{A} \alpha}} \mathrm{e}^{-\mathrm{i} \phi} 
         + \frac{\partial t_{\vec{R}_l^0  \text{A}}^{\vec{R}_l^0  - \vec{a}_2   \text{B}}}{\partial u_{l  \text{A} \alpha}} \mathrm{e}^{\mathrm{i} \phi} \mathrm{e}^{\mathrm{i} \vec{q} \cdot \left(-\vec{a}_2 \right)}
          + \frac{\partial t_{\vec{R}_l^0  \text{A}}^{\vec{R}_l^0  - \vec{a}_1   \text{B}}}{\partial u_{l  \text{A} \alpha}}  \mathrm{e}^{\mathrm{i} \phi} e^{\mathrm{i} \vec{q} \cdot \left(-\vec{a}_1 \right)} 
         + \frac{\partial t_{\vec{R}_l^0  \text{A}}^{\vec{R}_l^0  - \vec{a}_1 -\vec{a}_2  B}}{\partial u_{l  \text{A} \alpha}} \mathrm{e}^{-\mathrm{i} \phi} \mathrm{e}^{\mathrm{i} \vec{q} \cdot \left(-\vec{a}_2 -\vec{a}_1\right)}
        \right) 
        c^{\dagger}_{\vec{q}'\text{A}}  c_{\vec{q} \text{B}}
        \right. \\
        & 
        \qquad\qquad
         \left. 
         + 
         \mathrm{e}^{\mathrm{i} \left(\vec{q}' -\vec{k}   -\vec{q} \right)\cdot\vec{R}_l^0 }\left(
         \frac{\partial t_{\vec{R}_l^0  \text{B}}^{\vec{R}_l^0  \text{A}}}{\partial u_{l  \text{A} \alpha}}
          \mathrm{e}^{\mathrm{i} \phi} + 
         \frac{\partial t_{\vec{R}_l^0 -\vec{a}_1 \text{B}}^{\vec{R}_l^0  \text{A}}}{\partial u_{l  \text{A} \alpha}}
         \mathrm{e}^{-\mathrm{i} \phi} \mathrm{e}^{-\mathrm{i} \vec{q} \cdot \left(-\vec{a}_1 \right)}
         + 
         \frac{\partial t_{\vec{R}_l^0 -\vec{a}_1 -\vec{a}_2 \text{B}}^{\vec{R}_l^0  \text{A}}}{\partial u_{l  \text{A} \alpha}} \mathrm{e}^{\mathrm{i} \phi} \mathrm{e}^{-\mathrm{i} \vec{q} \cdot \left(-\vec{a}_2 -\vec{a}_1 \right)} + 
          \frac{\partial t_{\vec{R}_l^0 -\vec{a}_2 \text{B}}^{\vec{R}_l^0  \text{A}}}{\partial u_{l  \text{A} \alpha}}
          \mathrm{e}^{-\mathrm{i} \phi}  \mathrm{e}^{-\mathrm{i} \vec{q} \cdot \left(-\vec{a}_2\right)}
         \right) c^{\dagger}_{\vec{q} \text{B}}  c_{\vec{q}' \text{A}}
        \right] \\
        & = -\sum_{\vec{q}}
        \left(
        \frac{\partial t_{\vec{R}_l^0 \text{A}}^{\vec{R}_l^0  \text{B}}}{\partial u_{l  \text{A} \alpha}} \mathrm{e}^{-\mathrm{i} \phi} 
         + \frac{\partial t_{\vec{R}_l^0  \text{A}}^{\vec{R}_l^0  - \vec{a}_2   \text{B}}}{\partial u_{l  A \alpha}} \mathrm{e}^{\mathrm{i} \phi} \mathrm{e}^{-\mathrm{i} 
         \left(\vec{q} + \vec{k} \right)\cdot \vec{a}_2 }
          + 
          \frac{\partial t_{\vec{R}_l^0  \text{A}}^{\vec{R}_l^0  - \vec{a}_1   \text{B}}}{\partial u_{l  \text{A} \alpha}}  \mathrm{e}^{\mathrm{i} \phi} \mathrm{e}^{-\mathrm{i} 
         \left(\vec{q} + \vec{k} \right)\cdot \vec{a}_1 }
         + \frac{\partial t_{\vec{R}_l^0  \text{A}}^{\vec{R}_l^0  - \vec{a}_1 -\vec{a}_2  \text{B}}}{\partial u_{l  \text{A} \alpha}} \mathrm{e}^{-\mathrm{i} \phi} \mathrm{e}^{-\mathrm{i} 
         \left(\vec{q} + \vec{k} \right)\cdot \left(\vec{a}_2 + \vec{a}_1\right)}
        \right) 
        c^{\dagger}_{\vec{q} \text{A}}  c_{\vec{q} + \vec{k} \text{B}}
         \\
        & 
         \qquad + 
         \left(
         \frac{\partial t_{\vec{R}_l^0 \text{B}}^{\vec{R}_l^0  \text{A}}}{\partial u_{l  \text{A} \alpha}}
          \mathrm{e}^{\mathrm{i} \phi} + 
         \frac{\partial t_{\vec{R}_l^0 -\vec{a}_1 \text{B}}^{\vec{R}_l^0  \text{A}}}{\partial u_{l  \text{A} \alpha}}
         \mathrm{e}^{-\mathrm{i} \phi} \mathrm{e}^{\mathrm{i} \vec{q} \cdot \vec{a}_1}  
         + \frac{\partial t_{\vec{R}_l^0 -\vec{a}_1 -\vec{a}_2 \text{B}}^{\vec{R}_l^0  \text{A}}}{\partial u_{l  \text{A} \alpha}} \mathrm{e}^{\mathrm{i} \phi} \mathrm{e}^{\mathrm{i} \vec{q} \cdot \left(\vec{a}_2 +\vec{a}_1 \right)} + 
          \frac{\partial t_{\vec{R}_l^0 -\vec{a}_2 \text{B}}^{\vec{R}_l^0  \text{A}}}{\partial u_{l  \text{A} \alpha}}
          \mathrm{e}^{-\mathrm{i} \phi}  \mathrm{e}^{\mathrm{i} \vec{q} \cdot \vec{a}_2}
         \right) c^{\dagger}_{\vec{q} \text{B}}  c_{\vec{q} + \vec{k} \text{A}} \\
         &=  
        \sum_{\vec{q}} \vec{\Psi}^{\dagger}_{\vec{q}} \mathcal{M}_{\vec{k} \text{A} \alpha} \vec{\Psi}_{\vec{q} + \vec{k}}
    \end{aligned}
\end{equation}
where we made use of $\vec{\tau}_{\text{A}}=0$ and defined $\vec{\Psi}_{\vec{q}} = \begin{pmatrix}
    c_{\vec{q} \text{A}} &  c_{\vec{q} \text{B}}
\end{pmatrix}^T$ 
and 
\begin{equation}
    \mathcal{M}_{\vec{k} \text{A} \alpha}  = 
    \begin{pmatrix}
        0 & \left[\mathcal{M}_{\vec{k} \text{A} \alpha}\right]_{12} \\
        \left[\mathcal{M} _{\vec{k} \text{A} \alpha}\right]_{21} & 0 
    \end{pmatrix}, 
\end{equation}
with 
\begin{equation}
\begin{aligned}
& \begin{aligned}
    \left[\mathcal{M}_{\vec{k} \text{A} \alpha}\right]_{12} 
    &= -\left(
        \frac{\partial t_{\vec{R}_l^0  \text{A}}^{\vec{R}_l^0  \text{B}}}{\partial u_{l  \text{A} \alpha}} \mathrm{e}^{-\mathrm{i} \phi} 
         + \frac{\partial t_{\vec{R}_l^0  \text{A}}^{\vec{R}_l^0  - \vec{a}_2   \text{B}}}{\partial u_{l  \text{A} \alpha}} \mathrm{e}^{\mathrm{i} \phi} \mathrm{e}^{-\mathrm{i} 
         \left(\vec{q} + \vec{k} \right)\cdot \vec{a}_2 }  
          + \frac{\partial t_{\vec{R}_l^0  \text{A}}^{\vec{R}_l^0  - \vec{a}_1   \text{B}}}{\partial u_{l  \text{A} \alpha}}  \mathrm{e}^{\mathrm{i} \phi} \mathrm{e}^{-\mathrm{i} 
         \left(\vec{q} + \vec{k} \right)\cdot \vec{a}_1 }
         + \frac{\partial t_{\vec{R}_l^0  \text{A}}^{\vec{R}_l^0  - \vec{a}_1 -\vec{a}_2 , \text{B}}}{\partial u_{l  \text{A} \alpha}} \mathrm{e}^{-\mathrm{i} \phi} \mathrm{e}^{-\mathrm{i} 
         \left(\vec{q} + \vec{k} \right)\cdot \left(\vec{a}_2 + \vec{a}_1\right)}\right), 
\end{aligned} \\
& \begin{aligned}
    \left[\mathcal{M}_{\vec{k} \text{A} \alpha}\right]_{21} &= 
    -\left(
    \frac{\partial t_{\vec{R}_l^0  \text{B}}^{\vec{R}_l^0  \text{A}}}{\partial u_{l  \text{A} \alpha}}
    \mathrm{e}^{\mathrm{i} \phi} + 
         \frac{\partial t_{\vec{R}_l^0 -\vec{a}_1 \text{B}}^{\vec{R}_l^0  \text{A}}}{\partial u_{l  \text{A} \alpha}}
         \mathrm{e}^{-\mathrm{i} \phi} \mathrm{e}^{\mathrm{i} \vec{q} \cdot \vec{a}_1} 
         + \frac{\partial t_{\vec{R}_l^0 -\vec{a}_1 -\vec{a}_2 \text{B}}^{\vec{R}_l^0  \text{A}}}{\partial u_{l  \text{A} \alpha}} \mathrm{e}^{\mathrm{i} \phi} \mathrm{e}^{\mathrm{i} \vec{q} \cdot \left(\vec{a}_2 +\vec{a}_1 \right)} + 
          \frac{\partial t_{\vec{R}_l^0 -\vec{a}_2 \text{B}}^{\vec{R}_l^0  \text{A}}}{\partial u_{l  \text{A} \alpha}}
          \mathrm{e}^{-\mathrm{i} \phi}  \mathrm{e}^{\mathrm{i} \vec{q} \cdot \vec{a}_2} 
          \right). 
\end{aligned}
\end{aligned}
\end{equation}
An identical analysis for displacements of ions belonging to the B sublattice leads to
\begin{equation}
\begin{aligned}
     M_{\vec{k} \text{B} \alpha} 
     &= 
     \sum_{l} \mathrm{e}^{-\mathrm{i} \vec{k} \cdot \vec{R}_{l \text{B}}^0 } \partial_{u_{l \text{B} \alpha}} H_{\text{el}}
     \\
     &
     = -
        \sum_{l} \mathrm{e}^{-\mathrm{i} \vec{k} \cdot \vec{R}_{l \text{B}}^0 }
        \left( 
        \frac{\partial t^{\vec{R}_l^0  \text{B}}_{\vec{R}_l^0  \text{A}}}{\partial u_{l  \text{B} \alpha}}c^{\dagger}_{\vec{R}_l^0  \text{A}} c_{l \text{B}} \mathrm{e}^{-\mathrm{i} \phi} +  \frac{\partial t^{\vec{R}_l^0  \text{B}}_{\vec{R}_l^0  + \vec{a}_2 \text{A}}}{\partial u_{l  \text{B} \alpha}}c^{\dagger}_{l + l_y \text{A}} c_{l \text{B}} \mathrm{e}^{\mathrm{i} \phi}  
        + 
        \frac{\partial t^{\vec{R}_l^0  \text{B}}_{\vec{R}_l^0  + \vec{a}_1 \text{A}}}{\partial u_{l \text{B} \alpha}}c^{\dagger}_{l  + 
        l_x \text{A}} c_{l \text{B}} \mathrm{e}^{\mathrm{i} \phi} +
        \frac{\partial t^{\vec{R}_l^0  \text{B}}_{\vec{R}_l^0  + \vec{a}_1 +\vec{a}_2 \text{A}}}{\partial u_{l  \text{B} \alpha}}c^{\dagger}_{l + l_y + l_x \text{A}} c_{l \text{B}} \mathrm{e}^{-\mathrm{i} \phi} 
        \right.\\
        &\qquad \qquad +
        \left.
        \frac{\partial t^{\vec{R}_l^0  \text{A}}_{\vec{R}_l^0  \text{B}}}{\partial u_{l  \text{B} \alpha}}c^{\dagger}_{l \text{B}} c_{l \text{A}} \mathrm{e}^{\mathrm{i} \phi} + 
        \frac{\partial t^{\vec{R}_l^0  +\vec{a}_1 \text{A}}_{\vec{R}_l^0  \text{B}}}{\partial u_{l \text{B} \alpha}}
         c^{\dagger}_{l \text{B}} c_{l +l_x \text{A}} \mathrm{e}^{-\mathrm{i} \phi} 
         +  
         \frac{\partial t^{\vec{R}_l^0  + \vec{a}_1 + \vec{a}_2 \text{A}}_{\vec{R}_l^0  \text{B}}}{\partial u_{l  \text{B} \alpha}}c^{\dagger}_{l  \text{B}} c_{l +l_x + l_y \text{A}} \mathrm{e}^{\mathrm{i} \phi} + 
         \frac{\partial t^{\vec{R}_l^0  + \vec{a}_2 \text{A}}_{\vec{R}_l^0  \text{B}}}{\partial u_{l  \text{B} \alpha}}
         c^{\dagger}_{l  \text{B}} c_{l +l_y \text{A}} \mathrm{e}^{-\mathrm{i} \phi}
        \right) \\
        & = -
        \frac{1}{N}
        \mathrm{e}^{-\mathrm{i} \vec{k} \cdot \vec{\tau}_{\text{B}}}
        \sum_{l} \sum_{\vec{q}, \vec{q}'} 
        \left[
        \mathrm{e}^{\mathrm{i} \left(\vec{q} - \vec{q}'-\vec{k} \right)\cdot \vec{R}_l^0 } 
        \left(
         \frac{\partial t^{\vec{R}_l^0  \text{B}}_{\vec{R}_l^0  \text{A}}}{\partial u_{l  \text{B} \alpha}} \mathrm{e}^{-\mathrm{i} \phi} +  \frac{\partial t^{\vec{R}_l^0  \text{B}}_{\vec{R}_l^0  + \vec{a}_2 \text{A}}}{\partial u_{l  \text{B} \alpha}} \mathrm{e}^{\mathrm{i} \phi}
         \mathrm{e}^{-\mathrm{i} \vec{q}'\cdot \vec{a}_2} 
        + 
        \frac{\partial t^{\vec{R}_l^0  \text{B}}_{\vec{R}_l^0  + \vec{a}_1 \text{A}}}{\partial u_{l   \text{B} \alpha}} \mathrm{e}^{\mathrm{i} \phi} \mathrm{e}^{-\mathrm{i} \vec{q}'\cdot \vec{a}_1} +
        \frac{\partial t^{\vec{R}_l^0  \text{B}}_{\vec{R}_l^0  + \vec{a}_1 +\vec{a}_2 \text{A}}}{\partial u_{l \text{B} \alpha}} \mathrm{e}^{-\mathrm{i} \phi} \mathrm{e}^{-\mathrm{i} \vec{q}'\cdot \left(\vec{a}_2 +\vec{a}_1 \right)}
        \right)
        \right. 
        c^{\dagger}_{\vec{q}' \text{A}} c_{\vec{q} \text{B}} 
        \\
        & \qquad \qquad +
        \mathrm{e}^{\mathrm{i} \left(\vec{q}' - \vec{q}-\vec{k} \right)\cdot \vec{R}_l^0 } 
        \left.
        \left(
         \frac{\partial t^{\vec{R}_l^0  \text{A}}_{\vec{R}_l^0  \text{B}}}{\partial u_{l \text{B} \alpha}} \mathrm{e}^{\mathrm{i} \phi} 
         +  
         \frac{\partial t^{\vec{R}_l^0  + \vec{a}_2 \text{A}}_{\vec{R}_l^0  \text{B}}}{\partial u_{l \text{B} \alpha}} \mathrm{e}^{-\mathrm{i} \phi}
         \mathrm{e}^{\mathrm{i} \vec{q}'\cdot \vec{a}_2}
        + \frac{\partial t^{\vec{R}_l^0  + \vec{a}_1 \text{A}}_{\vec{R}_l^0  \text{B}}}{\partial u_{l \text{B} \alpha}} \mathrm{e}^{-\mathrm{i} \phi} \mathrm{e}^{\mathrm{i} \vec{q}'\cdot \vec{a}_1} 
        +
        \frac{\partial t^{\vec{R}_l^0 + \vec{a}_1 +\vec{a}_2 \text{A}}_{\vec{R}_l^0  \text{B}}}{\partial u_{l \text{B} \alpha}} \mathrm{e}^{\mathrm{i} \phi} \mathrm{e}^{\mathrm{i} \vec{q}'\cdot \left(\vec{a}_2 +\vec{a}_1 \right)}
        \right) 
        c^{\dagger}_{\vec{q} \text{B}} c_{\vec{q}' \text{A}}
        \right] \\
        & 
        = 
        \sum_{\vec{q}} \vec{\Psi}^{\dagger}_{\vec{q}} \mathcal{M}_{\vec{k} \text{B} \alpha} \vec{\Psi}_{\vec{k}+\vec{q}}, 
\end{aligned}
\end{equation}
where 
\begin{equation}
\begin{aligned}
    \mathcal{M}_{\vec{k} \text{B} \alpha}
    &= 
    \begin{pmatrix}
        0 & \left[  \mathcal{M}_{\vec{k} \text{B} \alpha}  \right]_{12} \\
        \left[  \mathcal{M}_{\vec{k} \text{B} \alpha}  \right]_{21} & 0 
    \end{pmatrix}, 
    \\
    \left[\mathcal{M}_{\vec{k} \text{B} \alpha}  \right]_{12}
    &= 
    - \mathrm{e}^{-\mathrm{i} \vec{k} \cdot \vec{\tau}_{\text{B}}}\left(
    \frac{\partial t^{\vec{R}_l^0  \text{B}}_{\vec{R}_l^0  \text{A}}}{\partial u_{l  \text{B} \alpha}} \mathrm{e}^{-\mathrm{i} \phi} +  \frac{\partial t^{\vec{R}_l^0  \text{B}}_{\vec{R}_l^0  + \vec{a}_2 \text{A}}}{\partial u_{l  \text{B} \alpha}} \mathrm{e}^{\mathrm{i} \phi}
         \mathrm{e}^{-\mathrm{i} \vec{q}\cdot \vec{a}_2}   
        + \frac{\partial t^{\vec{R}_l^0  \text{B}}_{\vec{R}_l^0  + \vec{a}_1 \text{A}}}{\partial u_{l  \text{B} \alpha}} \mathrm{e}^{\mathrm{i} \phi} \mathrm{e}^{-\mathrm{i} \vec{q}\cdot \vec{a}_1} +
        \frac{\partial t^{\vec{R}_l^0  \text{B}}_{\vec{R}_l^0  + \vec{a}_1 +\vec{a}_2 \text{A}}}{\partial u_{l  \text{B} \alpha}} \mathrm{e}^{-\mathrm{i} \phi} \mathrm{e}^{-\mathrm{i} \vec{q}\cdot \left(\vec{a}_2 +\vec{a}_1 \right)}
        \right), 
        \\
        \left[\mathcal{M}_{\vec{k} \text{B} \alpha} \right]_{21} 
        &= - \mathrm{e}^{-\mathrm{i} \vec{k} \cdot \vec{\tau}_{\text{B}}}\left(
        \frac{\partial t^{\vec{R}_l^0  \text{A}}_{\vec{R}_l^0  \text{B}}}{\partial u_{l  \text{B} \alpha}} \mathrm{e}^{\mathrm{i} \phi} 
         +  
         \frac{\partial t^{\vec{R}_l^0  + \vec{a}_2 \text{A}}_{\vec{R}_l^0  \text{B}}}{\partial u_{l \text{B} \alpha}} \mathrm{e}^{-\mathrm{i} \phi}
         \mathrm{e}^{\mathrm{i} \left(\vec{q} + \vec{k} \right)\cdot \vec{a}_2} 
        + \frac{\partial t^{\vec{R}_l^0  + \vec{a}_1 \text{A}}_{\vec{R}_l^0  \text{B}}}{\partial u_{l \text{B} \alpha}} \mathrm{e}^{-\mathrm{i} \phi} \mathrm{e}^{\mathrm{i} \left(\vec{q}+\vec{k}\right)\cdot \vec{a}_1} 
        +
        \frac{\partial t^{\vec{R}_l^0 + \vec{a}_1 +\vec{a}_2 \text{A}}_{\vec{R}_l^0  \text{B}}}{\partial u_{l  \text{B} \alpha}} \mathrm{e}^{\mathrm{i} \phi} \mathrm{e}^{\mathrm{i} \left(\vec{q} + \vec{k}\right)\cdot \left(\vec{a}_2 +\vec{a}_1 \right)}
        \right)
        .
\end{aligned}
\end{equation}

To obtain explicit expressions for the coupling matrices in the $x$ and $y$ directions, we decompose the displacement vectors into components parallel and perpendicular to the corresponding bond. We further assume that, to linear order, only the parallel component modifies the bond length. Since the hopping amplitudes decrease as the bond length increases due to the reduced orbital overlap, their derivatives with respect to bond extension are negative (and positive for bond compression). Under these assumptions, the general expressions above reduce to the forms given in Eqs.~\eqref{eq: coupling elements A} and \eqref{eq: coupling elements B}.
\end{subsection}

\subsubsection{Optical Phonon Splitting at the \texorpdfstring{$\Gamma$}{Gamma} Point}
\label{AP: subsubsection: Gamma splitting}
We demonstrate here that the altermagnetic symmetries of the electron system constrain the MBC in such a way that does not allow for a splitting of the optical modes at the $\Gamma$ point. 
First, we recall the finding of App.~\ref{AP: subsec: MBC symmetries} that $\mathcal{T} \mathcal{C}_{4z, +}$ and $\mathcal{M}_{xy}$ impose the following constrains on a general element of the MBC-matrix 
\begin{equation}
    \begin{aligned}
        G_{\kappa'\beta}^{\kappa \alpha}\left( k_x, k_y \right) &= - \sum_{\overline{\alpha} \overline{\beta}} \left[\text{C}_{4z, +}\right]_{\overline{\alpha} \alpha}  \left[\text{C}_{4z, +}\right]_{\overline{\beta} \beta}  G_{\kappa'\overline{\beta}}^{\kappa \overline{\alpha}}\left(-k_y, k_x \right), \\
         G_{\kappa'\beta}^{\kappa \alpha}\left( k_x, k_y \right) & = \sum_{\overline{\alpha} \overline{\beta}} \left[\text{M}_{xy}\right]_{\overline{\alpha} \alpha}
          \left[\text{M}_{xy}\right]_{\overline{\beta} \beta} G_{\kappa'\overline{\beta}}^{\kappa \overline{\alpha}}\left( k_y, k_x\right), 
    \end{aligned}
    \label{AP: eq: symmetry relations for gamma splitting}
\end{equation}
where $\left[\text{C}_{4z, +}\right]_{\overline{\alpha} \alpha}$ and $\left[\text{M}_{xy}\right]_{\overline{\alpha} \alpha}$ are $\overline{\alpha}\alpha$ element of the rotation and mirroring matrices defined as $\text{C}_{4z, +} = \begin{pmatrix}
    0 & -1 \\
    1 & 0 
\end{pmatrix}$ and 
$\text{M}_{xy} = \begin{pmatrix}
    0 & 1 \\
    1 & 0 
\end{pmatrix}$ respectively. 

We start off considering the full MBC-matrix at the $\Gamma$ point, reading 
\begin{equation}
    \overline{G}_{\vec{k}=0} = 
    \begin{pmatrix}
        \overline{G}_{\text{A}x}^{\text{A}x}\left( \vec{k}=0\right)  & \overline{G}_{\text{A}y}^{\text{A}x} \left( \vec{k}=0 \right)& \overline{G}_{\text{B}x}^{\text{A}x}\left( \vec{k} =0\right)  & \overline{G}_{\text{B}y}^{\text{A}x}\left( \vec{k}=0 \right)  \\
        -\overline{G}_{\text{A}y}^{\text{A}x}\left( \vec{k}=0 \right)  & \overline{G}_{\text{A}y}^{\text{A}y}\left( \vec{k}=0 \right)  & \overline{G}_{\text{B}x}^{\text{A}y}\left( \vec{k}=0 \right)    & \overline{G}_{\text{B}y}^{\text{A}y}\left(\vec{k}=0 \right)  \\
        -\overline{G}_{\text{B}x}^{\text{A}x}\left( \vec{k}=0 \right) & - \overline{G}^{\text{A}y}_{\text{B}x} \left( \vec{k}=0\right) & \overline{G}^{\text{B}x}_{\text{B}x}\left( \vec{k} =0\right)    & \overline{G}^{\text{B}y}_{\text{B}x}\left( \vec{k}=0\right)  \\
        -\overline{G}^{\text{A}x}_{\text{B}y} \left( \vec{k}=0 \right) & - \overline{G}^{\text{A}y}_{\text{B}y}\left( \vec{k}=0 \right) & - \overline{G}_{\text{B}x}^{\text{B}y}\left( \vec{k} =0\right)  & \overline{G}_{\text{B}y}^{\text{B}y}\left( \vec{k}=0 \right) 
    \end{pmatrix}, 
    \label{AP: G-matrix checkerboard elements at Gamma}
\end{equation}
and gradually eliminate its elements. In writing Eq.~\eqref{AP: G-matrix checkerboard elements at Gamma} have already made use of the anti-Hermitian nature of the matrix and noticed that at the $\Gamma$ point the MBC is by definition real, thus leading to the relation of $G_{\kappa'\beta}^{\kappa \alpha}\left(\vec{k}=0 \right)= -G_{\kappa\alpha}^{\kappa' \beta}\left(\vec{k} = 0 \right)$, rendering $\overline{G}_{\vec{k}=0}$ an anti-symmetric matrix. 
\begin{enumerate}
    \item Diagonal Elements: 
    
    Since the $\overline{G}_{\vec{k}=0}$ is anti-symmetric, the diagonal elements are by definition zero. The matrix thus simplifies to 
    \begin{equation}
    \overline{G}_{\vec{k}=0} = 
    \begin{pmatrix}
        0  & \overline{G}_{\text{A}y}^{\text{A}x} \left( \vec{k}=0 \right)& \overline{G}_{\text{B}x}^{\text{A}x}\left( \vec{k} =0\right)  & \overline{G}_{\text{B}y}^{\text{A}x}\left( \vec{k}=0 \right)  \\
        -\overline{G}_{\text{A}y}^{\text{A}x}\left( \vec{k}=0 \right)  & 0  & \overline{G}_{\text{B}x}^{\text{A}y}\left( \vec{k}=0 \right)    & \overline{G}_{\text{B}y}^{\text{A}y}\left(\vec{k}=0 \right)  \\
        -\overline{G}_{\text{B}x}^{\text{A}x}\left( \vec{k}=0 \right) & - \overline{G}^{\text{A}y}_{\text{B}x} \left( \vec{k}=0\right) & 0    & \overline{G}^{\text{B}y}_{\text{B}x}\left( \vec{k}=0\right)  \\
        -\overline{G}^{\text{A}x}_{\text{B}y} \left( \vec{k}=0 \right) & - \overline{G}^{\text{A}y}_{\text{B}y}\left( \vec{k}=0 \right) & - \overline{G}_{\text{B}x}^{\text{B}y}\left( \vec{k} =0\right)  & 0
    \end{pmatrix}. 
    \label{AP: G-matrix checkerboard elements at Gamma, first step}
\end{equation}
\item $\overline{G}_{\text{A}y}^{\text{A}x} \left( \vec{k}=0 \right)$ and $\overline{G}_{\text{B}y}^{\text{B}x} \left( \vec{k}=0 \right)$: 

These two elements can be shown to be zero by combining the $\mathcal{C}_{4z,+}\mathcal{T}$ symmetry together with the anti-symmetric nature of $\overline{G}_{\vec{k}=0}$. Indeed, $\mathcal{C}_{4z,+}\mathcal{T}$ gives the constrain $\overline{G}_{\text{A}y}^{\text{A}x} \left( \vec{k}=0 \right) = \overline{G}_{\text{A}x}^{\text{A}y} \left( \vec{k}=0 \right)$ and together with $\overline{G}_{\text{A}y}^{\text{A}x} \left( \vec{k}=0 \right) = -\overline{G}_{\text{A}x}^{\text{A}y} \left( \vec{k}=0 \right)$ forces the element to vanish. The same relations apply for the B sublattice, and consequently both elements are zero. Our matrix now reads as 
  \begin{equation}
    \overline{G}_{\vec{k}=0} = 
    \begin{pmatrix}
        0  & 0& \overline{G}_{\text{B}x}^{\text{A}x}\left( \vec{k} =0\right)  & \overline{G}_{\text{B}y}^{\text{A}x}\left( \vec{k}=0 \right)  \\
       0  & 0  & \overline{G}_{\text{B}x}^{\text{A}y}\left( \vec{k}=0 \right)    & \overline{G}_{\text{B}y}^{\text{A}y}\left(\vec{k}=0 \right)  \\
        -\overline{G}_{\text{B}x}^{\text{A}x}\left( \vec{k}=0 \right) & - \overline{G}^{\text{A}y}_{\text{B}x} \left( \vec{k}=0\right) & 0    & 0  \\
        -\overline{G}^{\text{A}x}_{\text{B}y} \left( \vec{k}=0 \right) & - \overline{G}^{\text{A}y}_{\text{B}y}\left( \vec{k}=0 \right) & 0 & 0
    \end{pmatrix}. 
    \label{AP: G-matrix checkerboard elements at Gamma, second step}
\end{equation}

\item $\overline{G}_{\text{B}x}^{\text{A}x}\left( \vec{k} =0\right)$ and $\overline{G}_{\text{B}x}^{\text{A}x}\left( \vec{k} =0\right)$:

These two elements can be shown to vanish identically due to the combination of the two aforementioned symmetries. This can be readily seen to be true as $\mathcal{C}_{4z,+}\mathcal{T}$ provides the relation $\overline{G}_{\text{B}x}^{\text{A}x}\left( \vec{k} =0\right) = - \overline{G}_{\text{B}y}^{\text{A}y}\left(\vec{k} =0\right)$ while $\mathcal{M}_{xy}$ gives $\overline{G}_{\text{B}x}^{\text{A}x}\left( \vec{k} =0\right) = \overline{G}_{\text{B}y}^{\text{A}y}\left(\vec{k} =0\right)$, making the two elements vanish. Consequently, the MBC-matrix at $\Gamma$ becomes

  \begin{equation}
    \overline{G}_{\vec{k}=0} = 
    \begin{pmatrix}
        0  & 0& 0 & \overline{G}_{\text{B}y}^{\text{A}x}\left( \vec{k}=0 \right)  \\
       0  & 0  & \overline{G}_{\text{B}x}^{\text{A}y}\left( \vec{k}=0 \right)    & 0 \\
        0 & - \overline{G}^{\text{A}y}_{\text{B}x} \left( \vec{k}=0\right) & 0    & 0  \\
        -\overline{G}^{\text{A}x}_{\text{B}y} \left( \vec{k}=0 \right) & 0 & 0 & 0
    \end{pmatrix}. 
    \label{AP: G-matrix checkerboard elements at Gamma, third step}
\end{equation}

\item $\overline{G}_{\text{A}y}^{\text{B}x}\left( \vec{k}=0 \right)$ and $\overline{G}_{\text{A}x}^{\text{B}y}\left( \vec{k}=0 \right)$:

We note that these two elements are made equal via either $\mathcal{C}_{4z,+}\mathcal{T}$ 
or $\mathcal{M}_{xy}$ since both symmetries would swap the $x$ and $y$ cartesian coordinates, without causing a sign difference. The matrix thus assumes the form
  \begin{equation}
    \overline{G}_{\vec{k}=0} = g
    \begin{pmatrix}
        0  & 0& 0 & 1  \\
       0  & 0  & 1   & 0 \\
        0 & - 1 & 0    & 0  \\
        -1 & 0 & 0 & 0
    \end{pmatrix},
    \label{AP: G-matrix checkerboard elements at Gamma, fourth step}
\end{equation}
where we have defined $g=\overline{G}_{\text{B}y}^{\text{A}x}\left( \vec{k}=0 \right).$
\end{enumerate}
The last, finite element $g$ cannot be shown based on symmetries to be zero. Nonetheless, the reduced form of $G_{\vec{k}=0}$ in Eq.~\eqref{AP: G-matrix checkerboard elements at Gamma, fourth step} cannot produce a finite splitting of the optical modes even for finite $g$. To see so, we diagonalize  $H^{\text{eff}}_{\vec{k}}$ in Eq.~\eqref{eq: effective H_ph} at $\vec{k} = 0$ and obtain the phonon frequencies
\begin{equation}
    \begin{aligned}
        & \omega_{\vec{k}=0}^1 = \omega_{\vec{k}=0}^2 = 0, \\
        & \omega_{\vec{k}=0}^3 = \omega_{\vec{k}=0}^4 = \sqrt{\omega_0^2 + 4g^2},
    \end{aligned}
    \label{APP:eq: MBC-corrected modes at Gamma}
\end{equation}
with $\omega_0 = \sqrt{\frac{6|n_{11}|}{M_{\text{B}}}}$ being the unperturbed frequency of the optical phonons at $\vec{k}=0$. 
Thus, the MBC can only shift the optical modes higher in energy but cannot split them. 

To gain more intuition as to why this is happening, we employ degenerate perturbation theory as was done in Ref.~\cite{Chiral_Phonons_Angular_Momentum} to compute the linear correction $\delta \omega_{\vec{k}=0}$ to the optical energies at $\Gamma$ caused by the MBC. To do so, we introduce the basis that diagonalizes the unperturbed dynamical matrix at the $\Gamma$ point
\begin{equation}
\begin{aligned}
    u_{+}^\text{A} & = 
    \begin{pmatrix}
        1 & 0 & 1 & 0 
    \end{pmatrix}^{\text{T}} / \sqrt{2}, 
    \\
 u_{+}^\text{B} & = 
    \begin{pmatrix}
        0 & 1 & 0 & 1 
    \end{pmatrix}^{\text{T}} / \sqrt{2},
 \\
u_{-}^\text{A} & = 
    \begin{pmatrix}
        1 & 0 & -1 & 0 
    \end{pmatrix}^{\text{T}} / \sqrt{2}, \\
u_{-}^\text{B} & = 
    \begin{pmatrix}
        0 & 1 & 0 & -1 
    \end{pmatrix}^{\text{T}} / \sqrt{2},
 \end{aligned}
 \label{AP:eq: symmetric/antisymmetric basis}
\end{equation}
\end{section}
where the states labeled with ``$+$'' correspond to the eigenstates of the acoustic sector while the ones with ``$-$'' correspond to the eigenstates of the optical sector. By building the unitary matrix 
\begin{equation}
    U = 
    \begin{pmatrix}
        u_-^{\text{A}} & u_-^{\text{B}} & u_+^{\text{A}} & u_+^{\text{B}} 
    \end{pmatrix},
\end{equation}
we find that 
\begin{equation}
    U^{\dagger} \overline{G}_{\vec{k}=0}U = \overline{G}_{\vec{k}=0}. 
\label{APP:eq:MBC at Gamma in new basis}
\end{equation}
Thus, $\overline{G}_{\vec{k}=0}$ does not couple the two degenerate optical modes in the unperturbed eigenbasis in Eq.~\eqref{AP:eq: symmetric/antisymmetric basis} (note that the first diagonal block of the matrix is zero). Consequently, $\delta \omega_{\vec{k}=0}$ is zero, and the optical modes stay degenerate. 

We note in passing that according to Fig.~\ref{fig:MBC numerics}, $g=\overline{G}_{\text{B}y}^{\text{A}x}\left( \vec{k}=0 \right)$ is zero. However, this result is a consequence of the simplicity of the electronic model considered here, rather than of symmetry. As shown above, even a finite value of $\overline{G}_{\text{B}y}^{\text{A}x}\left( \vec{k}=0 \right)$ cannot induce a finite phonon splitting $\delta \omega_{\vec{k}=0}$. The absence of optical phonon splitting at the $\Gamma$ point is therefore dictated solely by the magnetic point group of the electronic system and is not an artifact of the simplified model.

\begin{section}{Phonon Dynamics with the Molecular Berry Curvature}
\label{AP:Phonon Dynamics with the Molecular Berry Curvature}
We present a detailed derivation of the phonon eigenstates and frequencies associated with a phonon Hamiltonian that incorporates the effects of electron-phonon coupling via the molecular Berry curvature.

We assume a two-dimensional system with a total of $n$ sublattices; for the checkerboard lattice in the main text one sets $n=2$. An extension to three dimensions is straightforward.
First, we recall that effects of the molecular Berry curvature can be incorporated in the phonon Hamiltonian by a minimal coupling, \cite{PhysRevX.15.011036, Chiral_Phonons_Angular_Momentum, MBP_Phonon_THE_Nagaosa, li2025phonondichroismsrevealingunusual, PhysRevLett.119.075301, PhysRevLett.134.206701, das2026antiferrochiralphononsmathcalpmathcaltsymmetricantiferromagnets,  dhakal2025theoryintrinsicphononthermal}
\begin{equation}
    \frac{1}{2}\sum_{l \kappa \alpha}\frac{1}{ M_\kappa} p^2_{l \kappa \alpha} \rightarrow 
      \frac{1}{2}\sum_{l \kappa \alpha}\frac{1}{ M_\kappa} \left(p_{l \kappa \alpha}- \hbar A_{l \kappa \alpha}\right)^2,   
\end{equation}
where the molecular Berry connection can be written using an anti-symmetric gauge in terms of the MBC  \cite{bfll-sdrb, MBP_Phonon_THE_Nagaosa, Chiral_Phonons_Angular_Momentum} as
\begin{equation}
    A_{l \kappa \alpha} = - 
    \frac{1}{2}\sum_{\beta, \kappa, l'}G_{\kappa'\beta}^{\kappa \alpha}\left( \vec{R}_l^0 - \vec{R}_{l'}^0\right) u_{l' \kappa'\beta}, 
    \label{AP:eq: gauge for molecular connection}
\end{equation}
where $\alpha, \beta =x,y$ are the Cartesian coordinates and $\kappa, \kappa'$ are sublattice indices.  
The individual terms of the phonon Hamiltonian can be grouped into the following compact form in momentum space \cite{Chiral_Phonons_Angular_Momentum}
\begin{equation}
    H_{\text{ph}} = \frac{1}{2}\sum_{\vec{k}}
    \begin{pmatrix}
        \vec{p}_{\vec{k}} \\
       \vec{u}_{\vec{k}} 
    \end{pmatrix}^{\dagger}
    \begin{pmatrix}
        I_{2n} &  \overline{G}_{\vec{k}} \\
        \overline{G}^{\dagger}_{\vec{k}} & \overline{D}_{\vec{k}} + \overline{G}^{\dagger}_{\vec{k}} \overline{G}_{\vec{k}}
    \end{pmatrix}
    \begin{pmatrix}
        \vec{p}_{\vec{k}} \\
       \vec{u}_{\vec{k}}
    \end{pmatrix}.
    \label{AP:eq: phonon Hamiltonian with MBC}
\end{equation}
Here, $I_{2n}$ is a unitary matrix of order $2n$, $\overline{G}_{\vec{k}}$ is the MBC-matrix with elements given by $\overline{G}_{\kappa'\beta}^{\kappa \alpha}\left( \vec{k} \right) = \frac{\hbar}{2}\frac{1}{\sqrt{M_{\kappa}M_{\kappa'}}}G_{\kappa'\beta}^{\kappa \alpha}\left( \vec{k} \right)$, and  $\overline{D}_{\vec{k}}$ is the phonon dynamical matrix with elements 
\begin{equation}
    \overline{D}_{\kappa' \beta}^{\kappa \alpha}\left(\vec{k} \right) = \frac{1}{N} \frac{1}{\sqrt{M_\kappa M_{\kappa'}}}\sum_{l l'} \mathrm{e}^{-\mathrm{i} \vec{k} \cdot \left(\vec{R}_l^0  - \vec{R}_{l'}^0 \right)} \Phi_{\kappa'\beta}^{\kappa \alpha}\left(\vec{R}_l^0  - \vec{R}_{l'}^0\right), 
\end{equation}
where $\Phi_{\kappa'\beta}^{\kappa \alpha}\left(\vec{R}_l^0 - \vec{R}_{l'}^0\right)$ is the element of the force constant matrix corresponding to the displacement of the $\kappa$-sublattice site in the $l$-th unit cell along the $\alpha$-direction, and the $\kappa'$-sublattice site in the $l'$-th unit cell along the $\beta$-direction. The $\vec{p}_{\vec{k}}$ and $\vec{u}_{\vec{k}}$ vectors are given by  
\begin{equation}
\begin{aligned}
        \vec{p}_{\vec{k}} &=
        \begin{pmatrix}
            p_{\vec{k} \kappa_1 x} / \sqrt{M_{\kappa_1}} &  p_{\vec{k} \kappa_1 y}/ \sqrt{M_{\kappa_1}}
            \dots
            & p_{\vec{k} \kappa_n x}/ \sqrt{M_{\kappa_n}} 
            & 
            p_{\vec{k} \kappa_n y}/ \sqrt{M_{\kappa_n}} 
        \end{pmatrix}, \\
        \vec{u}_{\vec{k}} &=
        \begin{pmatrix}
           u_{\vec{k} \kappa_1 x }  \sqrt{M_{\kappa_1}} &  u_{\vec{k} \kappa_1 y} \sqrt{M_{\kappa_1}}
            \dots 
            & u_{\vec{k} \kappa_n x} \sqrt{M_{\kappa_1}} 
            & 
            u_{\vec{k} \kappa_n y} \sqrt{M_{\kappa_n}} 
        \end{pmatrix}.
\end{aligned}
\label{AP:eq: u-p vectors}
\end{equation}

We upgrade the momenta and displacements in Eq.~\eqref{AP:eq: u-p vectors} to operators that obey canonical commutation relations in real space 
\begin{equation}
\left[u_{l\alpha \kappa},  p_{l'\beta \kappa'}\right] = \mathrm{i} \hbar  \delta_{\beta, \alpha} \delta_{l, l'} \delta_{\kappa, \kappa'}.
\label{AP:eq: real space commutation relations for u-p}
\end{equation} 
Using Eq.~\eqref{AP:eq: real space commutation relations for u-p}, we obtain the corresponding commutation relations in momentum space as 
\begin{equation}
\begin{aligned}
    \left[ u_{\vec{k}\kappa \alpha }, p_{\vec{k}'\kappa' \beta }  \right] &= \frac{1}{N} \sum_{l,l'}\left[u_{l \kappa \alpha} , p_{l \kappa' \beta }   \right] \mathrm{e}^{-\mathrm{i} \vec{R}_{l \kappa}^0  \cdot \vec{k}}  \mathrm{e}^{-\mathrm{i} \vec{R}_{l'\kappa'}^0 \cdot \vec{k}'} \\
    &= \frac{1}{N} \sum_{l,l'} \mathrm{i}\hbar \delta_{l, l'} \delta_{\beta, \alpha} \delta_{\kappa, \kappa'}\mathrm{e}^{-\mathrm{i} \vec{R}_{l \kappa}^0  \cdot \vec{k}}  \mathrm{e}^{-\mathrm{i} \vec{R}_{l' \kappa'}^0 \cdot \vec{k}'}  \\
    &= 
    \mathrm{i}\hbar \frac{1}{N} \mathrm{e}^{-\mathrm{i}\vec{\tau}_{\kappa} \cdot \left(\vec{k} +\vec{k}' \right)} \sum_{l} \delta_{\beta, \alpha}  \delta_{\kappa, \kappa'}\mathrm{e}^{-\mathrm{i} \vec{R}_l^0  \cdot\left( \vec{k} + \vec{k}'\right)} \\
    &= 
    \mathrm{i}\hbar \delta_{\vec{k}, -\vec{k}'} \delta_{\beta, \alpha}  \delta_{\kappa, \kappa'}.
\end{aligned}
\label{AP:eq: k-space commutation relations for u-p}
\end{equation}
In terms of the vectors in eq.~\eqref{AP:eq: u-p vectors}, the commutation relations are expressed as 
\begin{equation}
\begin{aligned}
    \left[\vec{u}_{\vec{k}}, \vec{p}_{\vec{k}'} \right] 
    &\equiv 
    \begin{pmatrix} \left[u_{\vec{k}\kappa_1x}, p_{\vec{k}'\kappa_1 x} \right] & \left[u_{\vec{k}\kappa_1 x}, p_{\vec{k}'\kappa_1 y} \right] &\dots &
\left[u_{\vec{k}\kappa_1x}, p_{\vec{k}'\kappa_n y} \right]
    \\
\left[u_{\vec{k}\kappa_1 y}, p_{\vec{k}'\kappa_1 x} \right] & \left[u_{\vec{k}\kappa_1 y}, p_{\vec{k}'\kappa_1 y} \right]
    & \cdots & \left[u_{\vec{k}\kappa_1 y}, p_{\vec{k}'\kappa_n y} \right]  \\
     \vdots & \vdots & \ddots & 
     \vdots \\
     \left[u_{\vec{k}\kappa_n y}, p_{\vec{k}'\kappa_1 x} \right] & \left[u_{\vec{k}\kappa_n y}, p_{\vec{k}'\kappa_1 y} \right] & \dots & \left[u_{\vec{k}\kappa_n y}, p_{\vec{k}'\kappa_n y} \right]
    \end{pmatrix} \\
    &= \mathrm{i} \hbar \delta_{\vec{k}, -\vec{k}'} I_{2n}.
\end{aligned}
\label{AP:eq: k-space commutation relatiions for vectors}
\end{equation}
A set of equations of motion for the displacement and momentum operators is obtained from Heisenberg's equation of time evolution, 
\begin{equation}
\begin{aligned}
&
    \dot{u}_{\vec{k} \kappa \alpha} = \frac{\mathrm{i}}{\hbar} \left[H_{\text{ph}}, u_{\vec{k} \kappa \alpha}\right], \\
& 
\dot{p}_{\vec{k} \kappa \alpha} = \frac{\mathrm{i}}{\hbar} \left[H_{\text{ph}}, p_{\vec{k} \kappa \alpha} \right]. 
\end{aligned}
\end{equation}
To better organize the algebra that follows we split $H_{\text{ph}}$ into three parts 
\begin{equation}
    H_{\text{ph}} = H_{\text{kin}} + H_{\text{el}} + H_{\text{RM}}, 
\end{equation}
with
\begin{equation}
\begin{aligned}
    H_{\text{kin}} &= 
    \frac{1}{2} \sum_{\vec{k}} \vec{p}_{\vec{k}}^{\dagger} \vec{p}_{\vec{k}}, \\
    H_{\text{el}} &= 
    \frac{1}{2} \sum_{\vec{k}} \vec{u}_{\vec{k}}^{\dagger} \overline{D}_{\vec{k}}\vec{u}_{\vec{k}}, \\
    H_{\text{RM}} & = \frac{1}{2} \sum_{\vec{k}} 
    \left(
    \vec{u}_{\vec{k}}^{\dagger} \overline{G}_{\vec{k}}^{\dagger} \vec{p}_{\vec{k}} + \vec{p}_{\vec{k}}^{\dagger} \overline{G}_{\vec{k}} \vec{u}_{\vec{k}}
    \right), 
\end{aligned}
\end{equation}
being the kinetic part of the Hamiltonian, the elastic potential, and the ``Raman-like'' terms coupling momenta and displacements stemming from the MBC,  respectively.  We note here already that $\left[H_{\text{el}}, u_{\vec{k} \kappa \alpha} \right] = 0$ and $\left[H_{\text{kin}}, p_{\vec{k} \kappa \alpha} \right] = 0$ since $H_{{\text{el}}}$  and $H_{{\text{kin}}}$ respectively contain only displacement and momentum operators. We find for the displacements that 
\begin{equation}
\begin{aligned}
\bigl[H_{\text{kin}}, u_{\vec{k} \kappa \alpha} \bigr] &= 
\frac{1}{2} \sum_{\vec{q}} 
\left[
\vec{p}^{\dagger}_{\vec{q}} \vec{p}_{\vec{q}}, u_{\vec{k} \kappa \alpha}
\right] 
\\
& =   \frac{1}{2} \sum_{\beta, \kappa'} \sum_{\vec{q}} \frac{1}{M_\beta}
\left[
p^{\dagger}_{\vec{q} \kappa' \beta} p_{\vec{q} \kappa'\beta}, u_{\vec{k} \kappa \alpha} 
\right] \\
& = 
\frac{1}{2} \sum_{\beta, \kappa'} \sum_{\vec{q}} \frac{1}{M_\beta}
\left(
\left[
p^{\dagger}_{\vec{q} \kappa' \beta} , u_{\vec{k} \kappa \alpha} 
\bigr] p_{\vec{q} \kappa'\beta} + 
p^{\dagger}_{\vec{q} \kappa' \beta} \bigl[
p_{\vec{q} \kappa'\beta}, u_{\vec{k} \kappa \alpha} 
\right]
\right) \\
& 
= 
\frac{1}{2} \sum_{\beta, \kappa'} \sum_{\vec{q}} \frac{1}{M_\beta}
\left(
\left[
p^{\dagger}_{\vec{q} \kappa' \beta} , u_{\vec{k} \kappa \alpha} 
\bigr] p_{\vec{q} \kappa'\beta} + 
p^{\dagger}_{\vec{q} \kappa' \beta} \bigl[
p_{\vec{q} \kappa'\beta}, u_{\vec{k} \kappa \alpha} 
\right]
\right) \\
& = -
\frac{\mathrm{i} \hbar}{2} \sum_{\beta, \kappa'} \sum_{\vec{q}} \frac{1}{M_\beta}
\left(
 p_{\vec{q} \kappa'\beta} \delta_{\vec{q}, \vec{k}}+ 
p_{-\vec{q} \kappa' \beta}  \delta_{-\vec{q}, \vec{k}}
\right) \delta_{\beta, \alpha} \delta_{\kappa, \kappa'} \\
& = - 
\frac{\mathrm{i} \hbar}{M_{\kappa}} p_{\vec{k} \kappa \alpha}, 
\end{aligned}
\end{equation}
and 
\begin{equation}
\begin{aligned}
    \left[H_{\text{RM}}, u_{\vec{k} \kappa \alpha} \right] &
    = 
    \frac{1}{2} \sum_{\vec{q}} 
     \left(
    \left[
    \vec{u}_{\vec{q}}^{\dagger} \overline{G}_{\vec{q}}^{\dagger} \vec{p}_{\vec{q}}, u_{\vec{k}\kappa \alpha} \right]
    + 
    \left[
    \vec{p}_{\vec{q}}^{\dagger} \overline{G}_{\vec{q}} \vec{u}_{\vec{q}}, u_{\vec{k}\kappa \alpha}
    \right]
    \right) \\
    & =
    \frac{1}{2} \sum_{\vec{q}} \sum_{\beta, \gamma} \sum_{ \kappa'', \kappa'} 
    \left( 
    \left[
    u_{-\vec{q} \kappa' \beta} \left[ \overline{G}_{\vec{q}}^{\dagger}\right] ^{\kappa' \beta}_{\kappa'' \gamma}
    p_{\vec{q} \kappa'' \gamma}, u_{\vec{k}\kappa \alpha}
    \right] \sqrt{\frac{M_{\kappa'}}{M_{\kappa''}}}
     + 
     \left[
    p_{-\vec{q} \kappa' \beta} \left[ \overline{G}_{\vec{q}}\right] ^{\kappa' \beta}_{\kappa'' \gamma}
    u_{\vec{q} \kappa'' \gamma}, u_{\vec{k}\kappa \alpha}
    \right] \sqrt{\frac{M_{\kappa''}}{M_{\kappa'}}}
    \right) \\
    & = -\frac{\mathrm{i}\hbar}{2} \sum_{\vec{q}} \sum_{\beta, \gamma} \sum_{ \kappa'', \kappa'} 
    \left( 
    \sqrt{\frac{M_{\kappa'}}{M_{\kappa''}}} 
    \left[
    \overline{G}_{\vec{q}}^{\dagger}
    \right] 
    ^{\kappa' \beta}_{\kappa'' \gamma} u_{-\vec{q} \kappa' \beta} \delta_{-\vec{q}, \vec{k}} \delta_{\kappa, \kappa''} \delta_{\gamma, \alpha}
    + 
    \sqrt{\frac{M_{\kappa''}}{M_{\kappa'}}} u_{\vec{q} \kappa'' \gamma}
    \delta_{\vec{q}, \vec{k}} \delta_{\kappa, \kappa'}
    \delta_{\beta, \alpha}
    \left[ 
    \overline{G}_{\vec{q}}
    \right] ^{\kappa' \beta}_{\kappa'' \gamma}
    \right) \\
    & 
    = -\frac{\mathrm{i}\hbar}{2} \sum_{\beta} \sum_{\kappa'} \sqrt{\frac{M_{\kappa'}}{M_{\kappa}}}\left(  
    \left[
    \overline{G}_{-\vec{k}}^{\dagger}
    \right] 
    ^{\kappa' \beta}_{\kappa \alpha} 
    + 
    \left[ 
    \overline{G}_{\vec{k}}
    \right] ^{\kappa \alpha}_{\kappa' \beta}
    \right) u_{\vec{k} \kappa' \beta} \\
    & = 
    -\mathrm{i}\hbar \sum_{\beta} \sum_{\kappa'} \sqrt{\frac{M_{\kappa'}}{M_{\kappa}}}  
    \left[ 
    \overline{G}_{\vec{k}}
    \right] ^{\kappa \alpha}_{\kappa' \beta}
    u_{\vec{k} \kappa' \beta}
    \\
    & =
    -\frac{\mathrm{i}\hbar}{M_{\kappa}} \frac{\hbar}{2} \sum_{\beta} \sum_{\kappa'}  
    G
     ^{\kappa \alpha}_{\kappa' \beta} \left(\vec{k} \right)
    u_{\vec{k} \kappa' \beta}. 
\end{aligned}
\end{equation}
In the above, we kept the brackets around the $\overline{G}_{\vec{k}}$ and $\overline{G}_{\vec{k}}^{\dagger}$ to make it clear that we are taking a specific element of either the MBC-matrix or of its hermitian conjugate and simplified the notation to $\left[\overline{G}_{\vec{k}}
\right] ^{\kappa \alpha}_{\kappa' \beta} = \overline{G}
     ^{\kappa \alpha}_{\kappa' \beta} \left(\vec{k} \right)$ when there was no danger of confusion. Additionally, we made use  of the anti-Hermitian nature of the MBC-matrix [recall eq.~\eqref{AP:eq: anti-hermiticity}] to get $\left[
    \overline{G}_{-\vec{k}}^{\dagger}
    \right] 
    ^{\kappa' \beta}_{\kappa \alpha} = \left[ -\overline{G}_{-\vec{k}}
    \right] 
    ^{\kappa' \beta}_{\kappa \alpha} = -\left(
    \left[\overline{G}_{\vec{k}}
    \right] 
    ^{\kappa' \beta}_{\kappa \alpha}
    \right)^* =  \left[ \overline{G}_{\vec{k}}
    \right] 
    ^{\kappa \alpha}_{\kappa' \beta}
    $, 
 as well as the fact that $u_{\vec{k} \kappa \alpha}^{\dagger} = u_{-\vec{k} \kappa \alpha}$ and $p_{\vec{k} \kappa \alpha}^{\dagger} = p_{-\vec{k} \kappa \alpha}$. 
 For the momenta, a similar calculation can be carried out, such that altogether we obtain 
 \begin{equation}
     \begin{aligned}
         & \dot{u}_{\vec{k} \kappa \alpha} = \frac{1}{M_\kappa}
         \left( 
         p_{\vec{k} \kappa \alpha} + 
         \frac{\hbar}{2}\sum_{\beta} \sum_{\kappa'}  
    G^{\kappa \alpha}_{\kappa' \beta} \left(\vec{k} \right)
    u_{\vec{k} \kappa' \beta}
         \right), \\
          & \dot{p}_{\vec{k} \kappa \alpha} =  \sum_{\beta} \sum_{\kappa'}
         \left(\left(-D^{\kappa \alpha}_{\kappa' \beta} \left(\vec{k} \right) + \frac{\hbar^2}{4}\sum_{\kappa'',\gamma}G^{\kappa \alpha}_{\kappa'' \gamma}\left(\vec{k} \right) G^{\kappa'' \gamma}_{\kappa' \beta}\left(\vec{k} \right) \right)
    u_{\vec{k} \kappa' \beta}
          +   
          \frac{1}{M_{\kappa'}} \frac{\hbar}{2}
    G^{\kappa \alpha}_{\kappa' \beta} \left(\vec{k} \right)
    p_{\vec{k} \kappa' \beta}
         \right), 
     \end{aligned}
 \end{equation}
 which we combine in matrix notation:
 \begin{equation}
     \begin{pmatrix}
        \dot{\vec{p}}_{\vec{k}} \\ \dot{\vec{u}}_{\vec{k}} 
     \end{pmatrix} = 
     \begin{pmatrix}
         \overline{G}_{\vec{k}} &-\overline{D}_{\vec{k}} - \overline{G}^{\dagger}\left(\vec{k}\right)\overline{G}\left(\vec{k}\right) \\
         I_{2n} & \overline{G}_{\vec{k}}
     \end{pmatrix}
     \begin{pmatrix}
         p_{\vec{k}} \\
         u_{\vec{k}}
     \end{pmatrix}. 
    \label{AP:eq: EOMS compact}
 \end{equation}
We proceed by assuming plain-wave solutions for the time-dependence $\vec{p}_{\vec{k}}\left(t\right)= \vec{\mu}_{\vec{k} \sigma} \mathrm{e}^{-\mathrm{i} \omega_{\vec{k} \sigma}t}$ and $\vec{u}_{\vec{k}}\left(t\right)= \vec{\epsilon}_{\vec{k} \sigma} \mathrm{e}^{-\mathrm{i} \omega_{\vec{k} \sigma}t}$ with $\vec{\mu}_{\vec{k} \sigma}$ being the momentum polarization vector, $\vec{\epsilon}_{\vec{k} \sigma}$ the displacement polarization vector, and $\sigma$ the band index that can take both positive and negative values corresponding to positive (physically meaningful) phonon frequencies and negative ones, respectively. Defining $\vec{\chi}_{\vec{k} \sigma}^{\text{R}} = \begin{pmatrix}
     \vec{\mu}_{\vec{k} \sigma} & \vec{\epsilon}_{\vec{k} \sigma}
 \end{pmatrix}^T$
 we arrive at the eigenvalue problem \cite{THE_Lifa_Zhang}
\begin{equation}
 \omega_{\vec{k} \sigma} \vec{\chi}_{\vec{k} \sigma}^{\text{R}}  = \mathrm{i}
\begin{pmatrix}
    \overline{G}_{\vec{k}} &-\overline{D}_{\vec{k}} - \overline{G}^{\dagger}\left(\vec{k}\right)\overline{G}\left(\vec{k}\right)\\
    I_{2n} & \overline{G}_{\vec{k}}
\end{pmatrix}
     \vec{\chi}_{\vec{k} \sigma}^{\text{R}} 
     = H_{\text{eff}} \hspace{0.1cm} \vec{\chi}_{\vec{k} \sigma}^{\text{R}},
 \end{equation}
 with the effective Hamilton matrix
 \begin{equation}
     H_{\text{eff}} = \mathrm{i}
     \begin{pmatrix}
         \overline{G}_{\vec{k}} &-\overline{D}_{\vec{k}} - \overline{G}^{\dagger}\left(\vec{k}\right)\overline{G}\left(\vec{k}\right)\\
    I_{2n} & \overline{G}_{\vec{k}}
     \end{pmatrix}.
     \label{eq:effective_ham_step}
 \end{equation} 
  To obtain the phonon eigenstates and frequencies one must diagonalize $H_{\text{eff}}$ by considering both the right and left eigenvectors since $H_{\text{eff}}$ is not Hermitian. The formalism for this task has already been developed in \cite{THE_Lifa_Zhang}, which we follow from here on. The left eigenvectors are defined through the eigenvalue equation 
 \begin{equation}
     \vec{\chi}^{\text{L}}_{\vec{k} \sigma} H_{\text{eff}} = \vec{\chi}^{\text{L}}_{\vec{k} \sigma} \omega_{\vec{k} \sigma}. 
 \end{equation}
 Since $H_{\text{eff}}^{\dagger}\left(\vec{\chi}_{\vec{k} \sigma}^{\text{L}}\right)^{\dagger} = \left(\vec{\chi}_{\vec{k} \sigma}^{\text{L}}\right)^{\dagger} \omega_{\vec{k} \sigma}$ and $H_{\text{eff}} = U^{\dagger} H^{\dagger}_{\text{eff}} U$, with the unitary matrix 
 $U= \begin{pmatrix}
     0 & I_{2n} \\
     -I_{2n} & 0 
 \end{pmatrix}$, 
 one can establish a relation between the left and right eigenvectors and construct the right eigenvector using the components of the left eigenvector. 
Here, we defined the left eigenvector as
\begin{equation}
    \vec{\chi}_{\vec{k} \sigma}^{\text{L}} = \begin{pmatrix}
        \vec{\epsilon}_{\vec{k}\sigma}^* & - \vec{\mu}_{\vec{k}\sigma}^* 
    \end{pmatrix} / \left(-2 \mathrm{i} \omega_{\vec{k}} \right),
\end{equation}
where the factor of $-2 \mathrm{i} \omega_{\vec{k} \sigma}$ in the denominator will later ensure that we can smoothly proceed with the second quantization of $\vec{u}_{\vec{k}}$ and $\vec{p}_{\vec{k}}$.
Additionally,  orthonormality and completeness hold between the left and right eigenvectors, which are explicitly given by 
\begin{equation}
    \begin{aligned}
        & \vec{\chi}_{\vec{k}\sigma}^{\text{L}} \cdot \vec{\chi}_{\vec{k} \sigma'}^{\text{R}} = \delta_{\sigma, \sigma'}, \\
        & 
        \sum_{\sigma}  \vec{\chi}_{\vec{k} \sigma}^{\text{R}} \otimes  \vec{\chi}_{\vec{k} \sigma}^{\text{L}} = I_{4n}. 
    \end{aligned}
\label{eq: R-L complete and orthogonal basis}
\end{equation}
We proceed by normalizing the left and right eigenvectors by 
\begin{equation}
    \sqrt{\vec{\chi}_{\vec{k} \sigma}^{\text{L}} \cdot \vec{\chi}_{\vec{k} \sigma}^{\text{R}}} = \sqrt{\vec{\epsilon}_{\vec{k} \sigma}^\dagger \vec{\epsilon}_{\vec{k} \sigma} - \mathrm{i} \frac{1}{\omega_{\vec{k} \sigma}} \vec{\epsilon}_{\vec{k} \sigma}^{\dagger} \overline{G}_{\vec{k}} \vec{\epsilon}_{\vec{k} \sigma}}. 
\label{eq: left-right norm}
\end{equation} 
We denote the normalized polarization vectors by 
\begin{equation}
\begin{aligned}
    \tilde{\vec{\epsilon}}_{\vec{k} \sigma} 
    & = \frac{\vec{\epsilon_{\vec{k} \sigma}}}{\sqrt{\vec{\chi}_{\vec{k} \sigma}^{\text{L}} \cdot \vec{\chi}_{\vec{k} \sigma}^{\text{R}}}}, 
    \\
    \tilde{\vec{\mu}}_{\vec{k} \sigma}
    & = 
    \frac{\vec{\mu_{\vec{k} \sigma}}}  {\sqrt{\vec{\chi}_{\vec{k} \sigma}^{\text{L}} \cdot \vec{\chi}_{\vec{k} \sigma}^{\text{R}}}}, 
\end{aligned}
\end{equation}
and the normalized left and right eigenvectors by
\begin{equation}
\begin{aligned}
    \tilde{\vec{\chi}}_{\vec{k} \sigma}^{\text{R}} 
    & = \frac{\vec{\chi}_{\vec{k} \sigma}^{\text{R}}}{\sqrt{\vec{\chi}_{\vec{k} \sigma}^{\text{L}} \cdot \vec{\chi}_{\vec{k} \sigma}^{\text{R}}}}, 
    \\
    \tilde{\vec{\chi}}_{\vec{k} \sigma}^{\text{L}} 
    & = \frac{\vec{\chi}_{\vec{k} \sigma}^{\text{L}}}{\sqrt{\vec{\chi}_{\vec{k} \sigma}^{\text{L}} \cdot \vec{\chi}_{\vec{k} \sigma}^{\text{R}}}}. 
\end{aligned}
\end{equation}
Since the norm in Eq.~\eqref{eq: left-right norm} has units of $\left[ \vec{\epsilon}_{\vec{k} \sigma}\right] = \left[\text{length} \right] \left[\sqrt{\text{mass}} \right]$ the  normalized displacement polarization vector $\tilde{\vec{\epsilon}}_{\vec{k} \sigma}$ is unitless, while the normalized momentum polarization vector $\tilde{\vec{\mu}}_{\vec{k} \sigma}$ has units of $\left[\text{momentum} \cdot \text{length}^{-1} \cdot \text{mass}^{-1}\right]$. 
We also require the following relations between the positive and negative bands for the polarization vectors and the frequencies:
\begin{equation}
\begin{aligned}
     & \omega_{\vec{k}\sigma} = -  \omega_{-\vec{k}-\sigma}, \\
     & \tilde{\vec{\epsilon}}_{\vec{k}\sigma}^* = \tilde{\vec{\epsilon}}_{-\vec{k}-\sigma}, \\
     & 
      \tilde{\vec{\mu}}_{\vec{k}\sigma}^* = \tilde{\vec{\mu}}_{-\vec{k}-\sigma}.
\end{aligned}
\end{equation}
We now switch to second quantization and introduce creation and annihilation operators that have the property 
\begin{equation}
    a_{-\vec{k} -\sigma} = a_{\vec{k} \sigma}^{\dagger},
\end{equation}
and the commutation relation 
\begin{equation}
    \left[a_{\vec{k} \sigma}, a_{\vec{k}' \sigma'}^{\dagger} \right] = \delta_{\vec{k}, \vec{k}'} \delta_{\sigma, \sigma'} \text{sign}\left( \sigma \right).
\label{eq: bosonic commutator}
\end{equation}
The first-quantized operators are related to their second-quantized counterparts by
\begin{equation}
\begin{aligned}
     & 
    \vec{u}_{\vec{k}} = \sqrt{\frac{\hbar}{2}}\sum_{\sigma} \tilde{\vec{\epsilon}}_{\vec{k}\sigma} \sqrt{\frac{1}{\left|\omega_{\vec{k} \sigma} \right|}} a_{\vec{k} \sigma}, \\
    & 
    \vec{p}_{\vec{k}} = \sqrt{\frac{\hbar}{2}}\sum_{\sigma} \tilde{\vec{\mu}}_{\vec{k}\sigma} \sqrt{\frac{1}{\left|\omega_{\vec{k} {\sigma}} \right|}}a_{\vec{k} \sigma}.
\end{aligned}
\label{eq: second quantization}
\end{equation}
One can readily see that the commutation relations in Eq.~\eqref{AP:eq: k-space commutation relations for u-p} are still intact 
\begin{equation}
\begin{aligned}
    \left[ \vec{u}_{\vec{k}}, \vec{p}^{\dagger}_{\vec{k}'}\right] 
    &= 
    \frac{\hbar}{2} \sum_{\sigma \sigma'} \frac{1}{\sqrt{|\omega_{\vec{k} \sigma}|}} \frac{1}{\sqrt{|\omega_{\vec{k}' \sigma'}|}} \tilde{\vec{\epsilon}}_{\vec{k} \sigma} \otimes \tilde{\vec{\mu}}_{\vec{k}' \sigma'}^{\dagger}\left[a_{\vec{k} \sigma}, a_{\vec{k}' \sigma'}^{\dagger}\right] \\
    & = 
    \frac{\hbar}{2} \sum_{\sigma} \frac{1}{|\omega_{\vec{k} \sigma}|} \tilde{\vec{\epsilon}}_{\vec{k} \sigma} \otimes \tilde{\vec{\mu}}_{\vec{k} \sigma}^{\dagger}\delta_{\vec{k} \vec{k}'} \text{sign}\left(\sigma\right) \\
    & =\mathrm{i} \hbar \delta_{\vec{k},  \vec{k}'} I_{2n}. 
\end{aligned}
\end{equation}
With the main formalism laid out, we can tweak the original Hamiltonian in Eq.~\eqref{eq: phonon Hamiltonian with MBC} as follows
\begin{equation}
\begin{aligned}
      H_{\text{ph}} &= \frac{1}{2}\sum_{\vec{k}}  \begin{pmatrix}
          \vec{p}_{\vec{k}} \\
          \vec{u}_{\vec{k}}
      \end{pmatrix}^{\dagger}
    \begin{pmatrix}
        I & G_{\vec{k}} \\
       G_{\vec{k}}^{\dagger}  &  -G_{\vec{k}}^2 + \overline{D}_{\vec{k}}
    \end{pmatrix}
   \begin{pmatrix}
          \vec{p}_{\vec{k}} \\
          \vec{u}_{\vec{k}}
      \end{pmatrix} \\
      &= 
       \frac{1}{2}\sum_{\vec{k}}  \begin{pmatrix}
          \vec{u}_{\vec{k}} \\
          -\vec{p}_{\vec{k}}
      \end{pmatrix}^{\dagger}
    \begin{pmatrix}
      -G_{\vec{k}} & +G_{\vec{k}}^2 - \overline{D}_{\vec{k}} \\
        -I  &  -G_{\vec{k}}
    \end{pmatrix}
   \begin{pmatrix}
          \vec{p}_{\vec{k}} \\
          \vec{u}_{\vec{k}}
      \end{pmatrix}  \\
      &=
      \frac{\mathrm{i}}{2}\sum_{\vec{k}}  \begin{pmatrix}
          \vec{u}_{\vec{k}} \\
          -\vec{p}_{\vec{k}}
      \end{pmatrix}^{\dagger}
    H_{\vec{k}}^{\text{eff}}
   \begin{pmatrix}
          \vec{p}_{\vec{k}} \\
          \vec{u}_{\vec{k}}
      \end{pmatrix}, 
\end{aligned}
\end{equation}
and perform the diagonalization in second quantization by replacing
\begin{equation}
\begin{aligned}
    & \begin{pmatrix}
          \vec{p}_{\vec{k}} \\
          \vec{u}_{\vec{k}}
      \end{pmatrix} =  \sqrt{\frac{\hbar}{2}}
      \sum_{\sigma} \frac{1}{\sqrt{\left| \omega_{\vec{k} \sigma} \right|}}
      \begin{pmatrix}
          \tilde{\vec{\mu}}_{\vec{k} \sigma} \\
          \tilde{\vec{\epsilon}}_{\vec{k} \sigma}
      \end{pmatrix} a_{\vec{k} \sigma},
      \\
    &
    \begin{pmatrix}
          \vec{u}_{\vec{k}} \\
          -\vec{p}_{\vec{k}}
      \end{pmatrix} = 
      \sqrt{\frac{\hbar}{2}}\sum_{\sigma}  \frac{1}{\sqrt{\left| \omega_{\vec{k} \sigma} \right|}}
      \begin{pmatrix}
         \tilde{\vec{\epsilon}}_{\vec{k} \sigma}  \\
         - \tilde{\vec{\mu}}_{\vec{k} \sigma}
      \end{pmatrix}  a_{\vec{k} \sigma}. 
\end{aligned}
\end{equation}
With these substitutions the Hamiltonian obtains a diagonal form
\begin{equation}
\begin{aligned}
    H_{\text{ph}} &=
      \frac{\mathrm{i}}{2}\sum_{\vec{k}}  \begin{pmatrix}
          \vec{u}_{\vec{k}} \\
          -\vec{p}_{\vec{k}}
      \end{pmatrix}^{\dagger}
    H_{\vec{k}}^{\text{eff}}
   \begin{pmatrix}
          \vec{p}_{\vec{k}} \\
          \vec{u}_{\vec{k}}
      \end{pmatrix} \\
      &= 
      \frac{\mathrm{i \hbar}}{4}\sum_{\vec{k}} \sum_{\sigma} \frac{1}{\left| \omega_{\vec{k} \sigma}\right|} \begin{pmatrix}
         \tilde{\vec{\epsilon}}_{\vec{k} \sigma}  \\
         -\tilde{\vec{\mu}}_{\vec{k} \sigma}
      \end{pmatrix}^{\dagger} H_{\vec{k}}^{\text{eff}} 
      \begin{pmatrix}
          \tilde{\vec{\mu}}_{\vec{k} \sigma} \\
          \tilde{\vec{\epsilon}}_{\vec{k} \sigma}
      \end{pmatrix}  a_{\vec{k} \sigma}^{\dagger} a_{\vec{k} \sigma} \\
      &=
       \frac{\mathrm{i \hbar}}{4}\sum_{\vec{k}} \sum_{\sigma} \frac{-2\mathrm{i} \omega_{\vec{k} \sigma}}{\left| \omega_{\vec{k} \sigma}\right|} 
       \tilde{\vec{\chi}}_{\vec{k} \sigma}^{\text{L}}  H_{\text{eff}} 
      \tilde{\vec{\chi}}_{\vec{k} \sigma}^{\text{R}}   a_{\vec{k} \sigma}^{\dagger} a_{\vec{k} \sigma} \\
      &= 
      \frac{\mathrm{\hbar}}{2}\sum_{\vec{k}} \sum_{\sigma} \text{sign}\left(\sigma \right)
       \omega_{\vec{k} \sigma}   a_{\vec{k} \sigma}^{\dagger} a_{\vec{k} \sigma} \\
       & = 
        \frac{\mathrm{\hbar}}{2}\sum_{\vec{k}} \sum_{\sigma} 
       \left|\omega_{\vec{k} \sigma}\right| a_{\vec{k} \sigma}^{\dagger} a_{\vec{k} \sigma} \\
       & = 
        \frac{\mathrm{\hbar}}{2}\sum_{\vec{k}} \sum_{\sigma>0} 
       \left|\omega_{\vec{k} \sigma}\right| a_{\vec{k} \sigma}^{\dagger} a_{\vec{k} \sigma} + \left|\omega_{\vec{k} -\sigma}\right| a_{\vec{k} -\sigma}^{\dagger} a_{\vec{k} -\sigma} \\
       &= 
        \frac{\mathrm{\hbar}}{2}\sum_{\vec{k}} \sum_{\sigma>0} 
       \left|\omega_{\vec{k} \sigma}\right| a_{\vec{k} \sigma}^{\dagger} a_{\vec{k} \sigma} + \left|\omega_{-\vec{k} -\sigma}\right| a_{-\vec{k} -\sigma}^{\dagger} a_{-\vec{k} -\sigma} \\
       & = 
        \frac{\mathrm{\hbar}}{2}\sum_{\vec{k}} \sum_{\sigma>0} 
       \left|\omega_{\vec{k} \sigma}\right| \left(a_{\vec{k} \sigma}^{\dagger} a_{\vec{k} \sigma} +  a_{\vec{k} \sigma}a_{\vec{k} \sigma}^{\dagger} \right) \\
       &= 
       \hbar\sum_{\vec{k}} \sum_{\sigma>0} 
       \left|\omega_{\vec{k}\sigma}\right| \left(a_{\vec{k} \sigma}^{\dagger} a_{\vec{k} \sigma} + \frac{1}{2} \right).
\end{aligned}
\label{eq:diagonalphonons}
\end{equation}

Any operator expressed in the 
$\vec{u}$-$\vec{p}$ basis can therefore be transformed to second quantization using eq.~\eqref{eq: second quantization} and its thermal expectation value computed using 
\begin{equation}
    \langle a_{\vec{k} \sigma}^{\dagger} a_{\vec{k} \sigma'}\rangle = \delta_{\sigma, \sigma'} f_{\vec{k} \sigma}, 
\end{equation}
where $f_{\vec{k} \sigma}= \left(\mathrm{e}^{ \beta \omega_{\vec{k} \sigma}} -1\right)^{-1}$ is the Bose-Einstein distribution function of the phonon mode with energy $\omega_{\vec{k} \sigma}$.
As shown in the Supplementary Material of Ref.~\cite{Lifa_phonon_De_Hass_effect}, by starting with the definition of the phonon angular momentum $\vec{J}$ given in Eq.~\eqref{eq: semiclassical OAM} and following the above logic, one arrives at Eq.~\eqref{eq: OAM thermal average}.
\end{section}

\begin{section}{Derivation of Phonon Dynamical Matrix}
\label{AP:Derivation of Phonon Dynamical Matrix}

Here, we derive the phonon dynamical matrix of the checkerboard lattice by utilizing the crystallographic symmetries of the lattice. 
 
To extract the phononic properties of a solid we consider a lattice in equilibrium described by the set of vectors $\{\vec{R}_{l\kappa}^0, l=0, 1, \dots, N; \kappa = \alpha_1, \alpha_2, \dots, \alpha_n\}$, where $\vec{R}_{l \kappa}^0 = \vec{R}_{l}^0 + \vec{\tau}_{\kappa}$, with $\vec{R}_{l}^0$  denoting the equilibrium position of the $l$-th unit cell and $\vec{\tau}_{\kappa}$ the position of the basis atom $\kappa$, and introduce deviations from the equilibrium configuration captured by the displacements $\vec{u}_{l \kappa}$ of each site. We then consider the potential of the lattice $V\left(\{\vec{R}_{l \kappa}^0 + \vec{u}_{l \kappa}\} \right)$ and expand it around the equilibrium set of coordinates. Keeping up to second-order terms in the displacements we arrive at the following expression for the potential 
\begin{equation}
    V = \frac{1}{2} \sum_{l, l'} \sum_{\kappa \kappa'} \sum_{\beta, \alpha} u_{l \kappa \alpha} \Phi_{\kappa' \beta }^{\kappa \alpha}\left(\vec{R}_l^0 - \vec{R}_{l'}^0 \right) u_{l' \kappa' \beta }. 
    \label{eq: phonon V in real space}
\end{equation}
We have dropped the constant term arising from the expansion and made use of the fact that the linear term in the displacements is zero in equilibrium. In Eq.~\eqref{eq: phonon V in real space}, $\alpha$ and $\beta$ are Cartesian coordinates, and $\Phi_{\kappa' \beta}^{\kappa \alpha}\left(\vec{R}_l^0 - \vec{R}_{l'}^0 \right) = \partial^2 V / \partial u_{l \kappa \alpha } \partial u_{l' \kappa' \beta}$ is the element of the force constant matrix $\Phi\left(\vec{R}_l^0 - \vec{R}_{l'}^0 \right)$ generated by the displacements $ u_{l \kappa \alpha }$ and $u_{l' \kappa' \beta}$. 
One then Fourier transforms the displacements in Eq.~\eqref{eq: phonon V in real space} to obtain the elements of the dynamical matrix as 
\begin{equation}
    \overline{D}_{\kappa' \beta}^{\kappa \alpha}\left(\vec{k} \right) = \frac{1}{N} \frac{1}{\sqrt{M_\kappa M_{\kappa'}}}\sum_{l l'} \mathrm{e}^{-\mathrm{i} \vec{k} \cdot \left(\vec{R}_{l \kappa}^0  - \vec{R}_{l' \kappa'}^0 \right)}
    \Phi_{\kappa'\beta}^{\kappa \alpha}\left(\vec{R}_l^0  - \vec{R}_{l'}^0\right). 
\label{eq: dynamical matrix definition}
\end{equation} 

\begin{figure}
\includegraphics[width=0.5\linewidth]{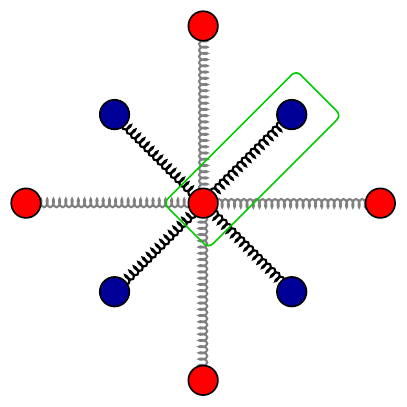}
\caption{Depiction of the blocks of the force constant matrix as springs for the first and second neighbors interactions considered in the text. The green rectangle encloses the chosen unit-cell with lattice vector $\vec{R}_l^0$, with $\vec{\tau}_{\text{A}} = 0$, $\vec{\tau}_{\text{B}} = \left(\vec{\hat{x}} + \vec{\hat{y}} \right)/2$ being the positions of the two atoms in the unit cell and $\vec{a}_1 = \vec{\hat{x}}$, $\vec{a}_2 = \vec{\hat{y}}$ the real lattice vectors. The spring inside the unit cell corresponds to the block $\Phi_{\text{B}}^{\text{A}}\left(\vec{R}_l^0, \vec{R}_l^0 \right)$ of the force constant matrix, while the rest of the blocks $\Phi_{\text{B}}^{\text{A}}\left(\vec{R}_l^0, \vec{R}_l^0 - \vec{a}_1 \right)$, $\Phi_{\text{B}}^{\text{A}}\left(\vec{R}_l^0, \vec{R}_l^0 - \vec{a}_2 \right)$, $\Phi_{\text{B}}^{\text{A}}\left(\vec{R}_l^0, \vec{R}_l^0 - \vec{a}_1 \right)$ and $\Phi_{\text{B}}^{\text{A}}\left(\vec{R}_l^0, \vec{R}_l^0 - \vec{a}_2 -\vec{a}_1\right)$ are related to $\Phi_{\text{B}}^{\text{A}}\left(\vec{R}_l^0, \vec{R}_l^0\right)$ via symmetries, as explained in the material. The blocks of the force constant matrix denoting second neighbors couplings between the A-atoms are denoted by gray springs. Since we have taken the second neighbors couplings between the B-atoms to be identical to these between the A ones, we have refrained from depicting them.
} 
\label{fig:phonon_springs}
\end{figure}

To construct the dynamical matrix, the explicit form of the force-constant matrix $\Phi \left(\vec{R}_l^0 - \vec{R}_{l'}^0 \right)$ must be specified. We follow the approach in Ref.~\cite{g4dl-1ff2} to derive constrains from crystallographic symmetries. Specifically, for a bond connecting the sites at $\vec{R}_{l \kappa}^0$ and $\vec{R}_{l' \kappa'}^0$, the crystallographic symmetries can be divided into two classes. The first, denoted by $G_{\mathrm{I}}$, consists of symmetries that leave the bond invariant, whereas the second, $G_{\mathrm{R}}$, contains symmetries that interchange the two sites. In both cases, the symmetry operation may be followed by a lattice translation. Consequently, the corresponding block of the force-constant matrix, $\Phi_{\kappa'}^{\kappa}\left(\vec{R}_l^0 - \vec{R}_{l'}^0 \right)$, satisfies \cite{g4dl-1ff2}
\begin{equation}
\begin{aligned}
    &  \Phi_{\kappa'}^{\kappa}\left(\vec{R}_l^0 - \vec{R}_{l'}^0 \right) = 
    \text{S}^\text{T} \Phi_{\kappa'}^{\kappa}\left(\vec{R}_l^0 - \vec{R}_{l'}^0 \right) \text{S} 
     \hspace{0.42 cm} \text{if $S \in G_\text{I}$}, \\
    & \Phi_{\kappa'}^{\kappa}\left(\vec{R}_l^0 - \vec{R}_{l'}^0 \right) = 
    \text{S}^\text{T} \Phi_{\kappa'}^{\kappa}\left(\vec{R}_l^0 - \vec{R}_{l'}^0 \right)^\text{T} \text{S} 
     \hspace{0.3 cm} \text{if $S \in G_\text{R}$}, 
\end{aligned}
\end{equation}
with $\text{S}$ being the representation of the considered symmetry.

On the checkerboard lattice with lattice vectors $\vec{a}_1 = a \vec{\hat{x}}$ and $\vec{a}_2 = a \vec{\hat{y}}$, the unit cell contains two atoms, $\text{A}$ and $\text{B}$, with basis positions $\vec{\tau}_A = 0$ and $\vec{\tau}_B = a \left(\vec{\hat{x}} + \vec{\hat{y}} \right) / 2$, respectively. The lattice constant $a$ is set to unity in the following. The point group of the lattice is $4/mmm$ and the following symmetries constrain the force matrix:
\begin{enumerate}
    \item $\mathcal{C}_{4z, \pm}$: a $90^{\circ}$ rotation around the $z$-axis perpendicular to the $xy$ plane,
    \item $\mathcal{C}_{2z, \pm}$: a $180^{\circ}$ rotation around the $z$-axis perpendicular to the $xy$ plane,
    \item $ \mathcal{M}_{xy}$: reflection about the line $x = y$,
    \item $\mathcal{M}_x$: reflection about the $x$-axis, 
\end{enumerate}
where the subscript ``$\pm$'' in the rotations refers to clockwise and counterclockwise for ``$+$'' and ``$-$'', respectively. Below, calligraphic symbols denote symmetry operations in the point group, while its corresponding representation is denoted by upright symbols. 

First, we consider the nearest-neighbor force constant matrix $\Phi^{\text{A}}_{\text{B}} \left(\vec{R}_l^0 - \vec{R}_{l'}^0 \right) \equiv \Phi^{\text{A}}_{\text{B}} \left(\vec{R}_l^0, \vec{R}_{l'}^0 \right)$ between sites on the $\text{A}$ and $\text{B}$ sublattice. The superscript refers to the atom on the A-sublattice residing in the unit cell at $\vec{R}_{l}^0$, while  the subscript refers to the atom on the B-sublattice residing in the unit cell at $\vec{R}_{l'}^0$. For the case that the sites reside in the same unit cell, $l' = l$, the corresponding block of the force constant matrix takes the general form  
    \begin{equation}
        \Phi_{\text{B}}^{\text{A}}\left(\vec{R}_l^0 , \vec{R}_l^0 \right) = 
        \begin{pmatrix}
            n_{11} & n_{12} \\
            n_{21} & n_{22} 
        \end{pmatrix}, 
        \label{eq: phonon AB general FM}
    \end{equation}
where $n_{11}, n_{12}, n_{21}$, and $n_{22}$ are parameters bearing spring-constant units. They are not independent.
Consider the $\mathcal{M}_{xy}$ mirror:
\begin{equation}
    \text{M}_{xy}^\text{T} \Phi_\text{B}^\text{A}\left(\vec{R}_l^0 , \vec{R}_l^0  \right)  \text{M}_{xy} = \Phi_\text{B}^\text{A}\left(\vec{R}_l^0 , \vec{R}_l^0  \right). 
    \label{eq: phonons Phi_AB constrains}
\end{equation}
It leads to the constraints $n_{11}=n_{22}$  and $n_{12} = n_{21}$. The remaining three nearest-neighbor force constant matrices can be obtained through rotations. 
One finds 
\begin{equation}
\Phi_\text{B}^\text{A}\left(\vec{R}_l^0 , \vec{R}_l^0  -\vec{a}_2 \right) 
 = \text{C}_{4z, +}^\text{T}  \Phi_\text{B}^\text{A}\left(\vec{R}_l^0 , \vec{R}_l^0 \right) \text{C}_{4z, +} = 
\begin{pmatrix}
 n_{11} & -n_{12} \\
-n_{12} & n_{11}
\end{pmatrix}, \\
\end{equation}
and 
\begin{equation}
    \begin{aligned}
        &
        \Phi_\text{B}^\text{A}\left(\vec{R}_l^0 , \vec{R}_l^0  -\vec{a}_1 \right) = \text{C}_{4z, -}^\text{T}  \Phi_\text{B}^\text{A}\left(\vec{R}_l^0 , \vec{R}_l^0 \right) \text{C}_{4z, -} = \Phi_\text{B}^\text{A}\left(\vec{R}_l^0 , \vec{R}_l^0  -\vec{a}_2 \right), \\
        & 
         \Phi_\text{B}^\text{A}\left(\vec{R}_l^0 , \vec{R}_l^0  -\vec{a}_1 -\vec{a}_2\right) = \text{C}_{2z, +}^\text{T}  \Phi_\text{B}^\text{A}\left(\vec{R}_l^0 , \vec{R}_l^0 \right) \text{C}_{2z, +} = \Phi_\text{B}^\text{A}\left(\vec{R}_l^0 , \vec{R}_l^0 \right).
    \end{aligned}
\end{equation}

Next, we consider the second-neighbor couplings that involve sites on the same sublattice. For simplicity, we assume that the force constant matrices between A-A and B-B second-neighbors are the same for bonds that lie in the same direction. Defining 
\begin{equation}
   \Phi_{\text{A}}^{\text{A}}\left(\vec{R}_l^0 , \vec{R}_l^0  + \vec{a}_1 \right)
   = \Phi_{\text{B}}^{\text{B}}\left(\vec{R}_l^0 , \vec{R}_l^0  + \vec{a}_1 \right)
   =
    \begin{pmatrix}
        \gamma_{11} & \gamma_{12} \\
        \gamma_{21} & \gamma_{22}
    \end{pmatrix} , 
    \label{eq: phonons FM AA-BB}
\end{equation}
one can show using the $\mathcal{M}_x$ mirror that the off-diagonal elements must vanish: $\gamma_{12} = \gamma_{21} = 0$.
 The remaining three  $\text{A-A}$ force constant matrices are related to $\Phi_{\text{A}}^{\text{A}}\left(\vec{R}_l^0 , \vec{R}_l^0  + \vec{a}_1 \right)$  via the $\mathcal{C}_{4z}$ and $\mathcal{C}_{2z}$ rotations
\begin{equation}
\begin{aligned}
   \Phi_{\text{A}}^{\text{A}}\left(\vec{R}_l^0 , \vec{R}_l^0  - \vec{a}_1 \right) 
    & = C_{2z,+}^{\text{T}}\Phi_{\text{A}}^{\text{A}}\left(\vec{R}_l^0 , \vec{R}_l^0  +  \vec{a}_1 \right) C_{2z,+} =
    \Phi_{\text{A}}^{\text{A}}\left(\vec{R}_l^0 , \vec{R}_l^0  + \vec{a}_1 \right), 
    \\
    \Phi_{\text{A}}^{\text{A}}\left(\vec{R}_l^0 , \vec{R}_l^0  + \vec{a}_2 \right) 
    & = C_{4z,-}^{\text{T}}\Phi_{\text{A}}^{\text{A}}\left(\vec{R}_l^0 , \vec{R}_l^0  +  \vec{a}_1 \right) C_{4z,-} =
    \begin{pmatrix}
        \gamma_{22} & 0 \\
        0 & \gamma_{11}
    \end{pmatrix}, 
    \\
    \Phi_{\text{A}}^{\text{A}}\left(\vec{R}_l^0 , \vec{R}_l^0  - \vec{a}_2 \right) 
    & = C_{2z,+}^{\text{T}}\Phi_{\text{A}}^{\text{A}}\left(\vec{R}_l^0 , \vec{R}_l^0  +  \vec{a}_2 \right) C_{2z,+} =
    \Phi_{\text{A}}^{\text{A}}\left(\vec{R}_l^0 , \vec{R}_l^0  + \vec{a}_2 \right). 
\end{aligned}
\end{equation}
This finding directly carries over to the $\Phi_{\text{B}}^{\text{B}}$ matrices. Finally, we compute also the part of the force constant matrix stemming from the self-forces to ensure the stability of the system. These can be obtained from the acoustic sum rule:
\begin{equation}
\begin{aligned}
     &\Phi_{\text{A}}^{\text{A}}\left(\vec{R}_l^0 , \vec{R}_l^0  \right) = -\sum_{\text{C}=\text{A,B}} \sum_{\vec{b}_j} \Phi_\text{C}^\text{A}\left(\vec{R}_l^0 , \vec{R}_l^0  -\vec{b}_j \right) = 
     \begin{pmatrix}
         -2 \left( \gamma_{11} + \gamma_{22} \right) - 4 n_{11} & 0 \\
         0 & -2 \left( \gamma_{11} + \gamma_{22} \right) - 4 n_{11}
     \end{pmatrix}, 
     \\
     & 
    \Phi_{\text{B}}^{\text{B}}\left(\vec{R}_l^0 , \vec{R}_l^0  \right) = -\sum_{\text{C}={A,B}} \sum_{\vec{b}_j} \Phi_\text{C}^\text{B}\left(\vec{R}_l^0 , \vec{R}_l^0  -\vec{b}_j \right) = 
    \Phi_{\text{A}}^{\text{A}}\left(\vec{R}_l^0 , \vec{R}_l^0  \right), 
\end{aligned}
\end{equation}
where $\vec{b}_j$ is the lattice vector that connects the site $\text{A} \left(\text{B} \right)$ to either any of its first neighbors $\text{B} \left(\text{A}\right)$ or to any of its second neighbors $\text{A} \left(\text{B}\right)$. 
The force constants blocks $\Phi_{\text{B}}^{\text{A}}\left(\vec{R}_l^0 , \vec{R}_l^0 - \vec{b}_j \right)$ and $\Phi_{\text{A}}^{\text{A}}\left(\vec{R}_l^0 , \vec{R}_l^0 - \vec{b}_j \right)$ are depicted as springs in Fig.~\ref{fig:phonon_springs}.
Using these force constant matrix elements in Eq.~\eqref{eq: dynamical matrix definition}, we obtain the elements of the dynamical matrix in Eqs.~\eqref{eq: D elements FN} and \eqref{eq: second neighbors D elements}. 
\end{section}

\end{document}